\documentclass[review]{elsarticle}

\usepackage{lineno,hyperref}
\usepackage{subcaption}
\usepackage{graphicx}
\usepackage{siunitx}
\usepackage{numprint}
\usepackage[]{xcolor}
\usepackage{amsmath}
\usepackage{amsthm}
\usepackage{amssymb}

\newcommand{\vect}[1]{\boldsymbol{\mathbf{#1}}}
\usepackage[utf8]{inputenc}
\modulolinenumbers[5]

\definecolor{darkgreen}{rgb}{0.0, 0.35, 0.0}
\newcommand{\reva}[1]{\textcolor{black}{#1}}
\newcommand{\revb}[1]{\textcolor{black}{#1}}

\journal{}

\begin{document}

\begin{frontmatter}

\title{Performance of the BGSDC integrator for computing fast ion trajectories in nuclear fusion reactors\tnoteref{mytitlenote}}
\tnotetext[mytitlenote]{This work has been carried out within the framework of the EUROfusion Consortium and has received funding from the Euratom research and training programme 2014-2018 and 2019-2020 under grant agreement No 633053. The views and opinions expressed herein do not necessarily reflect those of the European Commission.
This work was also supported by the Engineering and Physical Sciences Research Council EPSRC under grant EP/P02372X/1 ``A new algorithm to track fast ions in fusion reactors''.}

\author[leeds]{Krasymyr Tretiak\corref{cor}}
\ead{k.tretiak@leeds.ac.uk}

\author[ccfe]{James Buchanan}
\ead{james.buchanan@ukaea.uk}

\author[ccfe]{Rob Akers}
\ead{Rob.Akers@ukaea.uk}

\author[tuhh]{Daniel Ruprecht}
\ead{ruprecht@tuhh.de}

\address[leeds]{School of Mathematics, University of Leeds, United Kingdom}
\address[ccfe]{CCFE, Culham Science Centre, Abingdon, United Kingdom}
\address[tuhh]{Lehrstuhl Computational Mathematics, Institut für Mathematik, Technische Universität Hamburg, Germany}

\cortext[cor]{Corresponding author}

\begin{abstract}
Modelling neutral beam injection (NBI) in fusion reactors requires computing the trajectories of large ensembles of particles.
Slowing down times of up to one second combined with nanosecond time steps make these simulations computationally very costly.
This paper explores the performance of BGSDC, a new numerical time stepping method, for tracking ions generated by NBI in the DIII-D and JET reactors.
BGSDC is a high-order generalisation of the Boris method, combining it with spectral deferred corrections and the Generalized Minimal Residual method GMRES.
Without collision modelling, where numerical drift can be quantified accurately, we find that BGSDC can deliver higher quality particle distributions than the standard Boris integrator at comparable computational cost or comparable distributions at lower computational cost.
With collision models, quantifying accuracy is difficult but we show that BGSDC produces stable distributions at larger time steps than Boris.
\end{abstract}


\end{frontmatter}

\section{Introduction}
Computer simulations are a critical tool for the design and operation of fusion reactors~\cite{ArtaudEtAl2010}.
Numerical computation of the trajectories of fast ions generated, for example, from neutral beam injection\footnote{\revb{Since charged particles would be deflected by the magnetic field before reaching the plasma, a fusion reactor is heated by injecting neutral particles. Once inside the plasma, these neutral particles undergo collisions with the bulk plasma and become ionised.}} is important to minimise wall loads and energy loss from ions escaping magnetic confinement~\cite{HemsworthEtAl2009}.
At their core, particle trackers integrate the Lorentz equations
\begin{subequations}
\label{eq:lorentz}
\begin{align}
    \dot{\vect{x}}(t) &= \vect{v}(t) \\
    \dot{\vect{v}}(t) &= \alpha \left[ \vect{E}(\vect{x}) + \vect{v} \times \vect{B}(\vect{x}) \right] =: \vect{f}(\vect{x}, \vect{v}).
\end{align}
\end{subequations}
for a large ensemble of particles and use the resulting trajectories to generate statistical quantities like wall load.
Note that fast ions in fusion reactors are typically not fast enough to require consideration of relativistic effects and so the non-relativistic Lorentz equations can be used.

Fast ions interact with the plasma and deposit energy, thus heating it.
To compute steady-state distributions, trajectories need to be computed until the fast ions loose their energy through collisions and thermalise.
This slowing down time over which an ensemble of trajectories needs to be calculated depends on the energy of the ions when injected and the bulk plasma parameters.
For DIII-D, the required simulation time is around \SI{0.1}{\second}.
For the larger JET, it is around \SI{1}{\second}.
Because resolving gyro effects requires time steps of the order of nanoseconds, simulations involve many millions of time step, leading to substantial computational cost and thus long solution times.
Simulating a full ensemble of fast ions in JET until thermalisation takes several days, despite making use of modern GPU clusters.

Some models, for example NUBEAM~\cite{Goldston1981,Pankin2004} or OFMC~\cite{ShinoharaEtAl2016}, use a guiding centre approximation where gyro effects are neglected or only included once a particle is near the walls.
This allows \reva{taking} larger time step and thus reduces computational cost.
However, taking orbit effects into consideration is important to generate realistic wall loads~\cite{Snicker_2012}.
Other particle tracking codes, for example ASCOT~\cite{Hirvijoki2014} and LOCUST~\cite{AkersEtAl2012}, compute the full equations with gyro effects.
For numerical methods, the only choices available are the Boris integrator\revb{~\cite{Boris1970}, that is, a velocity Verlet or a Leapfrog scheme with a geometrical trick to resolve the implicit dependence on velocity\footnote{ASCOT seems to use the Verlet variant whereas LOCUST uses the Leapfrog version}}, and the Cash-Karp Runge-Kutta 4(5) method~\cite{CashKarp1990}.
LOCUST also features a  mover based on Strang splitting, which is very similar to Boris but avoids some issues around loss of accuracy in cylindrical coordinates~\cite{Delzanno2013}.
There seems to be agreement that due to its significant energy drift, RK4(5) is less \revb{accurate} than Boris and so the latter is typically used.
Although Boris is surprisingly efficient~\cite{QinEtAl2013}, long solution times remain an issue: \revb{the time for injected fast particles to reduce their energy to that of the thermal plasma via collisions, called the collisional slowing down time, is on the order of seconds for a reactor like ITER. 
Resolving cyclotron dynamics requires timesteps on the order of nanoseconds so that simulations require computing around $10^9$ steps. 
Even on CCFE's dedicated GPU cluster, such a simulation takes days to complete}.
While other algorithms have been proposed for solving the Lorentz equations~\cite{HeEtAl2015,QuandtEtAl2007,Tao2016,HeEtAl2016,HairerLubich2017,Umeda2018}, they have so far not been adopted for fast ion tracking.

Tretiak and Ruprecht~\cite{TretiakRuprecht2019} introduce BGSDC, a new high-order algorithm for solving the Lorentz equations based on a combination of spectral deferred corrections, a Generalized Minimal Residual (GMRES) iteration and the Boris integrator.
They show improvements in computational performance over Boris for individual particle trajectories in a mirror trap as well as trapped and passing particles in a Solev'ev equilibrium.
This paper extends their results by demonstrating performance for realistic test cases, studying practically relevant, aggregate quantities to assess quality of solutions instead of individual trajectories.
Instead of analytically given idealized magnetic fields, we use toroidally axisymmetric free-boundary equilibria that are more closely representative of the actual conditions on DIII-D and JET (i.e. with a poloidal flux map that is expressed as a bicubic spline so as to generate a divertor X-point).
In the non-collisional case, we track particle ensembles corresponding to real neutral beam injection (NBI) scenarios~\cite{HemsworthEtAl2009, AsuntaEtAl2015} and assess statistical distribution of particle drift in the ensemble in contrast to exploring accuracy of individual trajectories.
We show that BGSDC delivers better distributions with smaller \reva{mean and} standard deviation than Boris and, if \reva{means and} standard deviations of the order of micrometers are desired, can deliver them with less computational work.
Then, we investigate the case where models for collisions of fast ions with the plasma are active. 
While the stochastic nature of these results makes a quantitative assessment with respect to work versus precision difficult, we demonstrate that BGSDC can deliver similar results as Boris with larger time steps.
Delivering additional evidence to show that these gains are enough to also deliver computational gains in the collisional case will require a more developed framework to quantify accuracy for the generated statistical distributions as well as substantially more computational results and is left for future work.
Our results demonstrate that there can be a computational benefit from using BGSDC or other particle trackers with order of accuracy higher than two, in particular for high fidelity simulations with tight accuracy requirements.

All results were generated with the BGSDC implementation~\cite{CodeBGSDC} that is now part of the LOCUST code and the ITER Integrated Modelling \& Analysis Suite (IMAS)~\cite{IMAS_2015}.

\section{Methodology}

The GMRES-accelerated Boris-SDC algorithm (or BGSDC for short) is described in detail by Tretiak et al.~\cite{TretiakRuprecht2019} whereas its predecessor, without GMRES-acceleration, was introduced by Winkel et al.~\cite{WinkelEtAl2015}.
This original Boris-SDC combined the Boris algorithm introduced by Boris in 1970~\cite{Boris1970} with spectral deferred corrections (SDC) introduced by Dutt et al. in 2000~\cite{DuttEtAl2000} to generalize it to higher order.
BGSDC incorporates a GMRES-based convergence accelerator for SDC, introduced by Huang et al. in 2006~\cite{HuangEtAl2006} for first order problems, that leads to \revb{faster convergence to the collocation solution and thus} improved long-term energy stability.
All simulation results reported in this paper were generated using a BGSDC implementation in the GPU-accelerated LOCUST particle tracking code developed at CCFE.

\subsection{Collocation Methods}
In essence, BGSDC is an iterative solver for a collocation method.
Over one time step $[t_n, t_{n+1}]$, the Lorentz equations~\eqref{eq:lorentz} written in integral form become
\begin{subequations}
\label{eq:lorentz_integral}
\begin{align}
    \vect{x}(t) &= \vect{x}_0 + \int_{t_n}^t \vect{v}(s)~ds \\
    \vect{v}(t) &= \vect{v}_0 + \int_{t_n}^t \vect{f}(\vect{x}(s), \vect{v}(s))~ds
\end{align}
\end{subequations}
with $\vect{x}_0$, $\vect{v}_0$ being approximations of position and velocity at time $t_n$ brought forward from the previous step.
Note that we consider only the case where the electric and magnetic field vary in space but not in time, but the method can easily be generalised to the non-autonomous case.
Typically, the Boris method is based on a Leapfrog discretization of the differential form of the Lorentz equations~\eqref{eq:lorentz}.
Here, we use a variant based on the Velocity-Verlet method instead
\begin{subequations}
\begin{align}
    \vect{x}_{n+1} &= \vect{x}_n + \Delta t \left( \vect{v}_n + \frac{\Delta t}{2} \vect{f}(\vect{x}_n, \vect{v}_n) \right)\\
    \vect{v}_{n+1} &= \vect{v}_n + \frac{\Delta t}{2} \left( \vect{f}(\vect{x}_n, \vect{v}_n) +  \vect{f}(\vect{x}_{n+1}, \vect{v}_{n+1})  \right),
\end{align}
    \label{eq:velocity_verlet}
\end{subequations}
see Tretiak et al. for a discussion~\cite{TretiakRuprecht2019}.
A geometric trick was introduced by Boris in 1970~\cite{Boris1970} to avoid the seemingly implicit dependence on $\vect{v}_{n+1}$.
Birdsall and Langdon give a detailed description~\cite[Section 4-4]{Birdsall1985}.

Collocation methods discretise the integral form~\eqref{eq:lorentz_integral} using numerical quadrature with nodes $t_n \leq \tau_1 < \ldots <\tau_M \leq t_{n+1}$ instead of the differential form.
Approximate values $\vect{x}_{\textrm{new}}$ and $\vect{v}_{\textrm{new}}$ at time $t_{n+1}$ are computed via
\begin{subequations}
\begin{align}
    \vect{x}_{\textrm{new}} &= \vect{x}_0 + \sum_{m=1}^{M} q_{m} \vect{v}_m \\
    \vect{v}_{\textrm{new}} &= \vect{v}_0 + \sum_{m=1}^{M} q_m \vect{f}(\vect{x}_m, \vect{v}_m)
\end{align}
\end{subequations}
where $q_m$ are quadrature weights while $\vect{x}_m$, $\vect{v}_m$ are approximations to position and velocity at the quadrature nodes $\tau_m$.
These are equivalent to the stages of a collocation method, an implicit Runge-Kutta method with a dense Butcher tableau~\cite[Theorem 7.7]{hairer_nonstiff}, and can be computed or approximated by solving the stage equations
\begin{subequations}
\label{eq:quadrature}
\begin{align}
    \vect{x}_m &= \vect{x}_0 + \sum_{j=1}^{M} q_{m, j} \vect{v}_j \\
    \vect{v}_m &= \vect{v}_0 + \sum_{j=1}^{M} q_{m,j} \vect{f}(\vect{x}_j, \vect{v}_j).
\end{align}    
\end{subequations}
Depending on the choice of quadrature nodes, collocation methods can have a range of desirable properties.
\reva{When applied to Hamiltonian systems in canoncial coordinates,} they are symplectic for Gauss-Legendre nodes~\cite[Theorem 16.5]{hairer_nonstiff} and symmetric for Gauss-Lobatto nodes~\cite[Theorem 8.9]{hairer_geometric} as well as A-stable~\cite[Theorem 12.9]{hairer_stiff}.
\reva{Symplectic integration in non-canonical coordinates is extremely challenging.}
By combining the stages $\vect{x}_m$, $\vect{v}_m$ into one vector $\vect{U}$, the stage equations~\eqref{eq:quadrature} can compactly be written as a nonlinear system
\begin{equation}
    \label{eq:collocation}
    \vect{U} - \vect{Q} \vect{F}(\vect{U}) = \vect{U}_0
\end{equation}
with $\vect{Q}$ containing quadrature weights,
\begin{equation}
\vect{F}(\vect{U}) = \left(\vect{v}_1, \ldots, \vect{v}_M, \vect{f}(\vect{x}_1, \vect{v}_1), \ldots, \vect{f}(\vect{x}_M, \vect{v}_M) \right),
\end{equation}
and $\vect{U}_0$ containing repeated entries of $\vect{x}_{0}$ and $\vect{v}_0$. 
See Winkel et al. for details~\cite{WinkelEtAl2015}.

\subsection{Boris-GMRES-SDC (BGSDC)}
Spectral deferred corrections use an iteration based on a low order method to solve Eq.~\eqref{eq:collocation}.
For first order problems, this is typically an implicit or implicit-explicit Euler~\cite{DuttEtAl2000,Minion2003}.
For second order problems, velocity-Verlet integration or, in the special case of the Lorentz equations, the Boris integrator can be used~\cite{WinkelEtAl2015}.
If the collocation problem~\eqref{eq:collocation} is linear and thus, in a slight abuse of notation, reads
\begin{equation}
    \left( \vect{I} - \vect{Q} \vect{F} \right) \vect{U} = \vect{U}_0,
\end{equation}
 one can apply a preconditioned GMRES iteration instead of SDC to solve it~\cite{HuangEtAl2006}.
The key point is that GMRES does not require assembly of the system matrix but only a function that applies $\vect{I} - \vect{Q} \vect{F}$ to a given vector $\vect{U}$.
This simply means computing Eq.~\eqref{eq:quadrature} for $m=1, \ldots, M$.
However, to improve performance, it is advisable to use a preconditioner.
In the GMRES interpretation, the low order base method (Euler in the first order case, Boris in the second order case) can be understood as a preconditioner, modifying the original collocation system~\eqref{eq:collocation} to
\begin{equation}
    \left( \vect{I} - \vect{Q}_{\Delta} \vect{F} \right)^{-1} \left( \vect{I} - \vect{Q} \vect{F} \right) \vect{U} = \left( \vect{I} - \vect{Q}_{\Delta} \vect{F} \right)^{-1} \vect{U}_0,
\end{equation}
where $\vect{Q}_{\Delta t}$ has a block structure with each block being a lower triangular matrix~\cite{TretiakRuprecht2019}.
To apply GMRES to the preconditioned problem, a second function is required that can solve
\begin{equation}
    \label{eq:preconditioner}
    \left( \vect{I} - \vect{Q}_{\Delta} \vect{F} \right) \vect{U} = \vect{b}
\end{equation}
for a given right-hand side $\vect{b}$~\cite{Kelley1995}.
Because of the special structure of $\vect{Q}_{\Delta}$, this can be done in a sweep-like fashion, very similar to the sweeps in the original variant of SDC.
When using $M=3$ nodes, solving Eq.~\eqref{eq:preconditioner} amounts to a block-wise solve of
\begin{align*}
        \begin{bmatrix} \vect{x}_1 \\ \vect{x}_2 \\ \vect{x}_3 \end{bmatrix} - \begin{bmatrix} 0 & 0 & 0 \\ \Delta \tau_2 \vect{I} & 0 & 0 \\ \Delta \tau_2 \vect{I} & \Delta \tau_3 \vect{I} & 0  \end{bmatrix} \begin{bmatrix} \vect{v}_{1} \\ \vect{v}_{2} \\ \vect{v}_{3} \end{bmatrix} -
        \frac{1}{2} \begin{bmatrix} 0 & 0 & 0 \\ \Delta \tau_2^2 \vect{I} & 0 & 0 \\ \Delta \tau_2^2 \vect{I} & \Delta \tau_3^2 \vect{I} & 0 \end{bmatrix} \begin{bmatrix} \vect{F}(\vect{x}_1, \vect{v}_1) \\ \vect{F}(\vect{x}_2, \vect{v}_2) \\ \vect{F}(\vect{x}_3, \vect{v}_3) \end{bmatrix} = \begin{bmatrix} \vect{b}_{1} \\ \vect{b}_{2} \\ \vect{b}_{3} \end{bmatrix} 
\end{align*}
and
\begin{align*}
        \begin{bmatrix} \vect{v}_1 \\ \vect{v}_2 \\ \vect{v}_3 \end{bmatrix} - \frac{1}{2} \begin{bmatrix}  \Delta \tau_1 \vect{I} & 0 & 0 \\ (\Delta \tau_1 + \Delta \tau_2) \vect{I} & \Delta \tau_2 \vect{I} & 0 \\ (\Delta \tau_1 + \Delta \tau_2) \vect{I} & (\Delta \tau_2 + \Delta \tau_3) \vect{I} & \Delta \tau_3 \vect{I} \end{bmatrix} \begin{bmatrix} \vect{F}(\vect{x}_1, \vect{v}_1) \\ \vect{F}(\vect{x}_2, \vect{v}_2) \\ \vect{F}(\vect{x}_3, \vect{v}_3) \end{bmatrix} = \begin{bmatrix} \vect{b}_{4} \\ \vect{b}_{5} \\ \vect{b}_{6} \end{bmatrix}
\end{align*}
via first computing
\begin{subequations}
\label{eq:block_solve_1}
\begin{align}
        \vect{x}_1 &= \vect{b}_{1} \\
        \vect{v}_1 &= \vect{b}_{4} + \frac{1}{2} \Delta \tau_1 \vect{F}(\vect{x}_1, \vect{v}_1),
\end{align}        
\end{subequations}
then
\begin{subequations}
\label{eq:block_solve_2}
\begin{align}        
    \vect{x}_2 &= \vect{b}_{2} + \Delta \tau_2 \vect{v}_1 + \frac{1}{2} \Delta \tau_2^2 \vect{F}(\vect{x}_1, \vect{v}_1)\\
        \vect{v}_2 &= \vect{b}_{5} + \frac{1}{2} \left( \Delta \tau_1 + \Delta \tau_2 \right) \vect{F}(\vect{x}_1, \vect{v}_1) + \frac{1}{2} \Delta \tau_2 \vect{F}(\vect{x}_2, \vect{v}_2),
\end{align}        
\end{subequations}
and finally
\begin{subequations}
\label{eq:block_solve_3}
\begin{align}        \vect{x}_3 &= \vect{b}_{3} + \Delta \tau_2 \vect{v}_1 + \Delta \tau_3 \vect{v}_2 + \frac{1}{2} \Delta \tau_2^2 \vect{F}(\vect{x}_1, \vect{v}_1) + \frac{1}{2} \Delta \tau_3^2 \vect{F}(\vect{x}_2, \vect{v}_2) \\
        \vect{v}_3 &= \vect{b}_{6} + \frac{1}{2} \left( \Delta \tau_1 + \Delta \tau_2 \right) \vect{F}(\vect{x}_1, \vect{v}_1) + \frac{1}{2} \left( \Delta \tau_2 + \Delta \tau_3 \right) \vect{F}(\vect{x}_2, \vect{v}_2) + \frac{1}{2} \Delta \tau_3 \vect{F}(\vect{x}_3, \vect{v}_3)
\end{align}        
\end{subequations}
using Boris' trick to compute the velocities~\cite{TretiakRuprecht2019}.
Note that for a single particle, where $\vect{F} = \vect{f}$, with given initial values $\vect{x}_n$, $\vect{v}_n$ at the beginning of the time step, setting $\vect{b}_1 := \vect{x}_n + \Delta \tau_1 \left( \vect{v}_n + \frac{\Delta \tau_1}{2} \vect{F}(\vect{x}_n, \vect{v}_n) \right)$ and $\vect{b}_4 := \vect{v}_n + \frac{\Delta \tau_1}{2} \vect{F}(\vect{x}_n, \vect{v}_n)$ means that~\eqref{eq:block_solve_1} becomes identical to~\eqref{eq:velocity_verlet} with $\Delta t = \Delta \tau_1$.
For specific choices of $\vect{b}_2$, $\vect{b}_3$, $\vect{b}_5$ and $\vect{b}_6$, the steps~\eqref{eq:block_solve_1},~\eqref{eq:block_solve_2} and~\eqref{eq:block_solve_3} correspond to a total of three Boris steps~\eqref{eq:velocity_verlet} with step sizes $\Delta \tau_1$, $\Delta \tau_2$ and $\Delta \tau_3$.
However, to apply the GMRES procedure, the algorithm is modified to accept any input for $\vect{b}$.
This procedure is straightforward to generalize for any number of nodes $M$.

For nonlinear collocation problems, it is possible to apply a Newton iteration and use GMRES-SDC to solve the linear inner problems.
For the case where the nonlinearity is due to the magnetic field depending on $\vect{x}$, this was found not to be competitive, \revb{as the gains in convergence speed could not offset the significant added computational cost}~\cite{TretiakRuprecht2019}.
Instead, we linearize the collocation problem by freezing the magnetic field after the first sweep provides values $\vect{x}_m^0$ for $m = 1, \ldots, M$ by approximating
\begin{equation*}
    \vect{F}(\vect{U}) \approx \vect{F}_{\text{lin}}(\vect{X}^0)(\vect{U}) =: \begin{pmatrix} \vect{v}_1 \\ \vdots \\ \vect{v}_M \\ \vect{f}(\vect{x}_1^0, \vect{v}_1) \\ \vdots \\  \vect{f}(\vect{x}_M^0, \vect{v}_M) \end{pmatrix},    
\end{equation*}
that is, the electric and magnetic field are evaluated at the approximate position $\vect{x}_m^0$ from the first sweep instead of using the positions $\vect{x}_m$ in the argument $\vect{U}$.
The linearized system is preconditioned using $\left( \vect{I} - \vect{Q}_{\Delta} \vect{F}_{\text{lin}} \right)$ and solved with GMRES.
The result is then corrected to account for the nonlinearity by a small number of computationally cheap discrete Picard iterations
\begin{equation}
    \vect{U}^{k+1} = \vect{U}_0 + \vect{Q} \vect{F}(\vect{U}^{k}).
\end{equation}
If the fields only changes weakly over a single time step, the solution of the linearized collocation problem will be very close to the nonlinear solution and the Picard iteration converges quickly in very few iterations.
BGSDC($k$, $l$) refers to this combination of $k$ GMRES-SDC iterations for the linearized collocation problem followed by $l$ Picard iterations for the fully nonlinear problem.

\reva{Note that ideally we would want to fix an overall tolerance and have the algorithm set $k$ and $l$ to reach this tolerance.
However, the combination of GMRES iterations for the linearized collocation problem with nonlinear Picard iteration makes this difficult.
Currently we do not have a good heuristic how to set $k$ for a given nonlinear tolerance so that the Picard iteration converges reasonably fast but the GMRES iteration does not significantly oversolve the problem.
Finding a way to do that efficiently will require a more comprehensive numerical exploration and is left for future work.}

\subsection{LOCUST-GPU implementation}
LOCUST stands for "Lorentz Orbit Code for Use in Stellarators and Tokamaks"~\cite{AkersEtAl2012,locust_2016}.
It is a software platform for solving efficiently the Lorentz equations of motion in the presence of a collision operator that models small angle Coulomb scattering. 
Kernels are instantiated upon Nvidia GPU hardware as PGI CUDA Fortran kernels which allows millions of Monte Carlo markers to be tracked in a typical simulation. 
LOCUST is being used extensively to design plasma facing components, e.g. for MAST-U~\cite{MilnesEtAl2015}, and for studying the physics of fast ion distribution and loss due to Neoclassical Tearing Modes and the application of ELM Control Coils for ITER. 
It is part of the EUROfusion HALO programme, where it is used to study the implications of finite gyro-radius effects, e.g. for Toroidal Alfven Eigenmode (TAE) activity in Tokamak plasma~\cite{FitzgeraldEtAl2020}.

\section{Numerical Results}
We compare BGSDC against the Leapfrog-based staggered Boris method for tracking fast ions generated by NBI in both the DIII-D and Joint European Torus (JET) tokamak.
For both reactors, we study the deterministic case without models for the collision of fast ions with the plasma and the stochastic case with collosion models.
In the collionless case, we launch a particle ensemble corresponding to a NBI shot and use the \revb{mean and} standard deviation of the numerical drift distribution \revb{for individual runs} to compare the quality of both integrators.
In the presence of collisions with the plasma, stochasticity makes particle drift measurements of trajectories meaningless and a very large number of markers would be required to get statistically converged results.
Therefore, to focus on the impact of the numerical error and minimise the spread of ensembles, we instead study many trajectory realizations for the same particle with identical initial conditions and analyse the effect that time step size has on the resulting distribution function profiles.
\paragraph{Orbit drift}
\reva{LOCUST measures orbit drift by recording the maximum and minimum major radius at midplane crossing of the gyro-centre for the first 10\% of simulation time.
At this time, all orbits will have passed through the midplane multiple times.
The gyro-centre is computed for a uniform field, which leads to the presence of some residual gyro-oscillation.
This is the reason that multiple midplane traversals are necessary to find reliable values for minimum and maximum major radius.
The procedure is then repeated for the last 10\% of simulation time and drifts are computed by taking the differences of the measured maximum and minimum major radius.}

\subsection{Results for the DIII-D tokamak: non-collisional case}
\begin{figure}[t]
    \centering
    \includegraphics[width=0.95\textwidth]{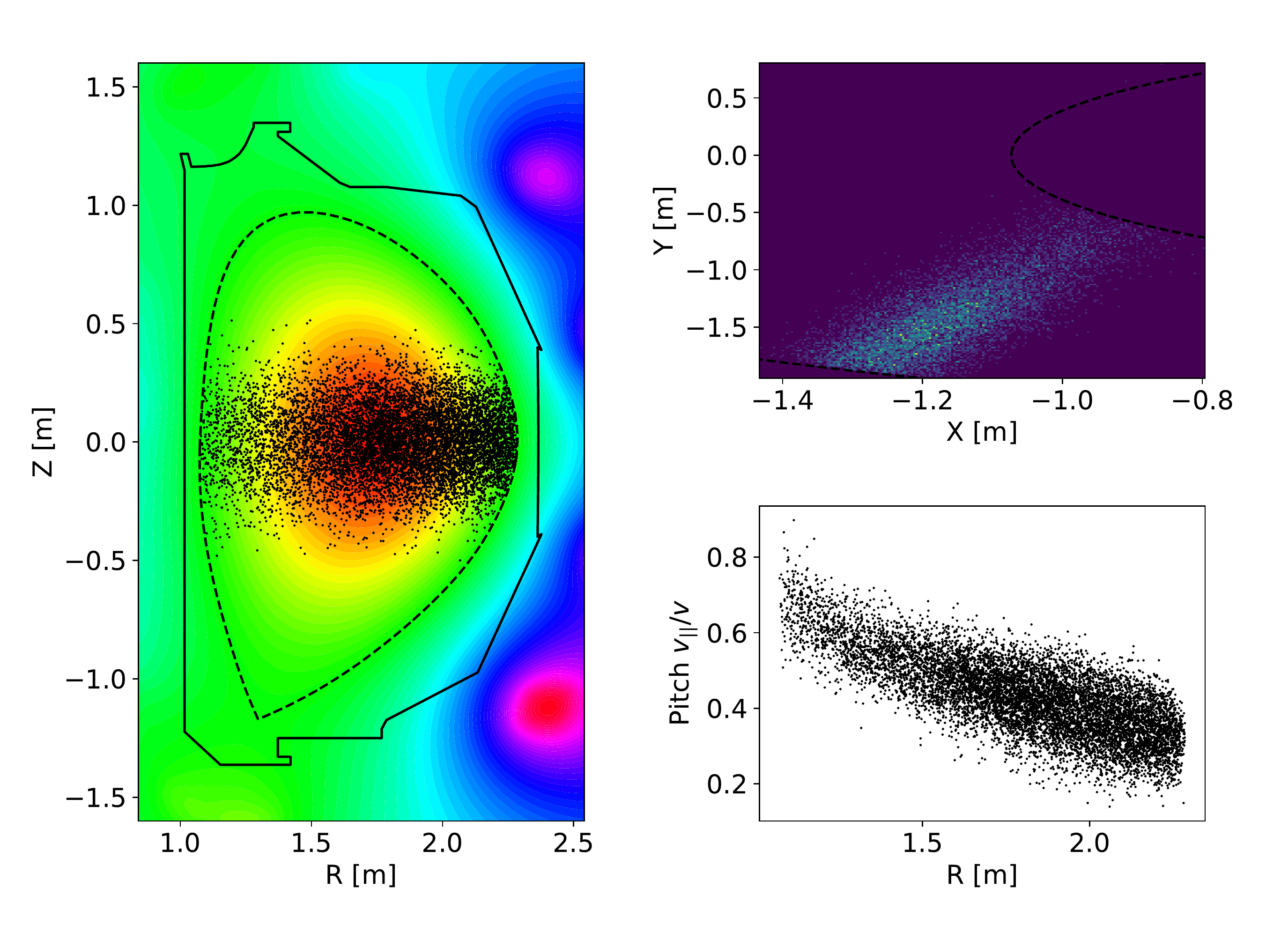}
    \caption{\reva{Configuration of the simulated NBI shot. Poloidal profile (left) and top-down view (upper right) of particles in birth list superimposed with the poloidal flux function. The dashed line is the last closed flux surface (separatrix) while the solid line is the DIII-D reactor wall. Distribution of pitch versus major radius (lower right).}}
    \label{fig:nbi_shot}
\end{figure}

\begin{figure}[t]
    \centering
    \includegraphics[width=0.95\textwidth]{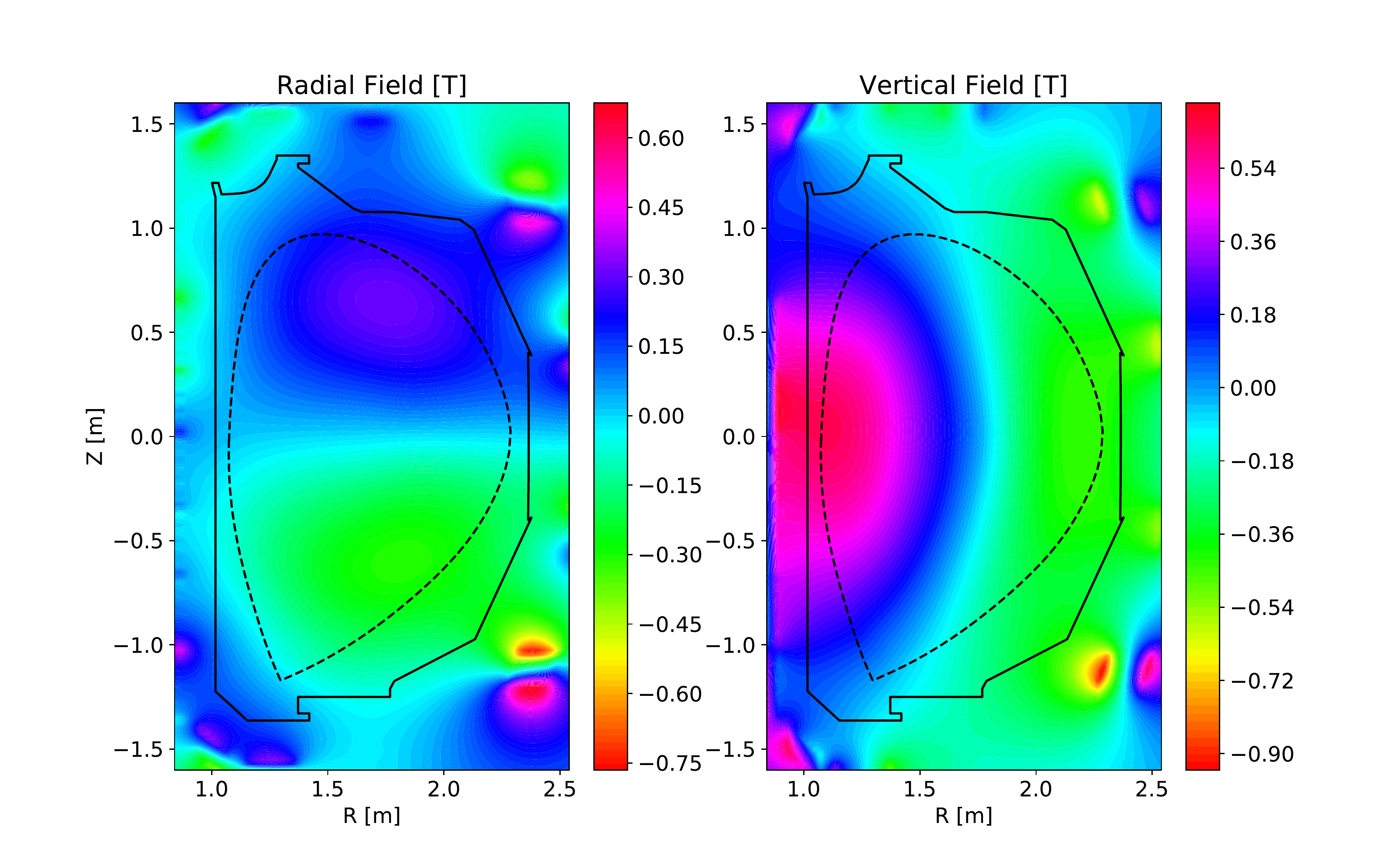}
    \caption{\reva{Radial (left) and vertical (right) component of the DIII-D magnetic field.}}
    \label{fig:magnetic_field}
\end{figure}

We compare Boris and BGSDC for full orbit simulations in a 2D equilibrium for the \text{DIII-D} NBI shot \reva{sketched in Fig.~\ref{fig:nbi_shot}}.
This setup has been used, for example, in fast ion transport studies for applied 3D magnetic perturbations in DIII-D~\cite{Van_Zeeland_2015}. 
The fast particle birth list contains \SI{10000}{} unique \reva{Deuterium ions} derived from an NBI source that is injected counter to the plasma current. 
\reva{If more particles are simulated, LOCUST creates copies of those particles to fill in information.
To match the GPU architecture that LOCUST runs on, we model $64 \times 64 \times 16 = \SI{65636}{}$ particles, since we run $64$ threads per block, with $64$ blocks per grid on $16$ GPUs.}
\reva{Fig.~\ref{fig:magnetic_field} shows the radial (left) and vertical (right) component of the DIII-D magnetic field.
LOCUST uses bicubic splines of the poloidal flux to calculate field components from the equilbrium data.}

The duration of all runs is $\SI{100}{\milli\second}$ and, for simplicity, we do not include plasma facing components (PFC) in our simulations \revb{and instead stop particles when they hit the wall of the equilibrium computational box.}
\revb{While inclusion of PFC makes the detection of where a particle hits the wall much more difficult, they don't impact performance of the numerical integration scheme.}

\begin{figure}[t]
	\begin{subfigure}{0.245\textwidth}	
	\includegraphics[width=\linewidth]{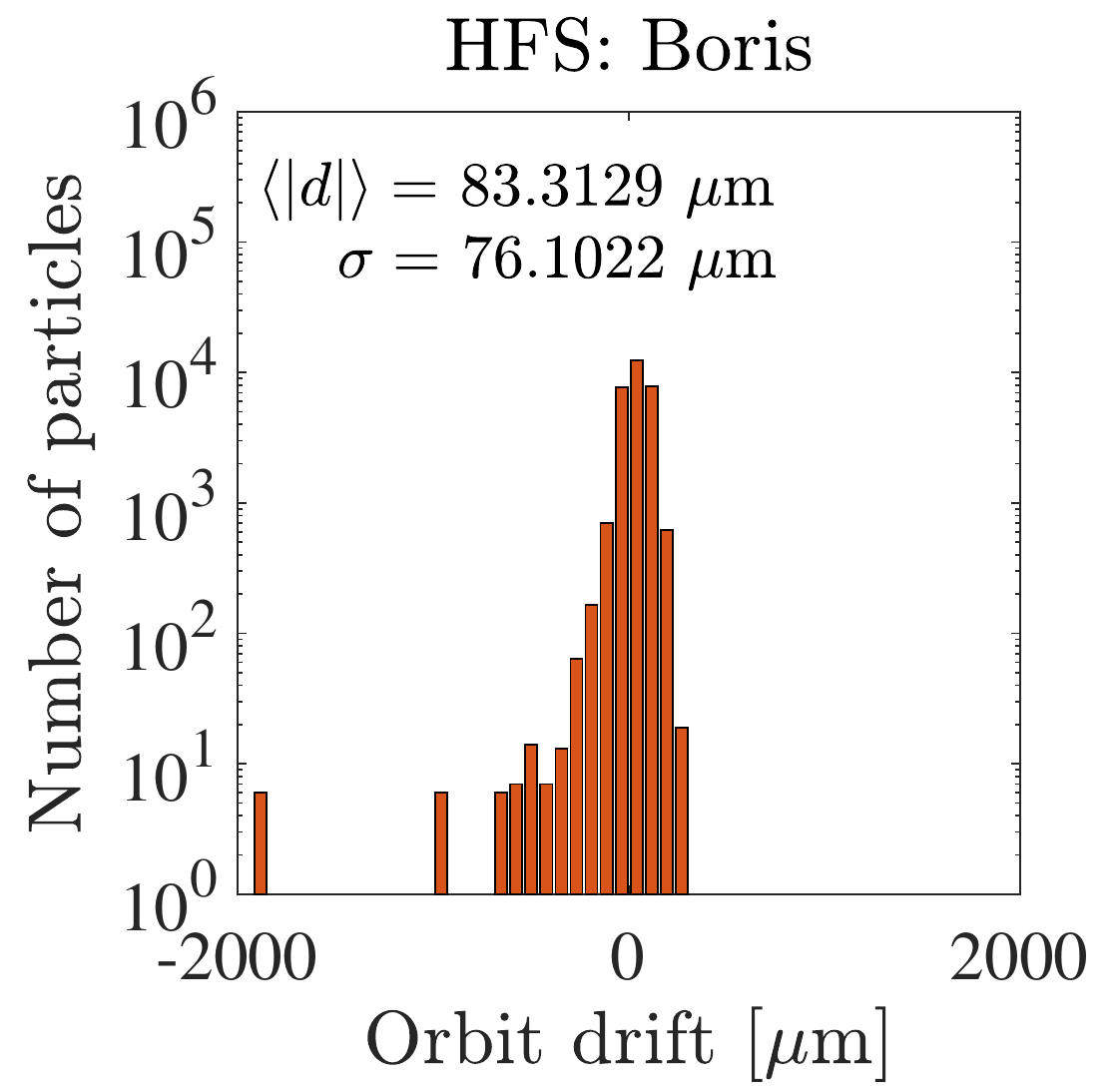}
	\end{subfigure}
	\begin{subfigure}{0.245\textwidth}
	\includegraphics[width=\linewidth]{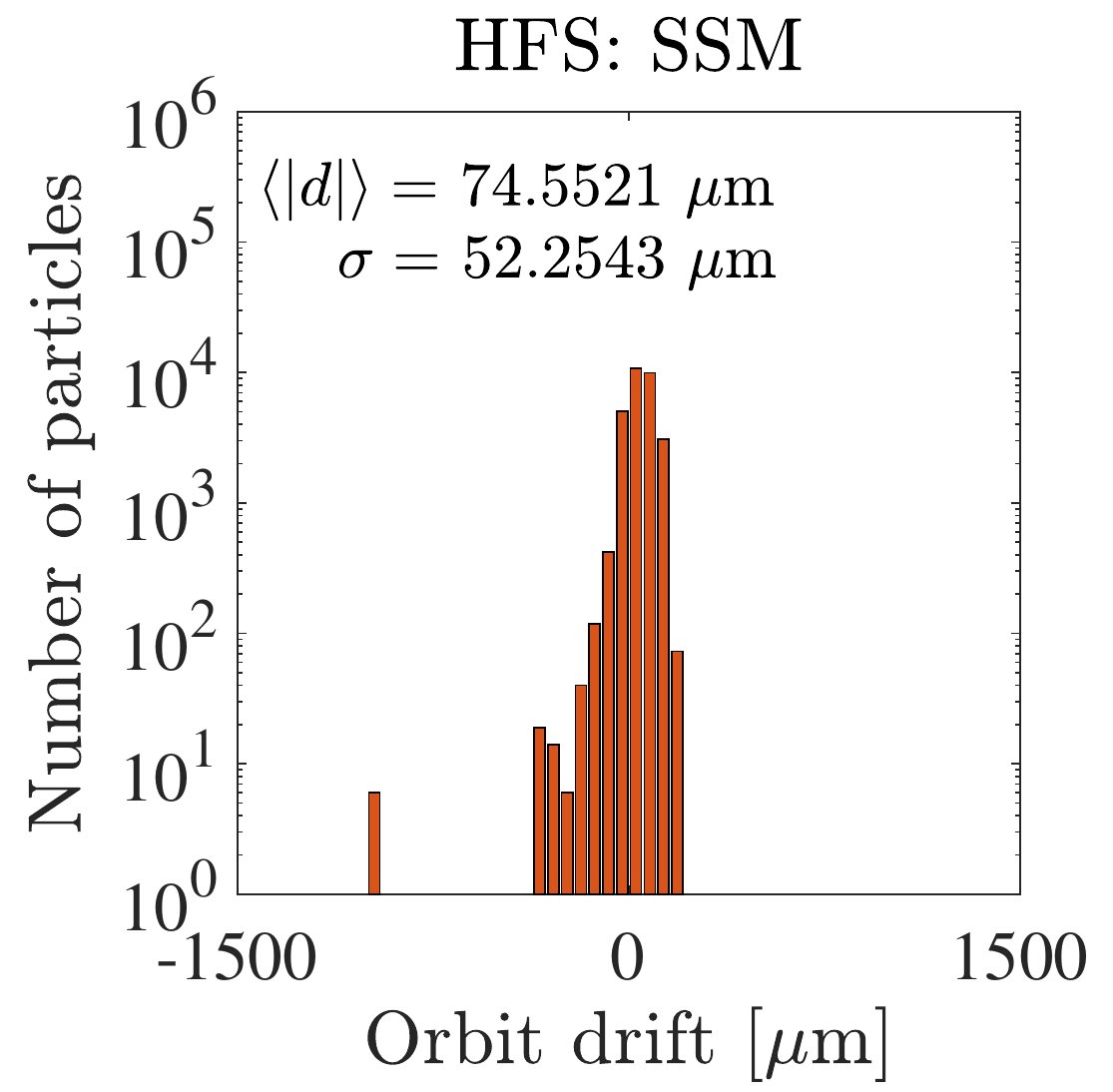}
	\end{subfigure}
	\begin{subfigure}{0.245\textwidth}
	\includegraphics[width=\linewidth]{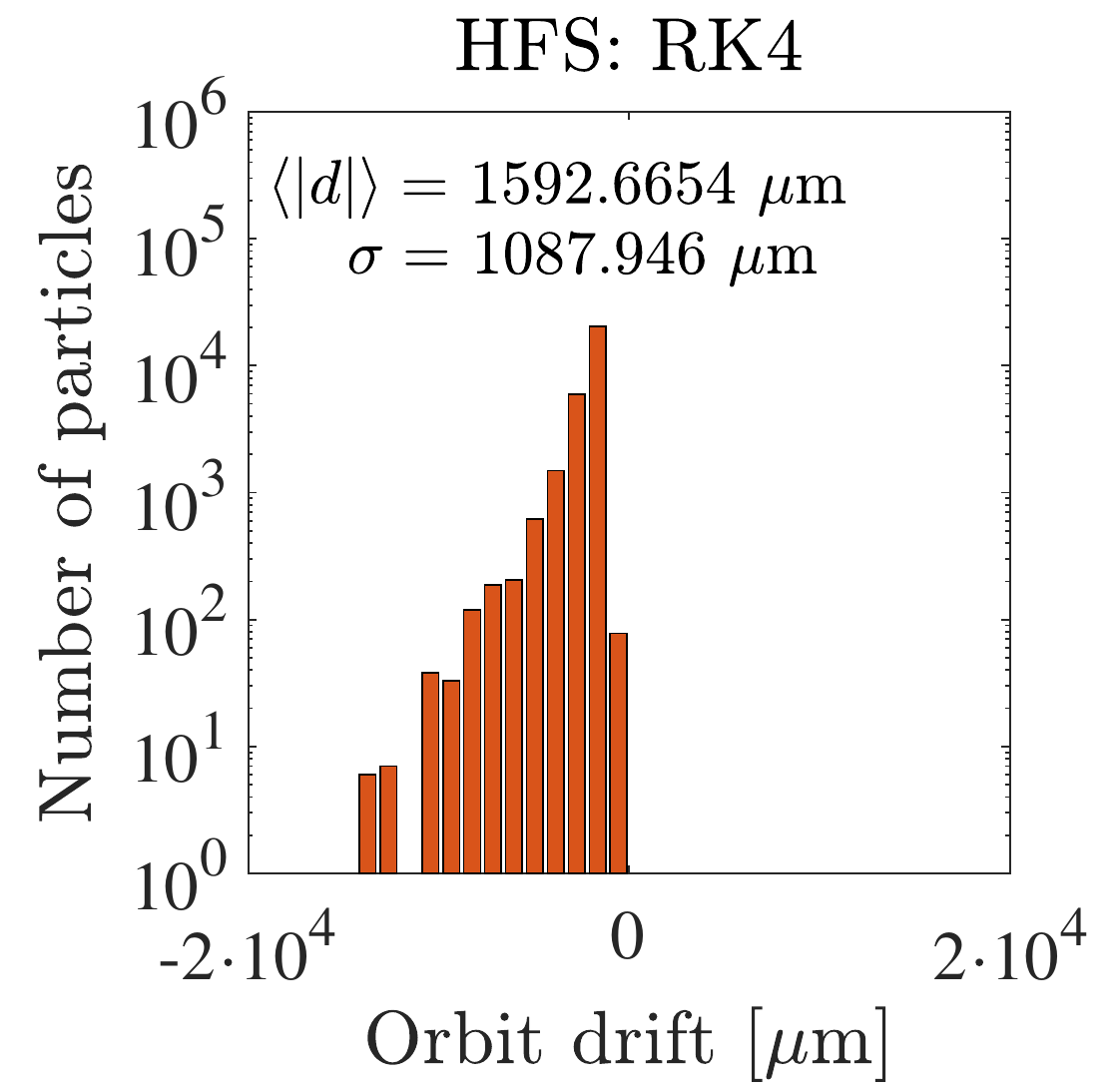}
	\end{subfigure}
	\begin{subfigure}{0.245\textwidth}
	\includegraphics[width=\linewidth]{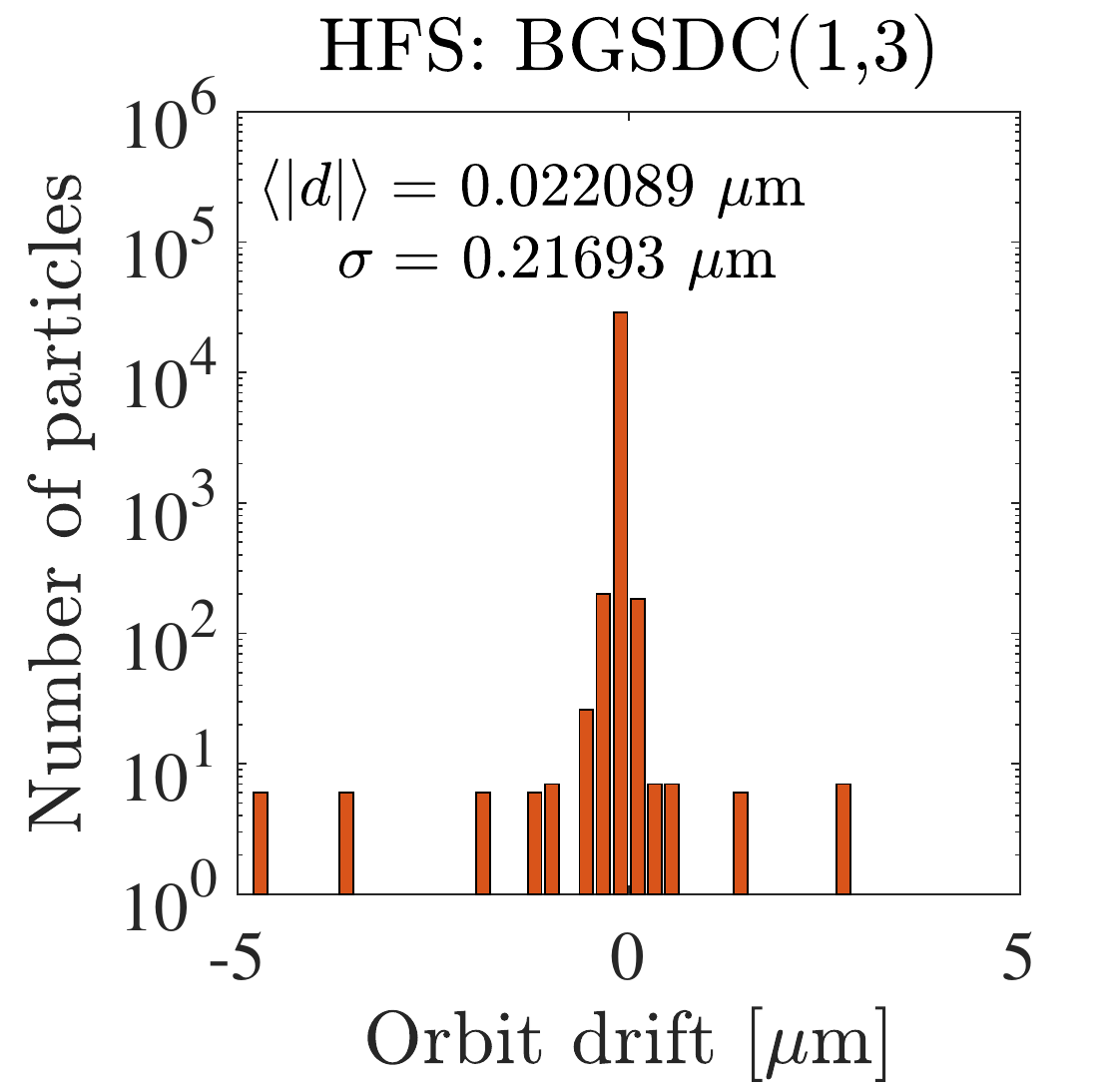}
	\end{subfigure}
	\par\bigskip 
	\begin{subfigure}{0.245\textwidth}	
	\includegraphics[width=\linewidth]{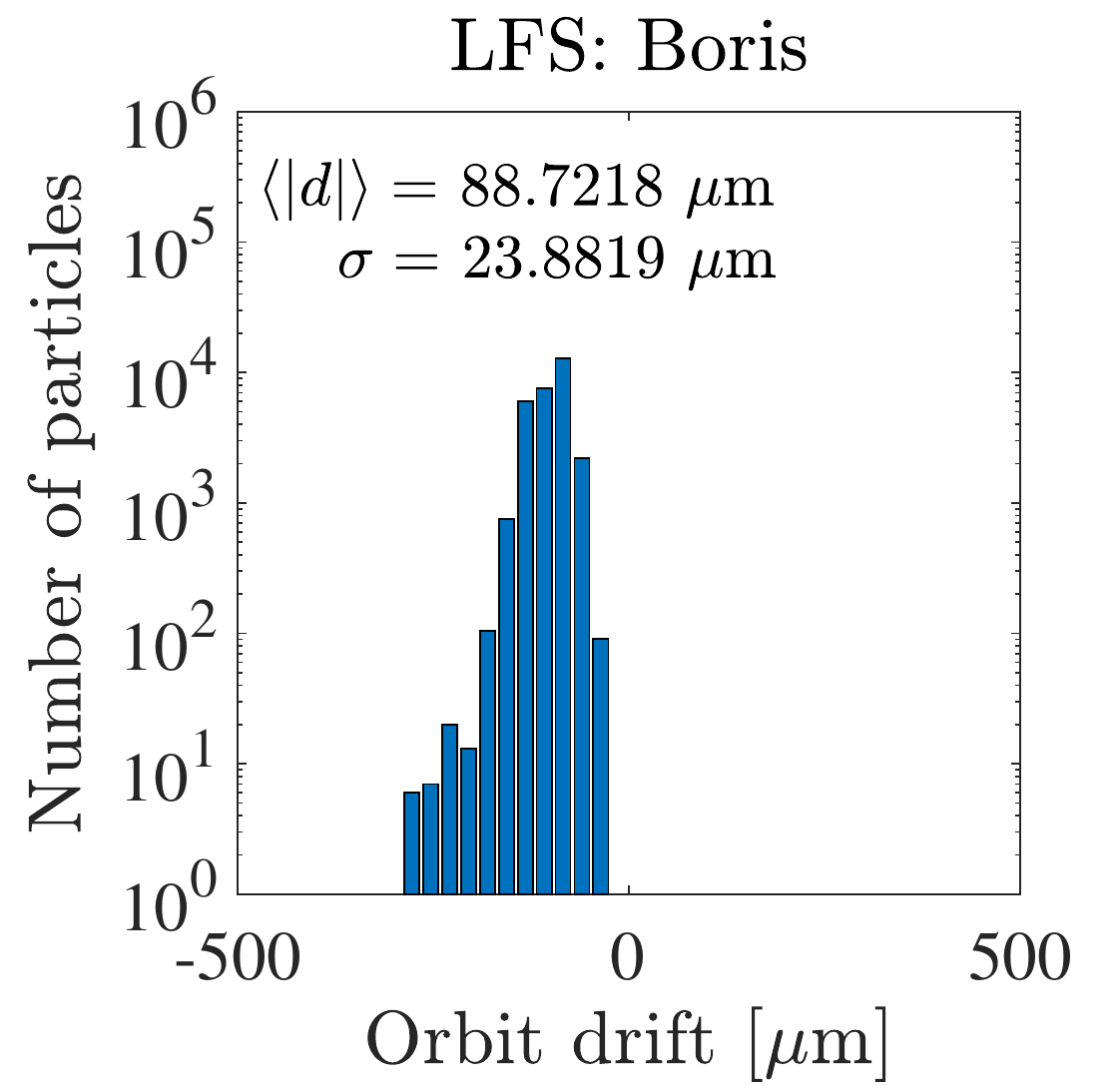}
	\end{subfigure}
	\begin{subfigure}{0.245\textwidth}
	\includegraphics[width=\linewidth]{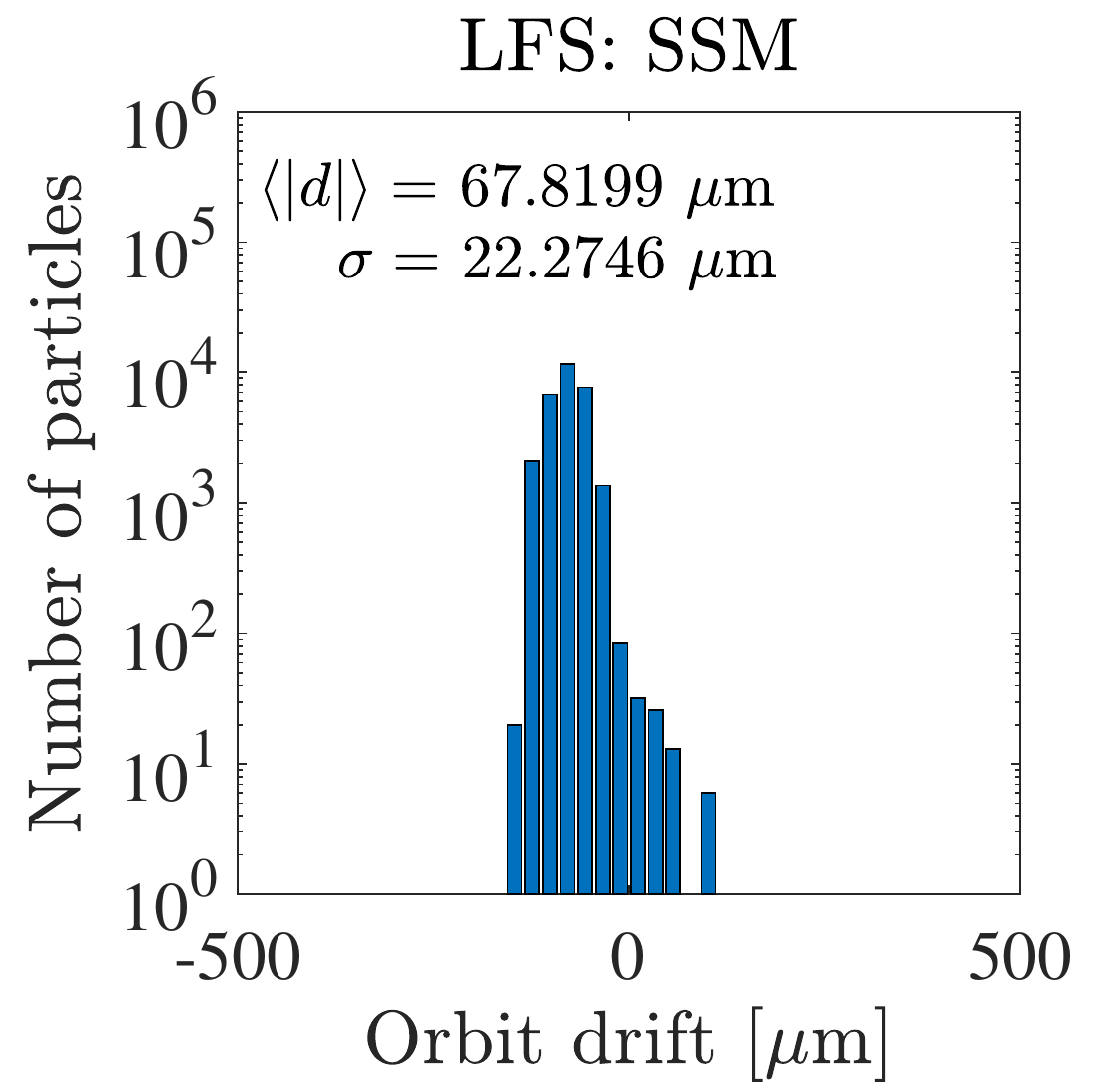}
	\end{subfigure}
	\begin{subfigure}{0.245\textwidth}
	\includegraphics[width=\linewidth]{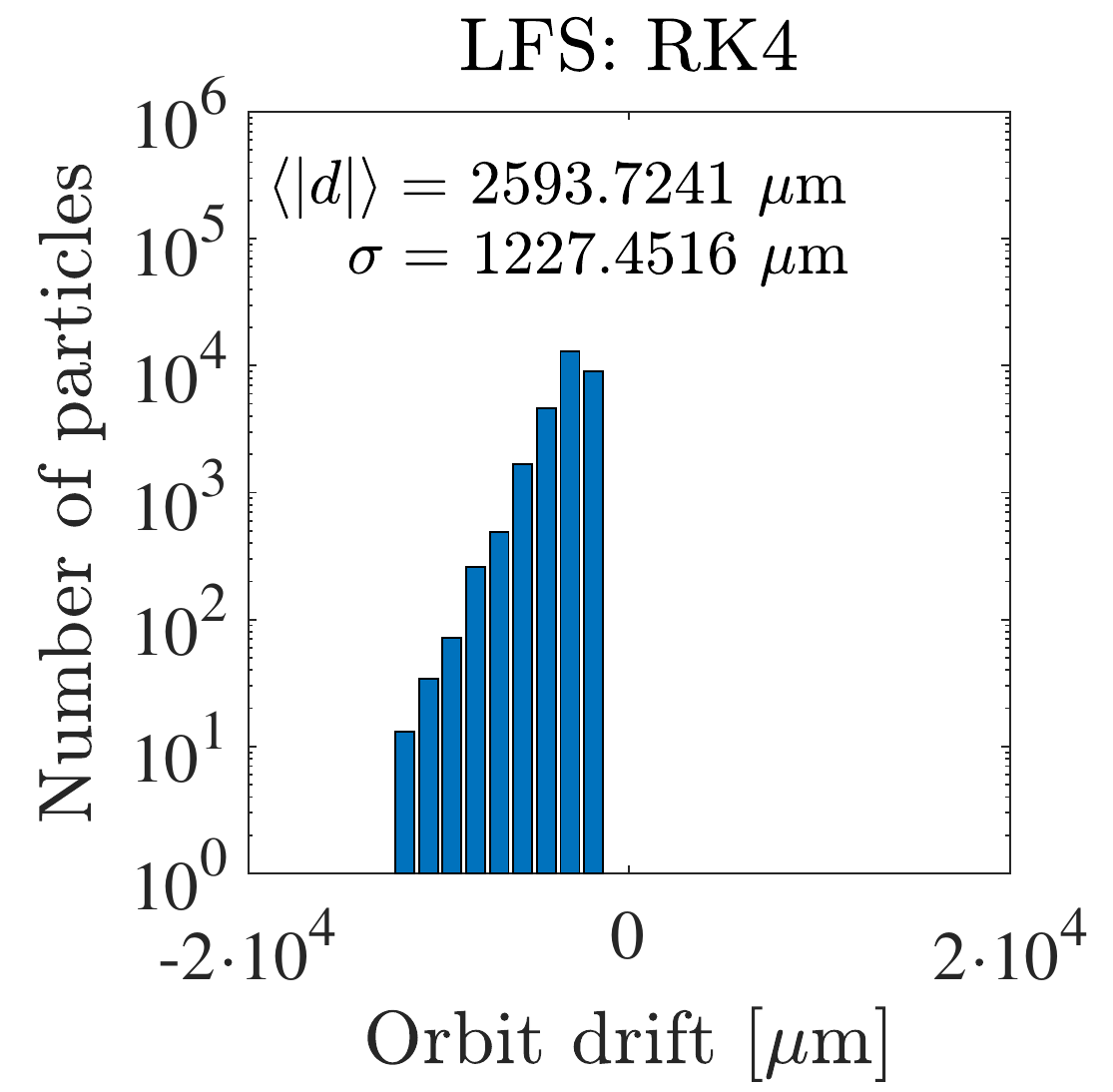}
	\end{subfigure}
	\begin{subfigure}{0.245\textwidth}
	\includegraphics[width=\linewidth]{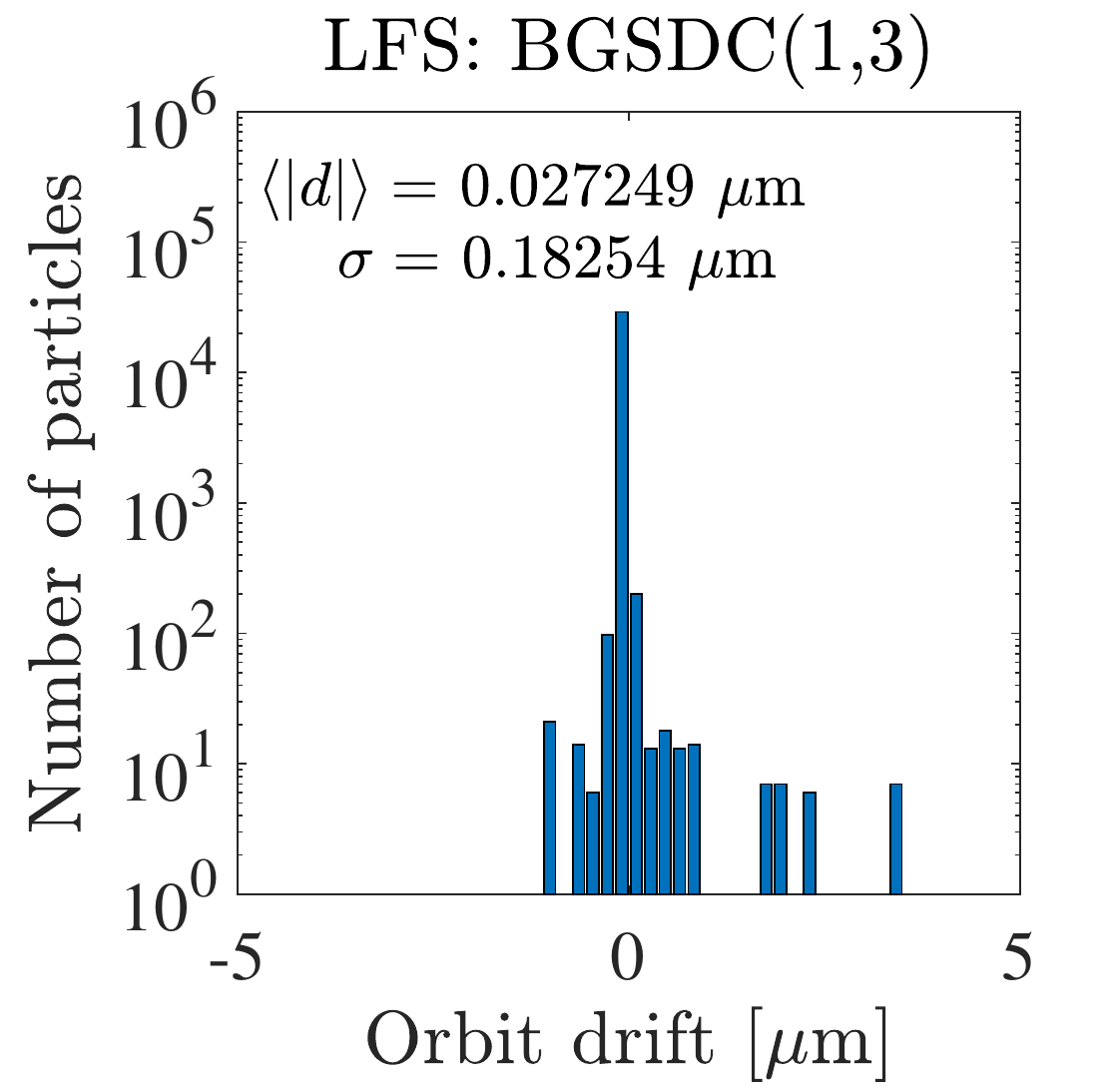}
	\end{subfigure}
	\caption{Accuracy comparison of classical Boris, Strang Splitting Mover, Runge-Kutta Cash \& Karp and BGSDC(1,3) methods with fixed time step  $\Delta t = \SI{1}{\nano\second}$ and $t_{end} = \SI{100}{\milli\second}$. Please note that the x-axes are scaled differently.}
	\label{fig:methods}
\end{figure}

\paragraph{Numerical drift}
Fig.~\ref{fig:methods} shows the distribution of numerical drift for four integrators available in LOCUST at final time $t_{end} = \SI{100}{\milli\second} $ for a particle ensemble in the non-collisional case.
All integrators use a fixed time step of $ \Delta t = \SI{1}{\nano\second}$ and the same initial conditions for particles. 
The High Field Side (HFS)\footnote{\revb{Because the magnetic field in a Tokamak changes radially like $1/R$ with $R$ being the major radius, the outboard side with large major radius is often referred to as Low Field Side (LFS) while the interior side, with small $R$, is called High Field Side (HFS)}} drift distributions are shown in the upper two graphs and $\sigma$ indicates the standard deviation. 
The lower graphs show the Low Field Side (LFS) drift distributions. 
The Strang splitting mover and classical Boris  deliver comparable accuracy at both sides of the plasma volume.
However, the Strang splitting mover requires more computational work than Boris \revb{and requires about 5\% longer runtime}.
RK4 shows unsatisfactory result with very large $\sigma$, most likely due to its inherent energy drift.
\revb{Since the Boris integrator performs better than SSM or RK4 we use it as a baseline to compare BGSDC against.}
BGSDC(1,3) is more accurate than Boris and Strang and for both LFS and HFS delivers a $\sigma$ two orders of magnitude smaller.
Of course, it also requires substantially more computational work per time step.
Below we will demonstrate that this additional work per time step can be offset by using larger time step sizes and thus computing fewer steps, leading to computational gains when values of $\sigma$ of around \SI{1}{\micro\meter} or below are required.

\begin{figure}[t]
    \begin{subfigure}{0.49\textwidth}
    \includegraphics[width=\linewidth]{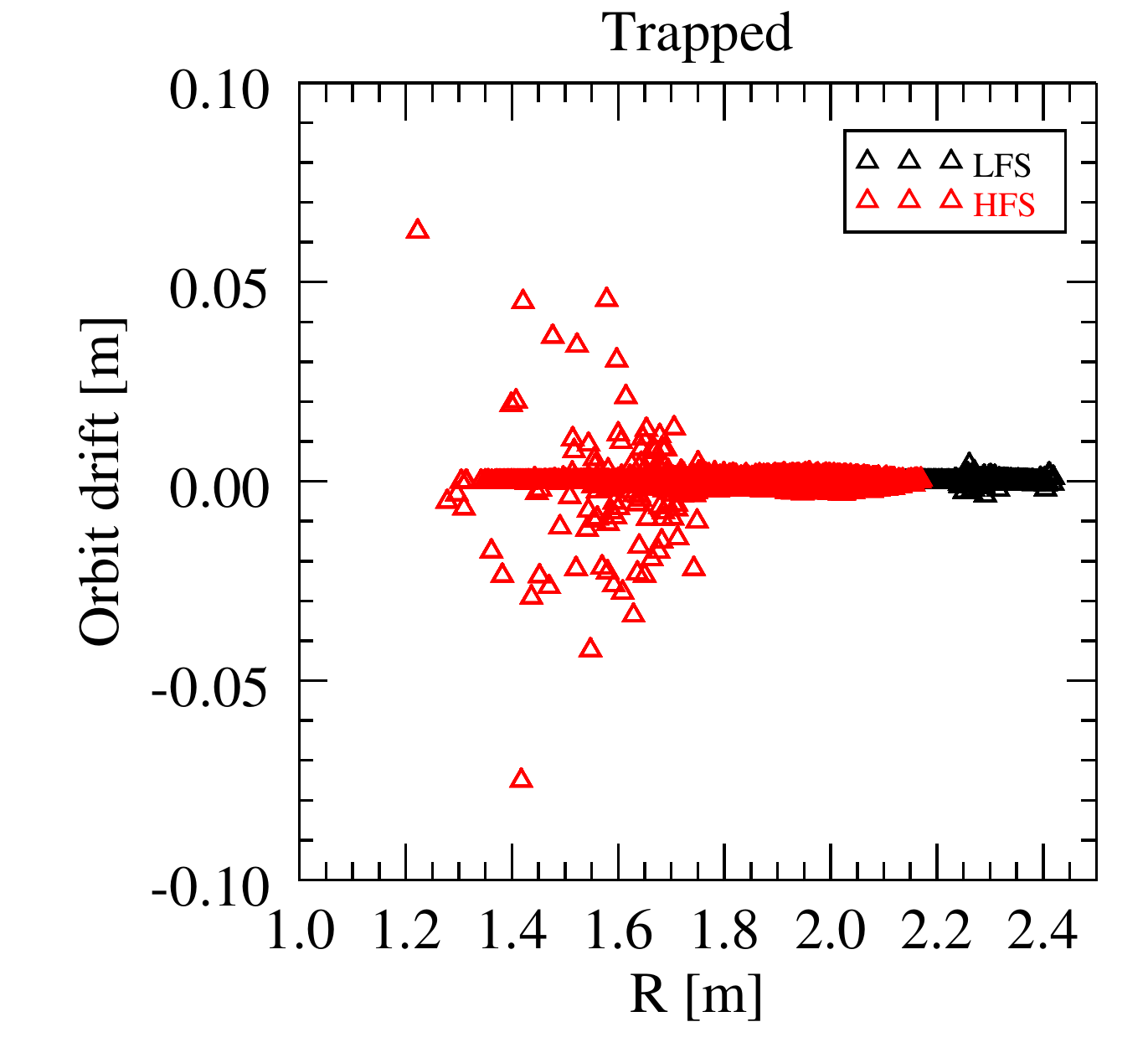}
    \end{subfigure}
    \begin{subfigure}{0.49\textwidth}
    \includegraphics[width=\linewidth]{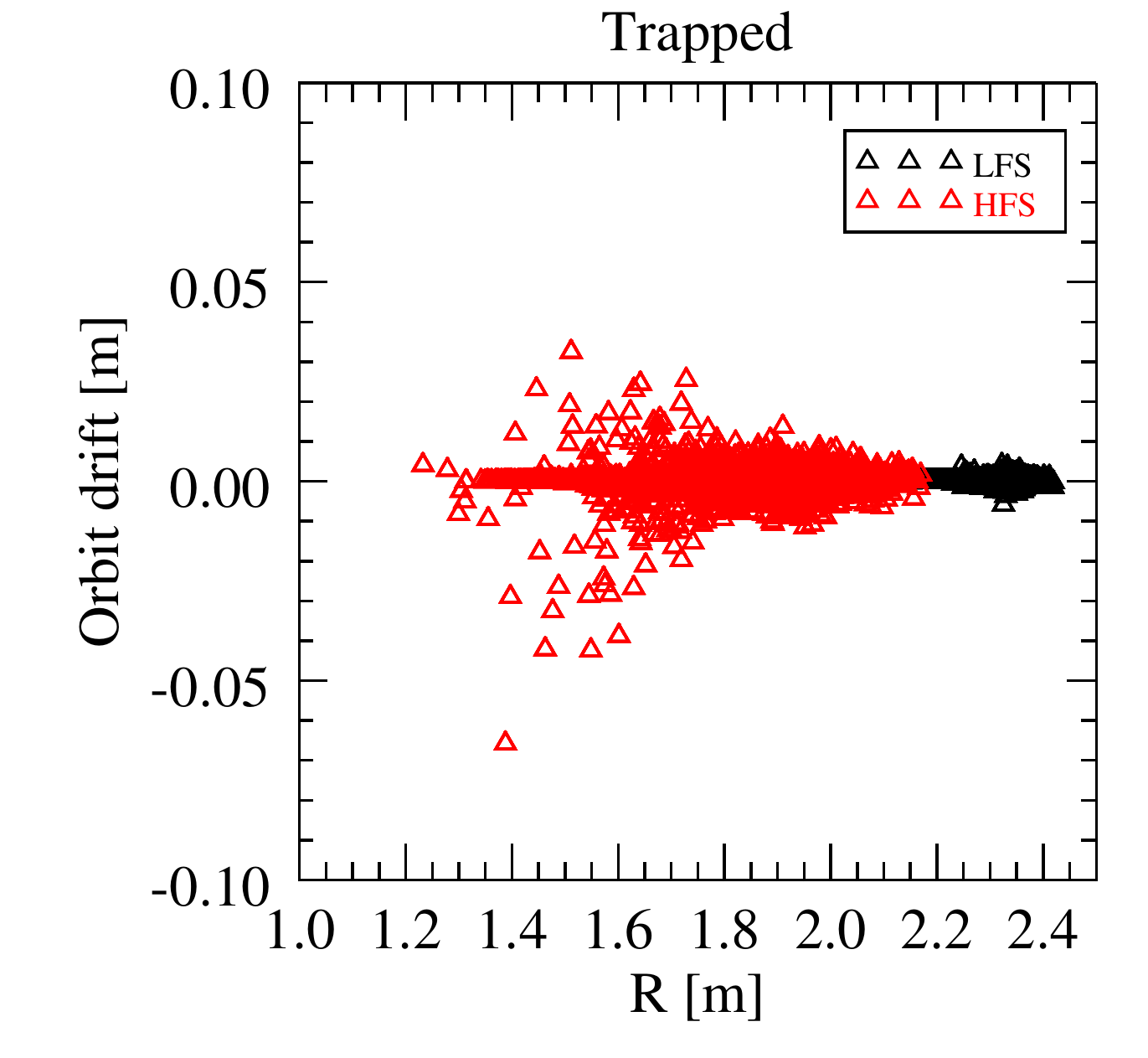}
    \end{subfigure}
    \begin{subfigure}{0.49\textwidth}
    \includegraphics[width=\linewidth]{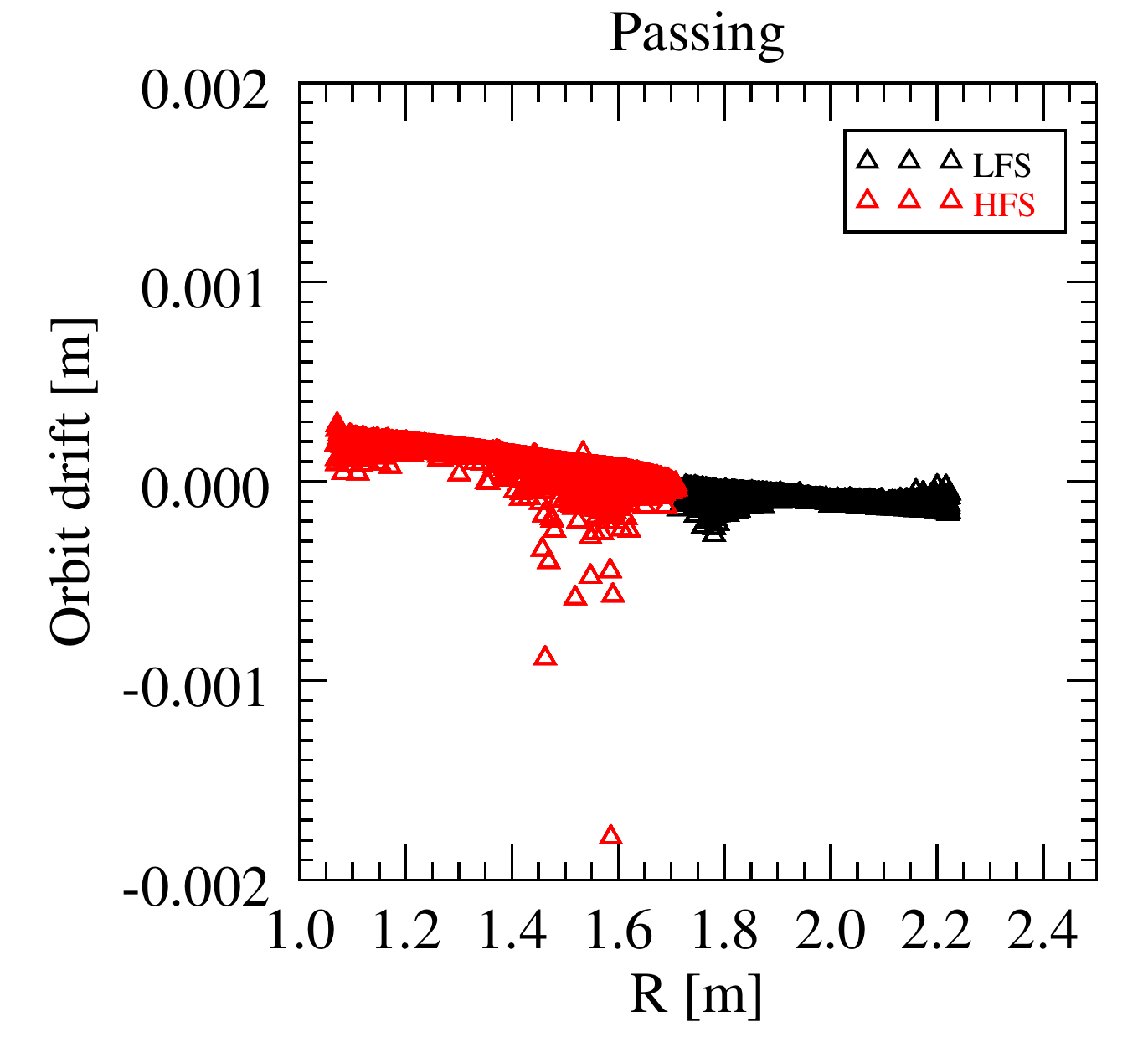}
    \end{subfigure}
    \begin{subfigure}{0.49\textwidth}
    \includegraphics[width=\linewidth]{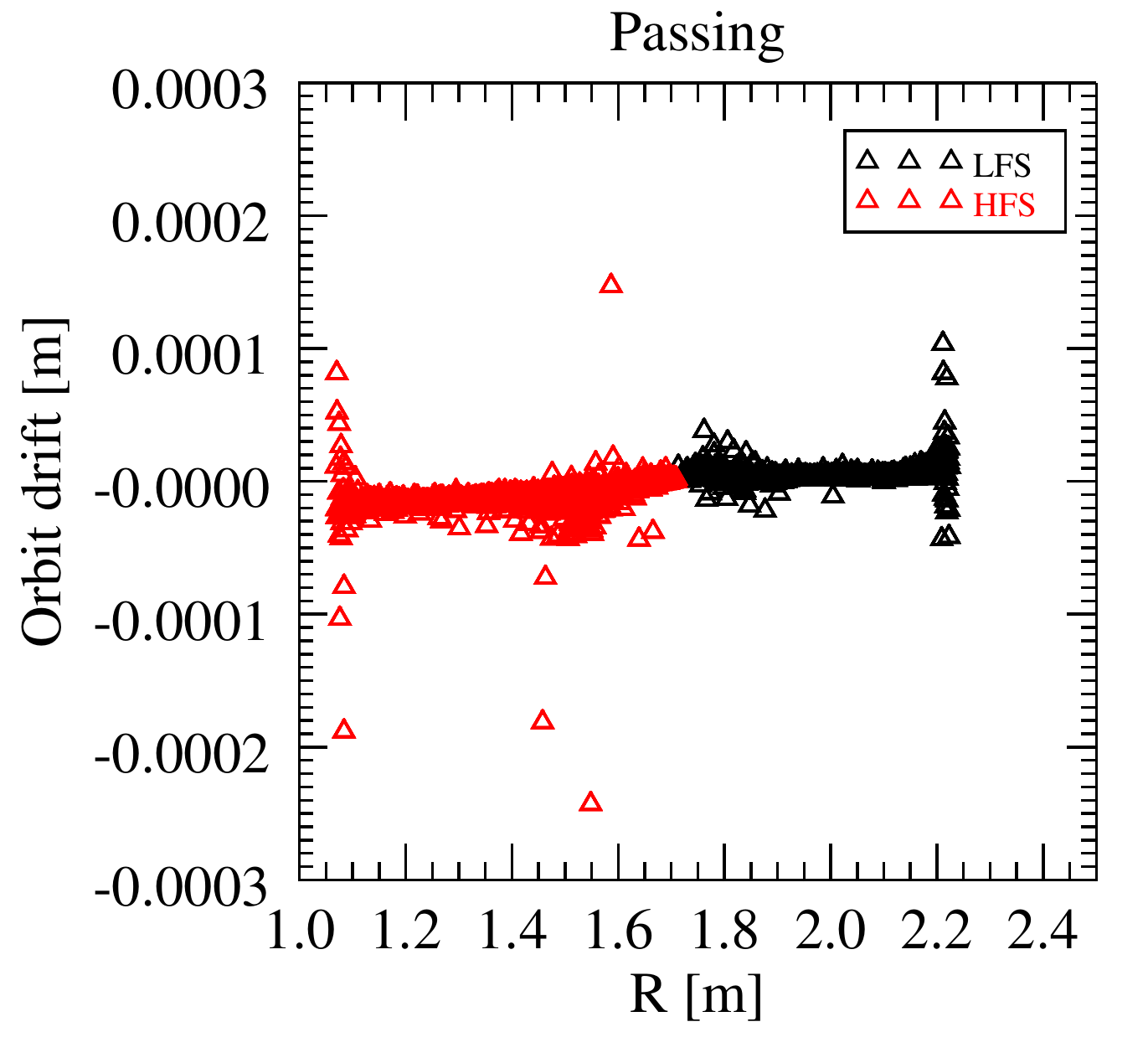}
    \end{subfigure}
    \caption{Particle drifts for DIII-D for Boris method with $\Delta t = \SI{1}{\nano\second}$ (left) and BGSDC(2,6) with $\Delta t = \SI{5}{\nano\second}$ (right).}
    \label{fig:diiid_ptcldrift}
\end{figure}

\revb{Charged particles in a magnetic field can experience mirror trapping when entering regions of higher magnetic field strength as a consequence of conservation of the magnetic moment.
These particles do not complete full orbits but instead experience a bounce point where their parallel velocity is zero and become trapped in the low field side.
Because these sharp turns and resulting large gradients can affect performance of the numerical integrator~\cite{TretiakRuprecht2019}, we show results separately for passing and trapped particles.}
Fig.~\ref{fig:diiid_ptcldrift} shows the resulting numerical orbital drift of each particle against the major radius $R$ at the end of simulation at $t_{end} = \SI{100}{\milli\second}$.
The left figure is for the Boris method with time step $\SI{1}{\nano\second}$ while the right shows results from BGSDC(2,6) with a time step of $ \SI{5}{\nano\second}$.
The upper graphs show trapped particles, the lower graphs passing particles where particles at the LFS are indicated by black markers and particles at HFS are marked in red.
Note that the $y$-axes in the two lower graphs are scaled differently.
For trapped particles, Boris and BGSDC deliver  comparable results with particle drifts of up to \SI{5}{\centi\meter}.
For passing particles, BGSDC drifts are about an order of magnitude smaller than Boris, despite a five times larger time step.

\begin{figure}[t]
    \centering
	\begin{subfigure}{0.49\textwidth}
	\includegraphics[width=\linewidth]{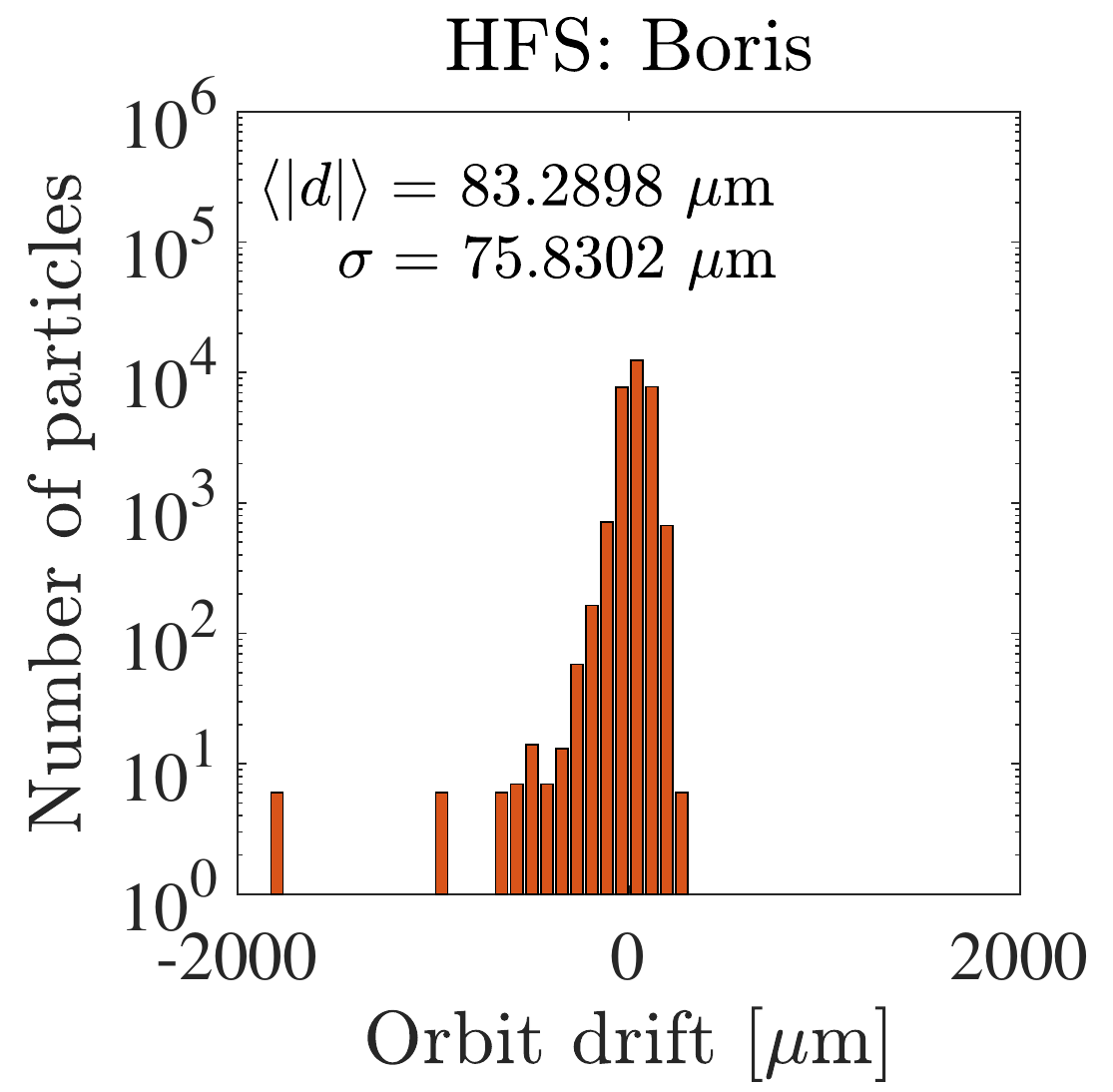}
	\end{subfigure}
	\begin{subfigure}{0.49\textwidth}
	\includegraphics[width=\linewidth]{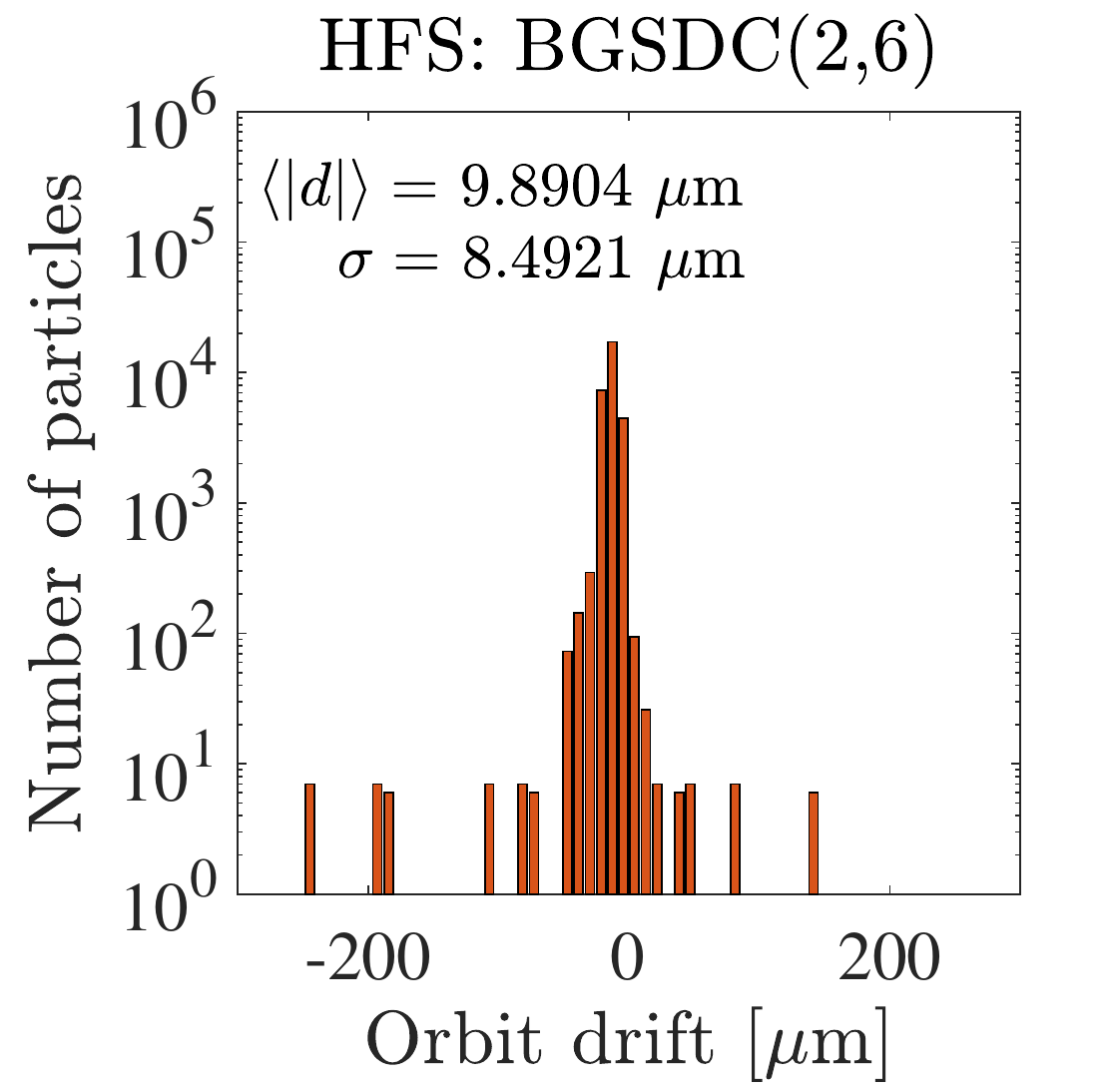}
	\end{subfigure}
	\begin{subfigure}{0.49\textwidth}
	\includegraphics[width=\linewidth]{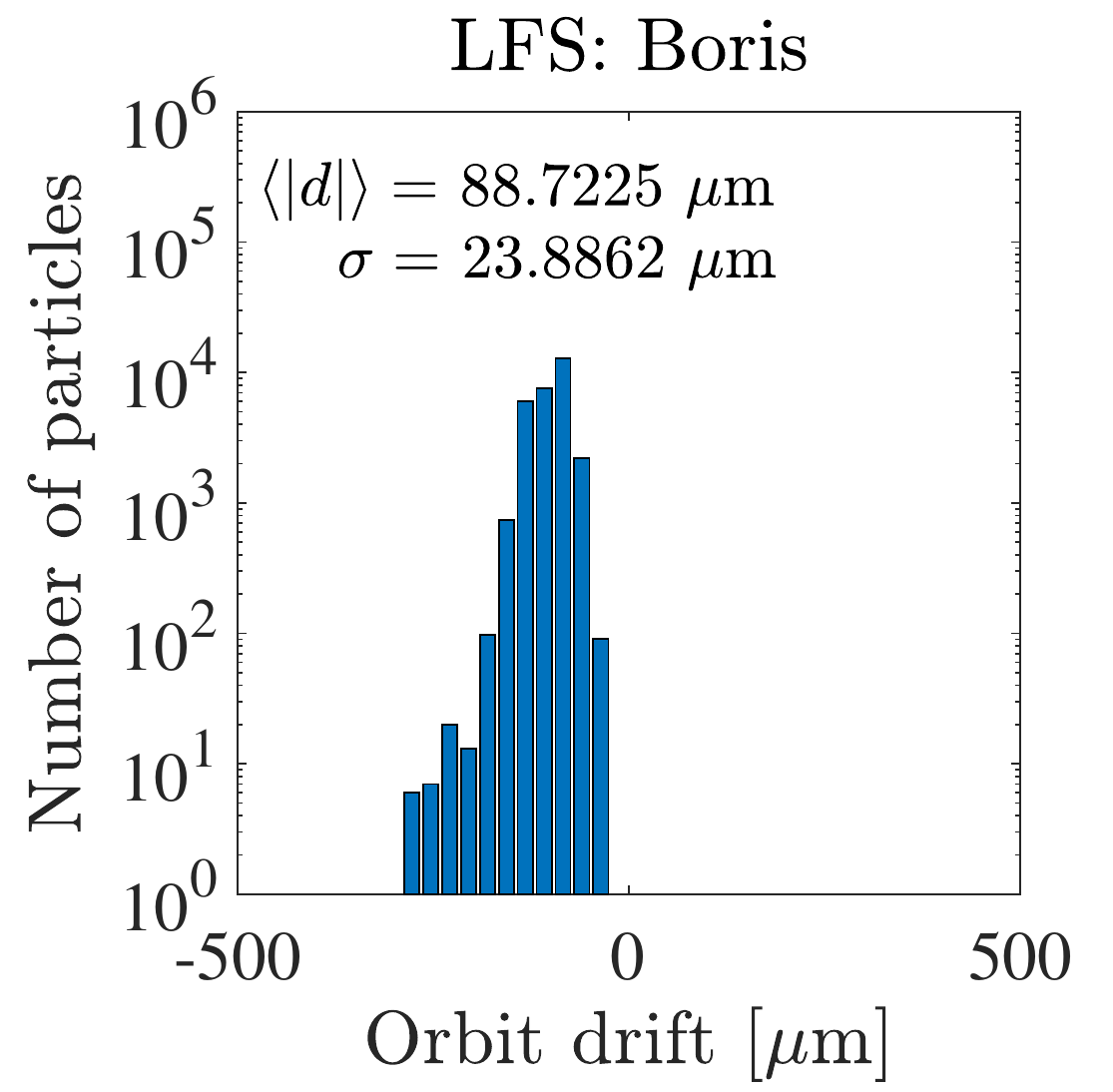}
	\end{subfigure}
	\begin{subfigure}{0.49\textwidth}
	\includegraphics[width=\linewidth]{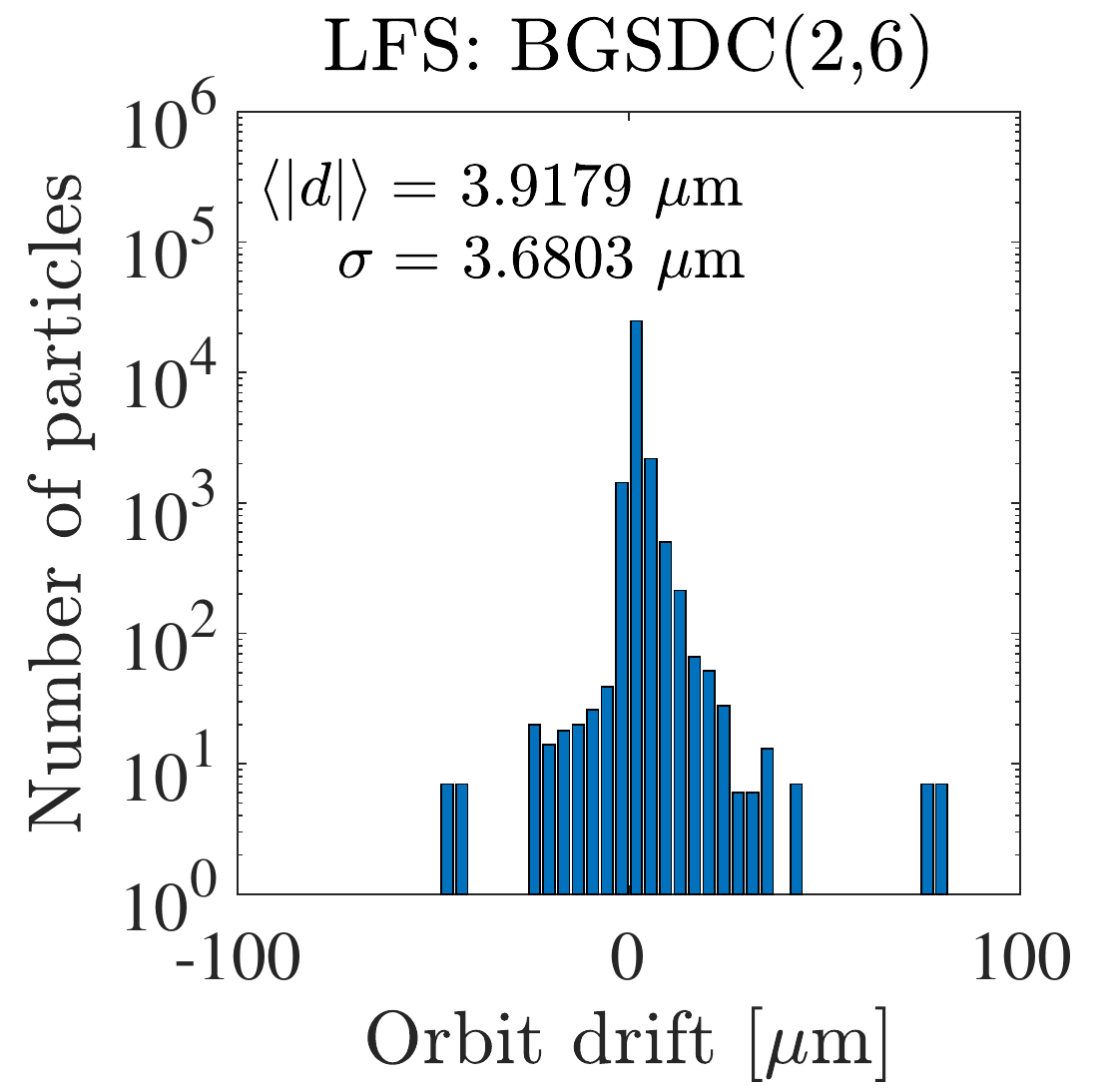}
	\end{subfigure}
	\caption{Drift distribution: Boris method $\Delta t = \SI{1}{\nano\second}$ (left) and BGSDC(2,6) $\Delta t = \SI{5}{\nano\second}$ (right).}
	\label{fig:diiid_driftdistr}
\end{figure}

Fig.~\ref{fig:diiid_driftdistr} shows the frequency distribution of orbital drift in \SI{}{\micro\meter} for all particles (both trapped and passing) shown in Fig.~\ref{fig:diiid_ptcldrift}.
The upper figures show values at the HFS, the lower figures at the LFS.
The $\sigma$ in each figure's title indicates the standard deviation for the particle cloud.
The smaller $\sigma$ is, the narrower and thus better is the distribution.
At HFS, the standard deviation for BGSDC(2,6) is about nine times smaller than for Boris while at LFS it is a factor of around six more accurate.

%
%
\paragraph{Magnetic moment}
%
%
\revb{While the Boris integrator conserves magnetic moment by design, for BGSDC this is only guaranteed for the underlying collocation.
For small iteration numbers, non-conservation of magnetic moment cannot be ruled out by theory.
However, in numerical experiments we confirmed that both Boris and BGSDC(2,6) conserve magnetic moment up to machine precision.}

%
%
\paragraph{Accuracy versus time step size}
\begin{figure}[t]
    \centering
	\begin{subfigure}{0.49\textwidth}
		\includegraphics[width=\linewidth]{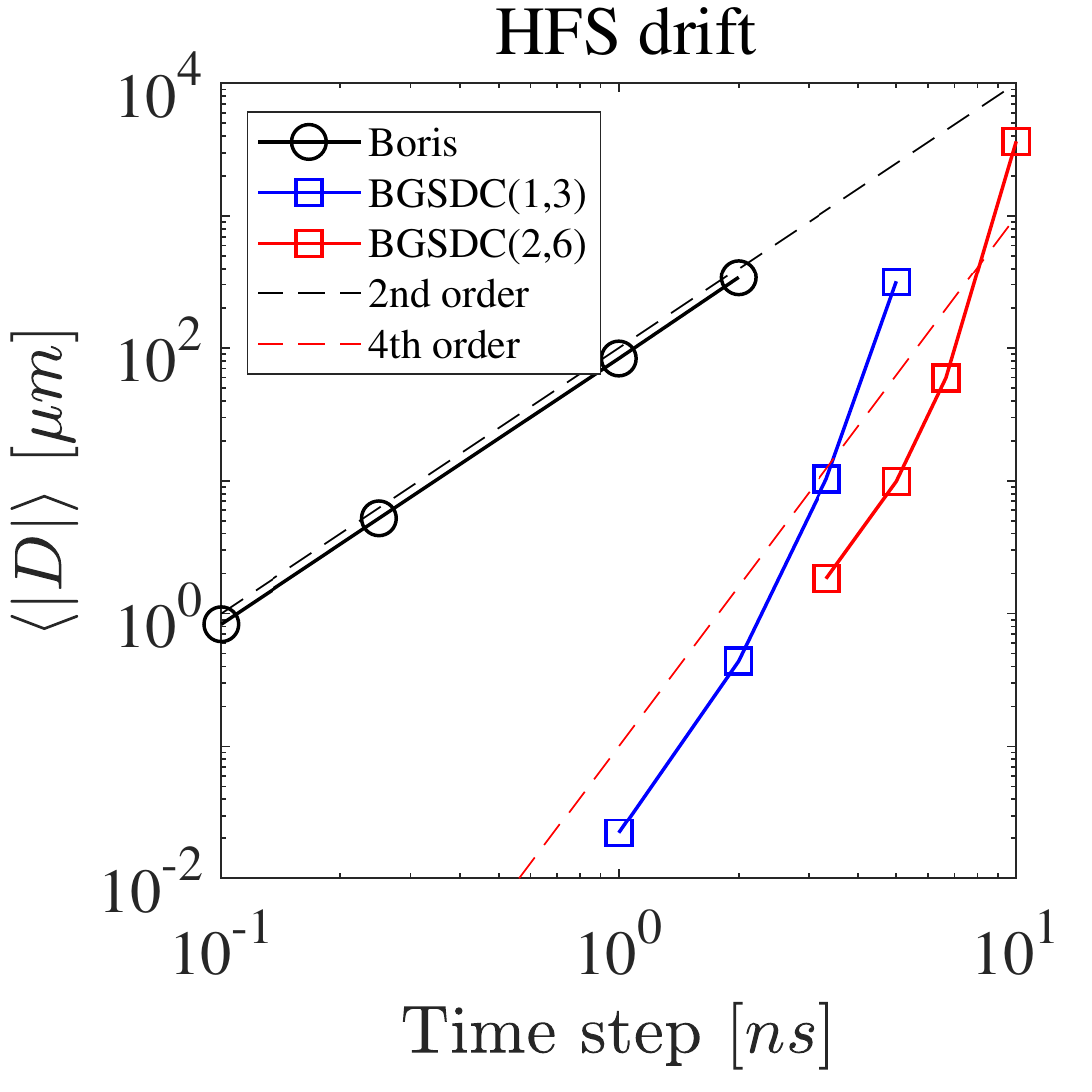}
	\end{subfigure}
	\begin{subfigure}{0.49\textwidth}
		\includegraphics[width=\linewidth]{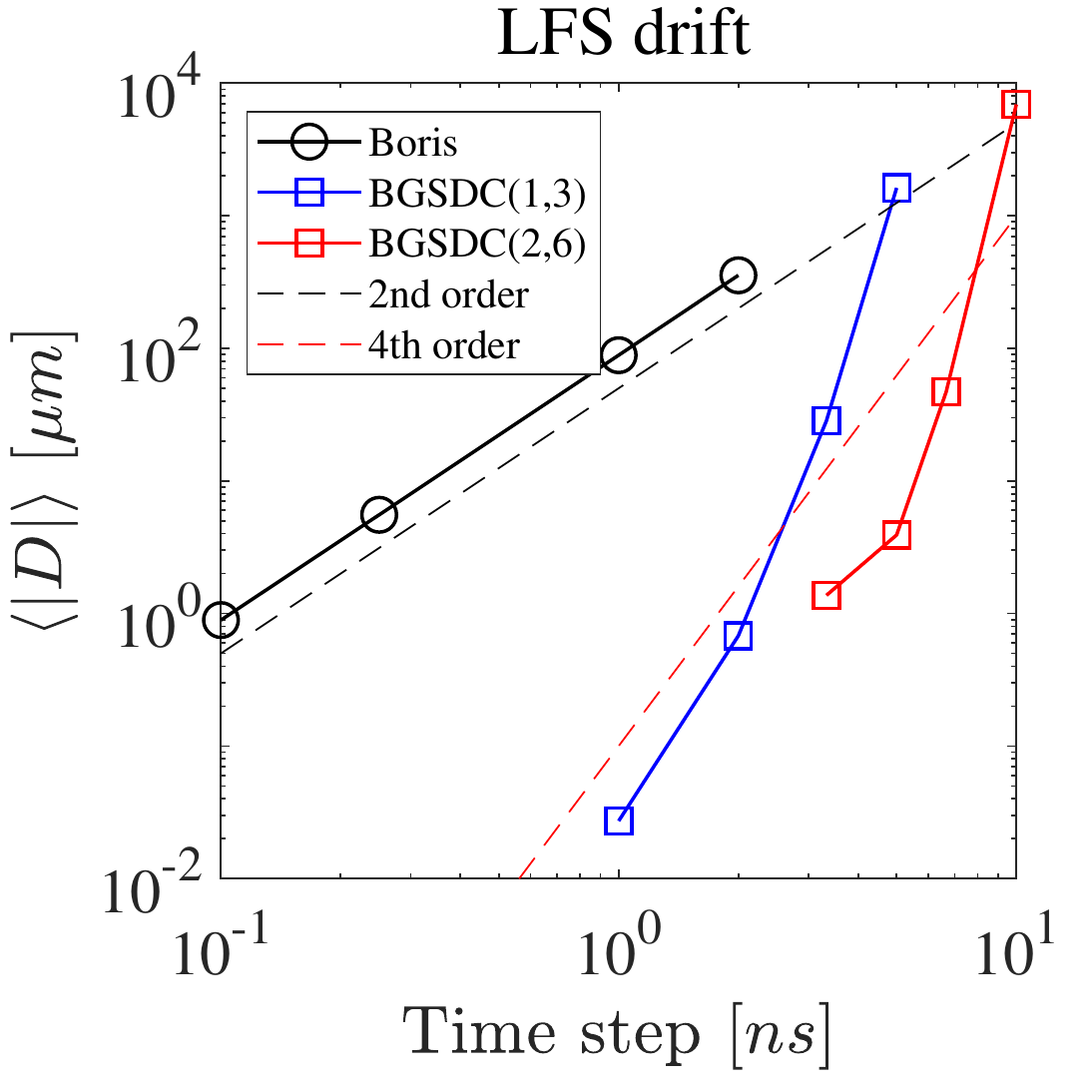}
	\end{subfigure}
	\newline
    \begin{subfigure}{0.49\textwidth}
		\includegraphics[width=\linewidth]{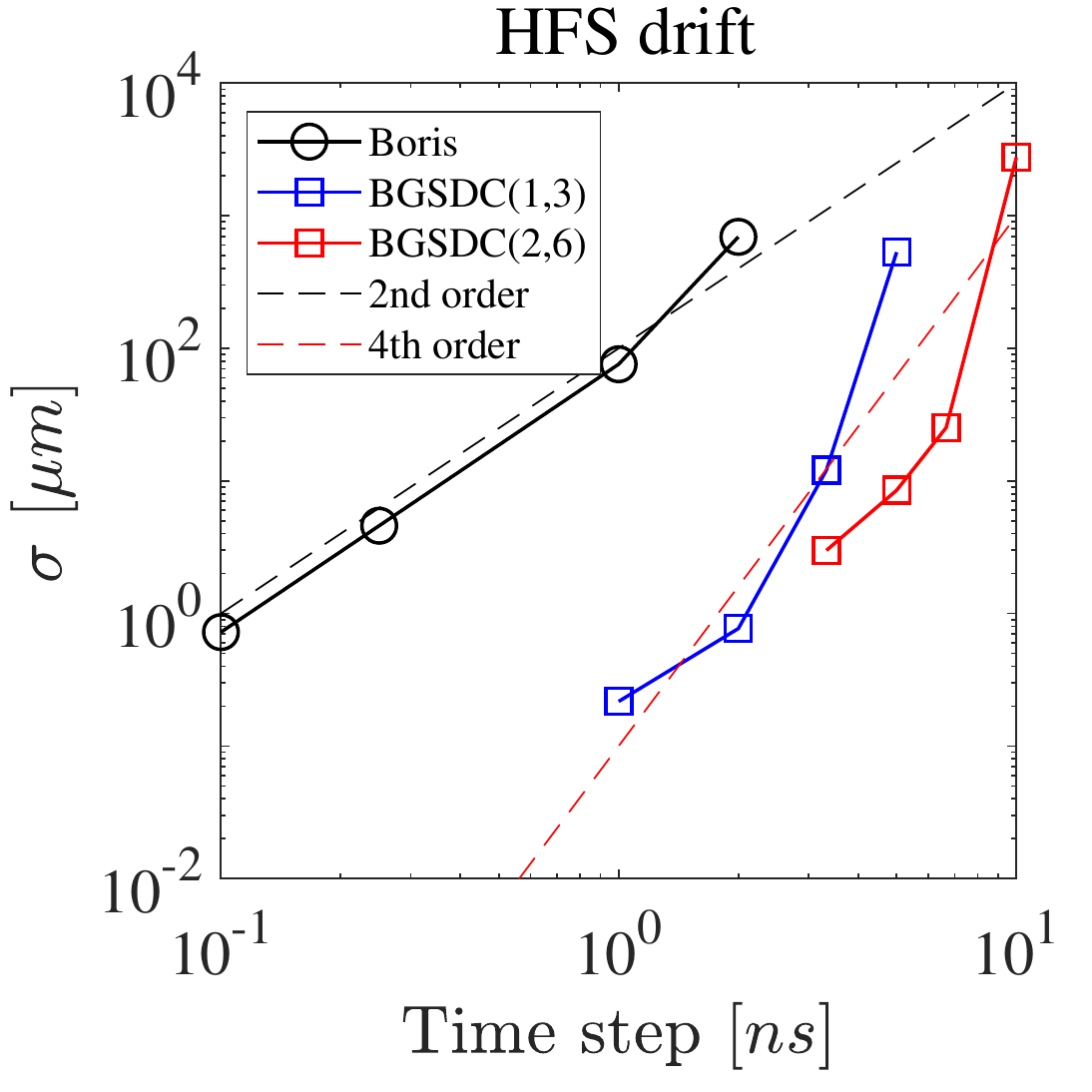}
	\end{subfigure}
	\begin{subfigure}{0.49\textwidth}
		\includegraphics[width=\linewidth]{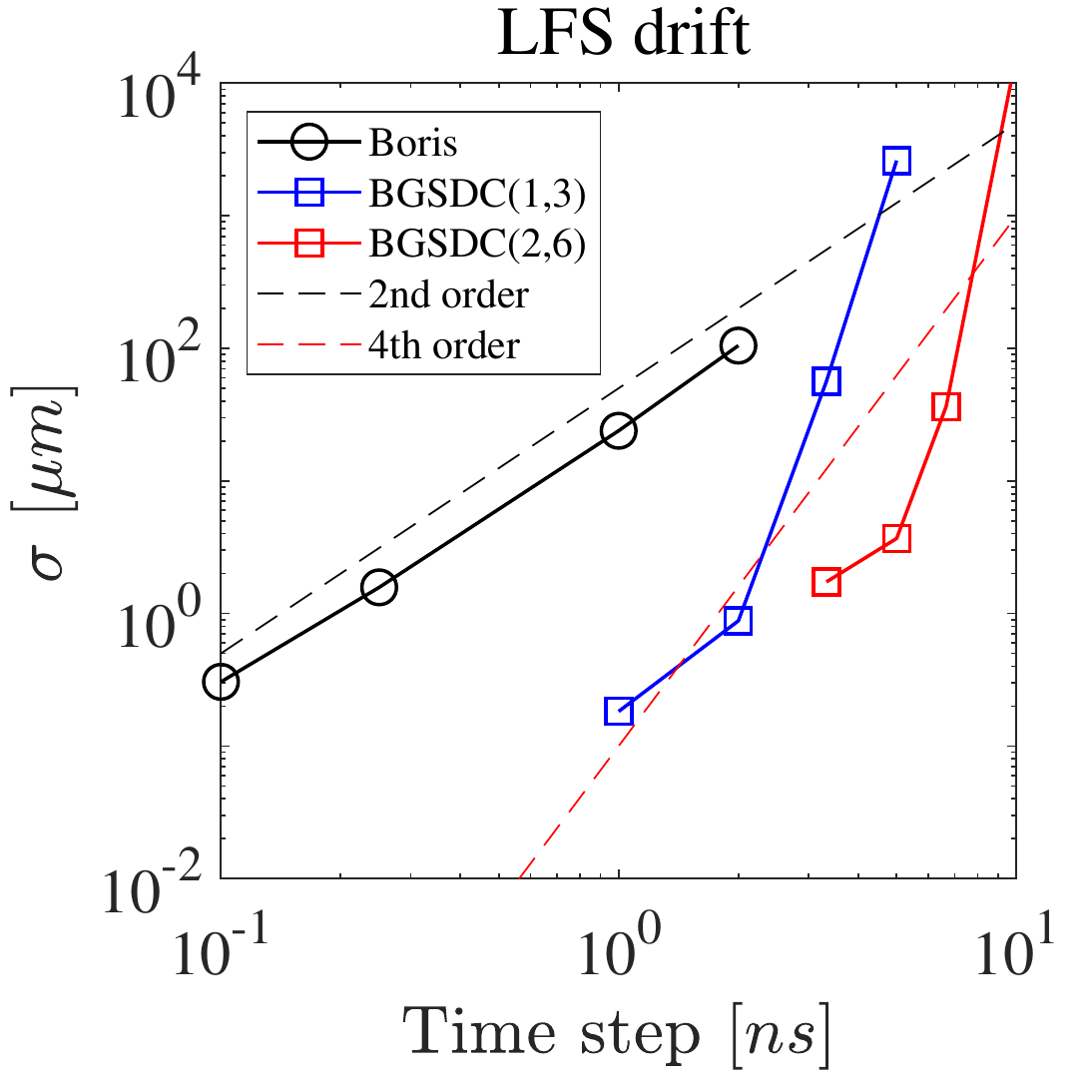}
	\end{subfigure}
	\caption{\revb{Mean (upper)} and standard deviation $\sigma$ \revb{(lower)} of drift distributions for classical Boris and BGSDC methods for different time steps in the non-collisional regime in DIII-D.}
	\label{fig:diiid_sigma}
\end{figure}
Fig.~\ref{fig:diiid_sigma} shows how the \reva{mean (upper)} and standard deviation $\sigma$ (lower) for the HFS (left) and LFS (right) changes with time step size.
The higher order of accuracy of both BGSDC(1,3) and BGSDC(2,6) translates into a faster drop of $\sigma$ with $\Delta t$ and thus a steeper slope of the curve.
This demonstrates that there is an accuracy gain from using integrators of higher orders, not only for individual trajectories but also for the full ensemble.
It also demonstrates that BGSDC can deliver a particle distribution with comparable accuracy with a much larger time step than Boris.
For a value of \reva{$\sigma = \SI{1}{\micro\meter}$} for example, Boris requires a time step of $\Delta t = \SI{0.1}{\nano\second}$ whereas for BGSDC a twenty times larger step size of $\Delta t = \SI{2}{\nano\second}$ suffices.

%
%
\paragraph{Work-precision}
\begin{figure}[t]
    \centering
    \begin{subfigure}{0.48\textwidth}
    \includegraphics[width=0.99\linewidth]{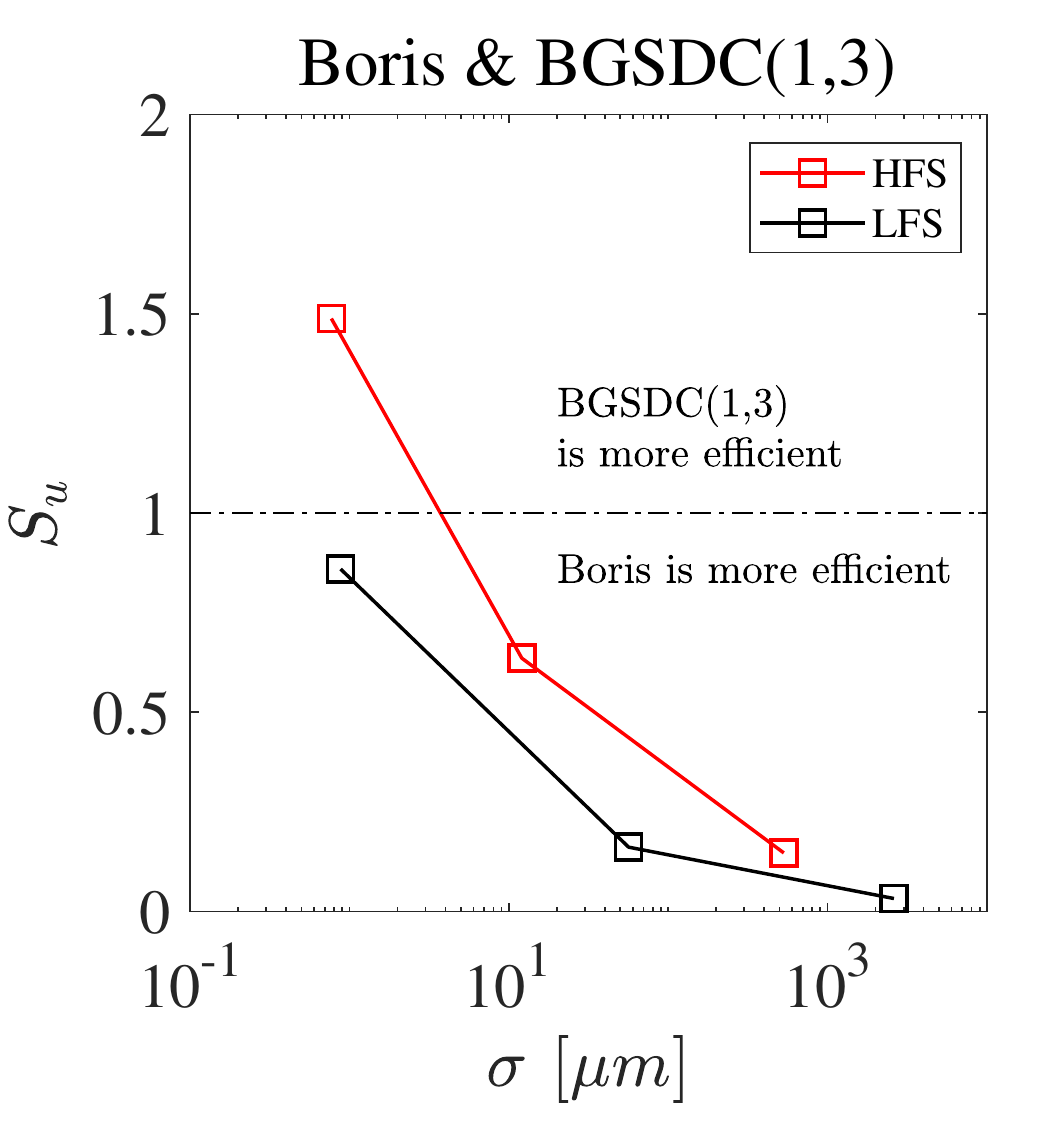}
    \end{subfigure}
    \begin{subfigure}{0.49\textwidth}
    \includegraphics[width=0.99\linewidth]{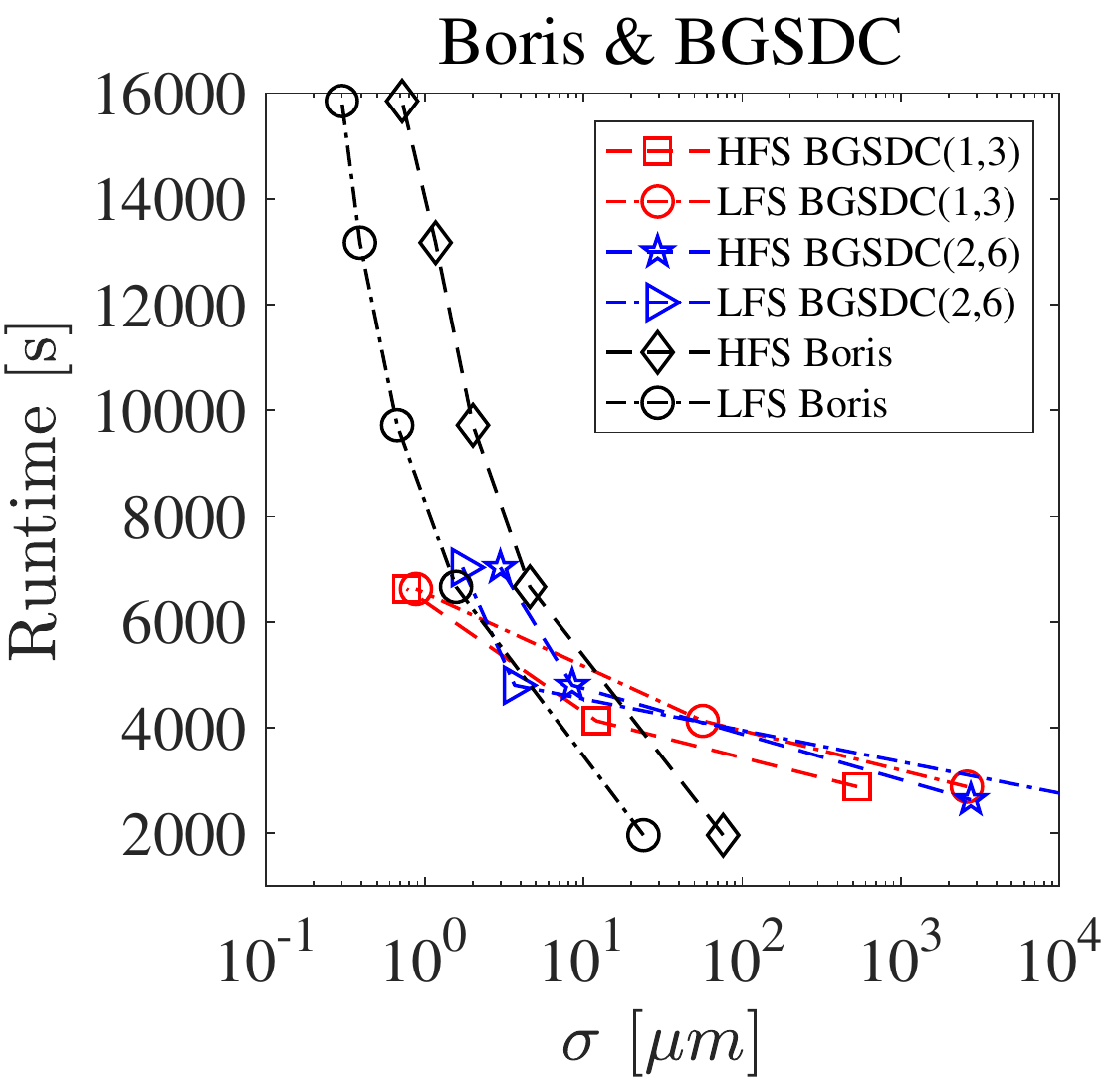}     
    \end{subfigure}
    \caption{Left: Ratio $S_u$ of field evaluations required by Boris to reach some standard deviation $\sigma$ divided by the number of evaluations required by BGSDC(1,3) for the DIII-D reactor. \reva{Right: Runtime to reach some standard deviation $\sigma$ for the Boris method, BGSDC(1,3) and BGSDC(2,6).}}
    \label{fig:diiid_speedup}
\end{figure}
BGSDC(2,6) with $M=3$ Gauss-Lobatto nodes means we perform $k=2$ iterations of GMRES-SDC on the linearised problem and $l = 6$ Picard iterations~\cite{TretiakRuprecht2019}. 
In contrast to Boris, which only needs one, BGSDC(2,6) therefore requires 19 right hand side ($RHS$) evaluations per time step.
However, as shown in Fig.~\ref{fig:diiid_ptcldrift} and~\ref{fig:diiid_driftdistr}, it can provide comparable accuracy with a much larger $\Delta t$ and will thus require fewer time steps. 
If the number of steps is sufficiently smaller to compensate for the increased work per step, BGSDC will be computationally more efficient.

To pinpoint where BGSDC starts to deliver computational gains, Fig.~\ref{fig:diiid_speedup} (left) shows the number of total RHS evaluations required by Boris to reach some standard deviation $\sigma$ divided by the number of evaluations required by BGSDC(1,3).
A value of $S_u > 1$ means that BGSDC(1,3) requires fewer evaluations and thus less computational work whereas $S_u < 1$ means Boris requires fewer evaluations.
We assume perfect second order convergence of Boris method to interpolate between data points, which is line with the behaviour we see in Fig.~\ref{fig:diiid_sigma}.
For the HFS, the break-even point for BGSDC is for accuracies slightly below $\sigma = \SI{1}{\micro\meter}$ and gains increase sharply from there.
A distribution with HFS $\sigma = \SI{0.5}{\micro\meter}$ will require only about $1 / 1.5 \approx 66\%$ as many evaluations as when using Boris.
Gains are less pronounced when looking at LFS values, where the break-even point seems to be slightly above $\sigma = \SI{0.1}{\micro\meter}$.

\reva{Fig.~\ref{fig:diiid_speedup} (right) shows the total runtime for LOCUST required to reach a given $\sigma$.
To match a HFS of $\sigma = 10^0$, BGSDC(1,3) requires \SI{5980}{\second} whereas Boris needs \SI{13169}{\second}.
This corresponds to a speedup of $2.2$, which is somewhat better than what we predict from counting RHS evaluations.}

\subsection{Results for the DIII-D tokamak: collisional case}
\begin{figure}[t]
	\begin{subfigure}{0.49\textwidth}
		\includegraphics[width=\linewidth]{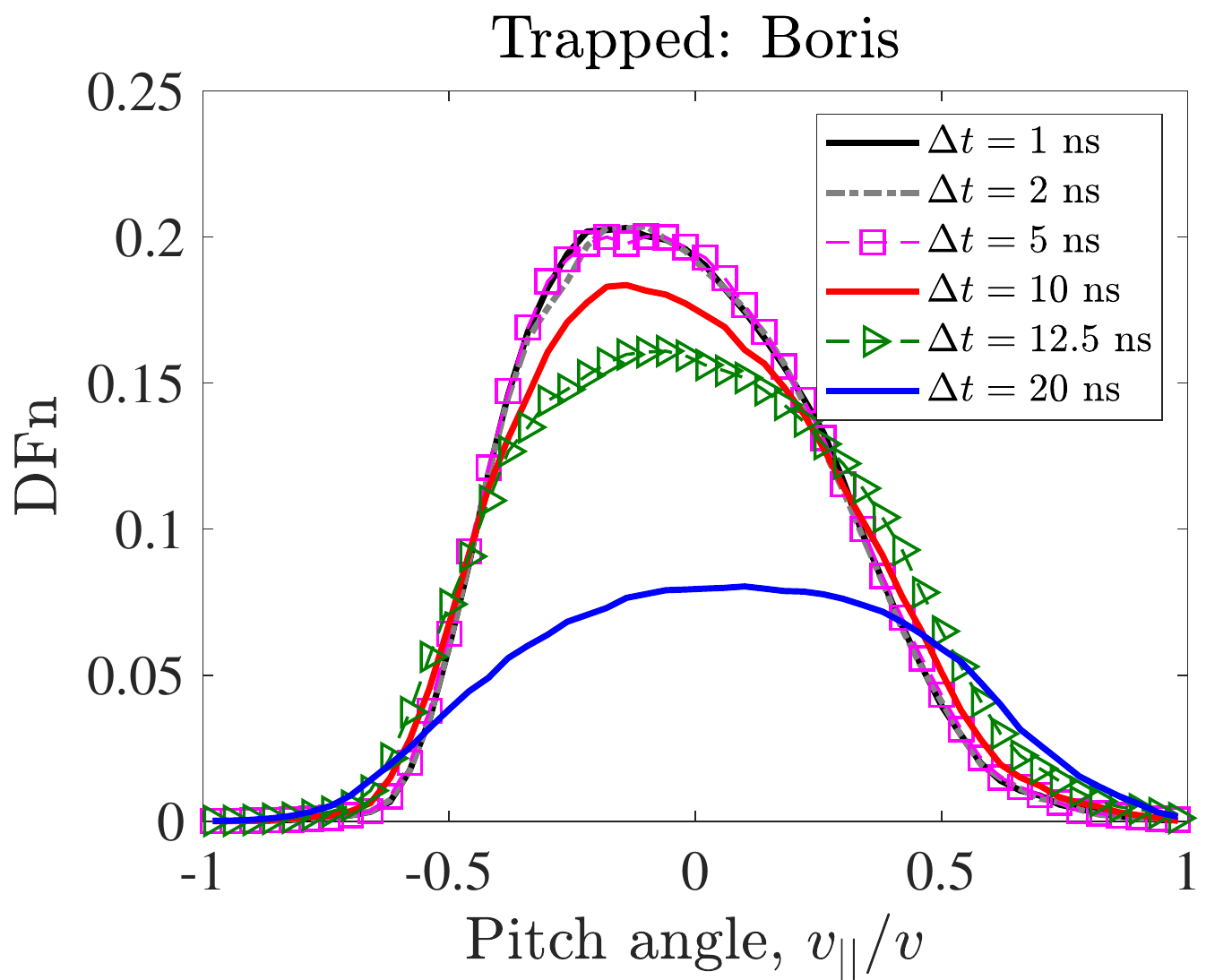}
	\end{subfigure}
	\hspace*{\fill} 
	\begin{subfigure}{0.49\textwidth}
		\includegraphics[width=\linewidth]{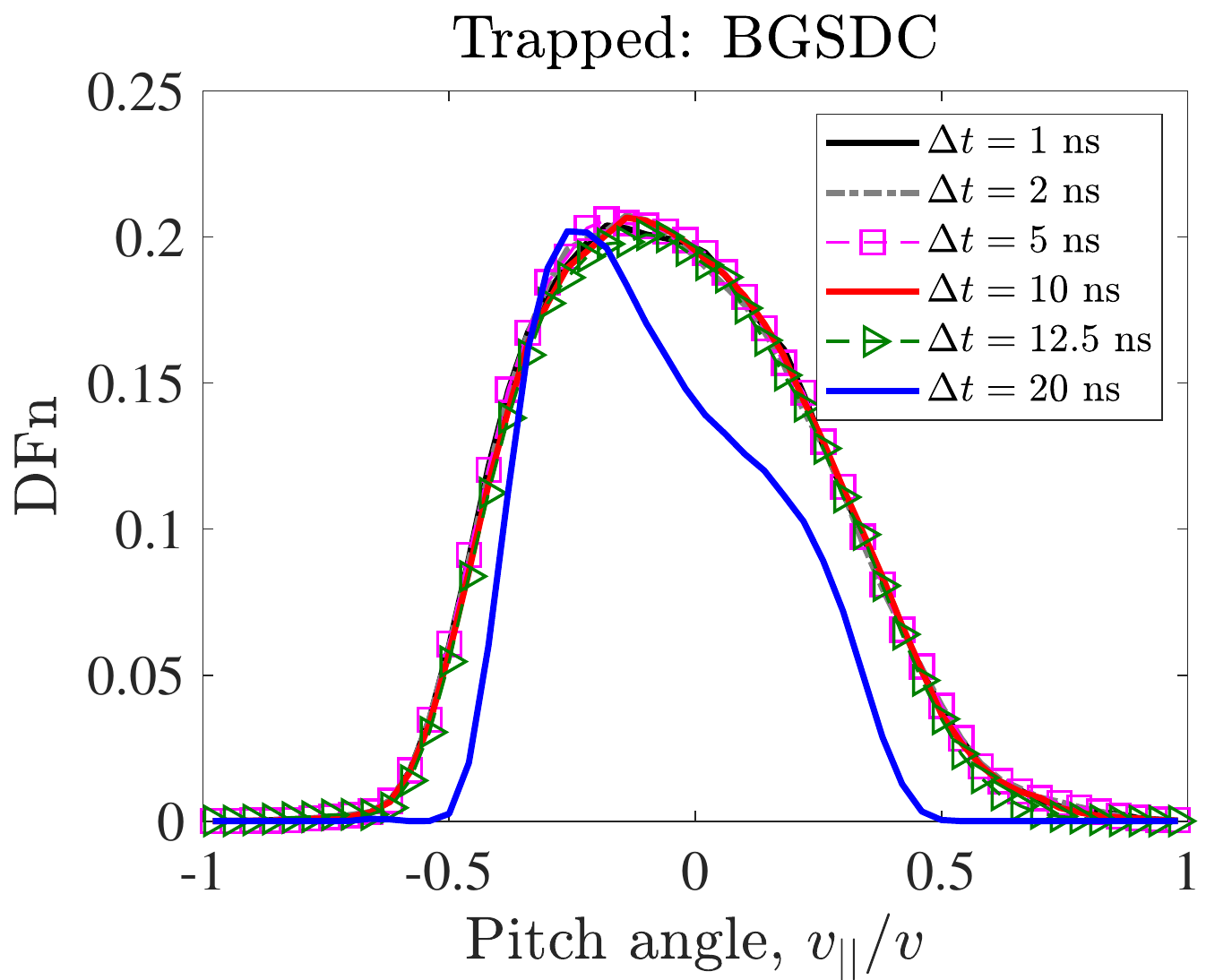}
	\end{subfigure}
	\par\bigskip 
	\begin{subfigure}{0.49\textwidth}
		\includegraphics[width=\linewidth]{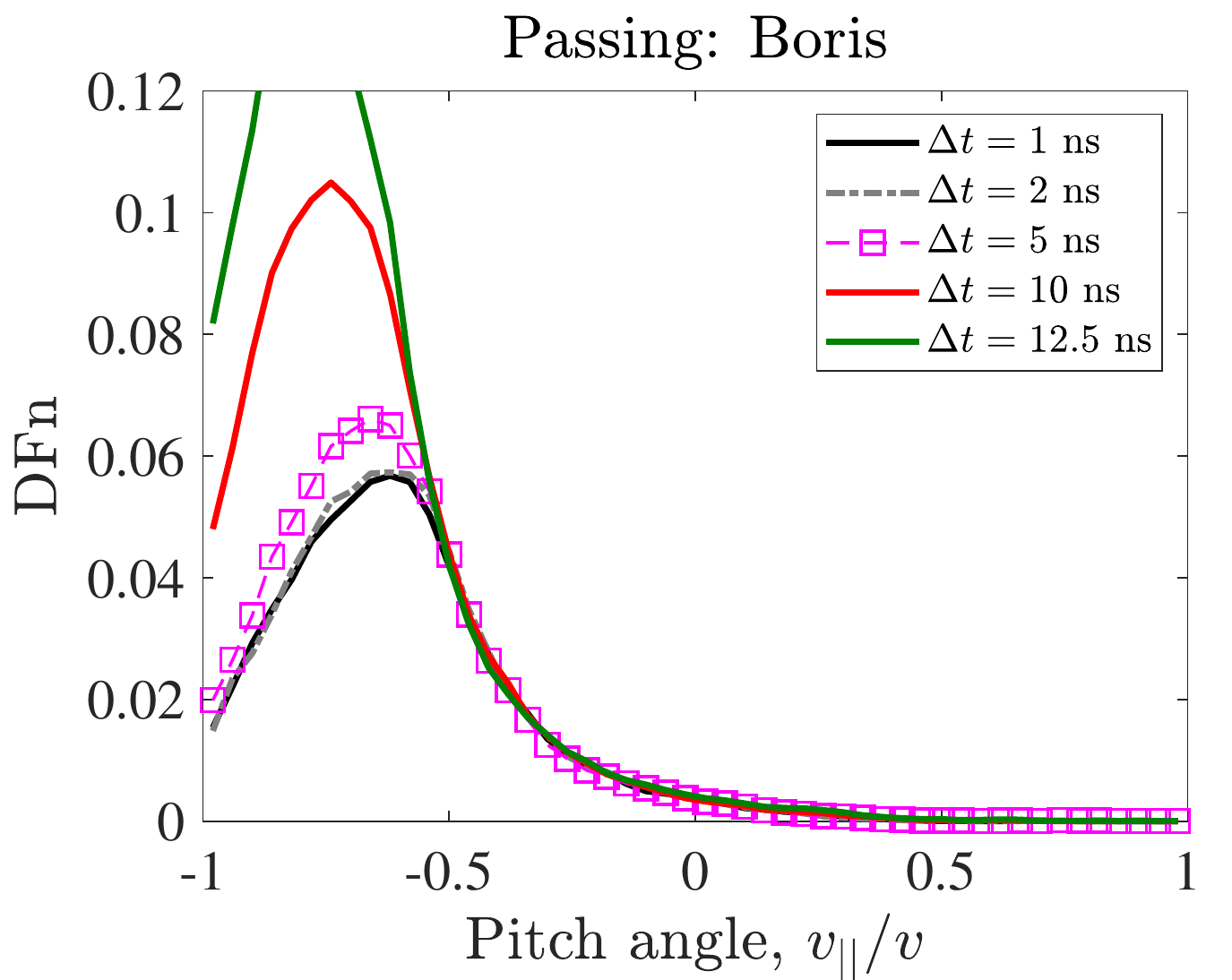}
	\end{subfigure}
	\hspace*{\fill} 
	\begin{subfigure}{0.49\textwidth}
		\includegraphics[width=\linewidth]{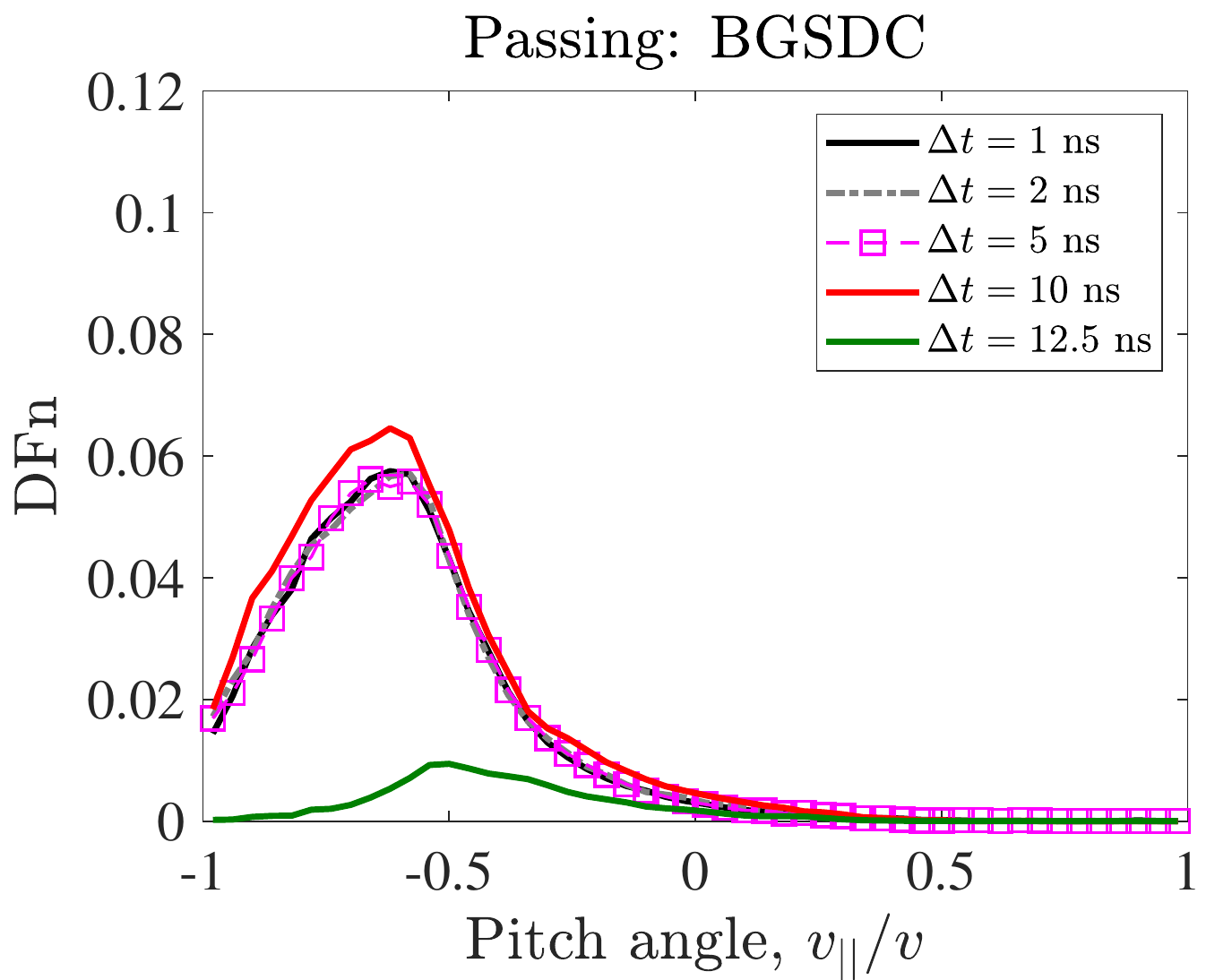}
	\end{subfigure}
	\caption{\reva{1D profiles of the distribution function for DIII-D for Boris method on the left and BGSDC(2,6) on the right.}}
	\label{fig:diiid_1dprofiles}
\end{figure}

\reva{LOCUST uses the same Fokker-Planck collision operator to model collisions as ASCOT.
A detailed description is provided by Hirvijoki et al.~\cite{HirvijokiEtAl2015}.}
In the collisional case, we launch one particle 131072 times from the same position. 
The duration of each run is $t_{end} = \SI{100}{\milli\second}$, the same as in the collisionless case, and we use the same 2D plasma equilibrium. 
\revb{We compare distribution functions generated by Boris and BGSDC.
Distribution functions have four arguments: velocity, coordinates $R$ and $Z$ and pitch angle $L$.
To visualize them in 2D plots, we fix velocity, $R$ and $Z$ and plot against the pitch angle $L$.}
Fig.~\ref{fig:diiid_1dprofiles} shows 1D cross-sections of the distribution function $F(v,L,R,Z)$ against pitch angle $ L = v_{\perp}/v_{\parallel}$ in a phase box for a trapped particle on the top with fixed velocity $v = \SI{1.8e6}{\meter/\second}$, major radius $ R = \SI{2}{\meter}$, axis $Z = \SI{0.5}{\meter}$ and for a passing particle at the bottom with $ v = \SI{1.8e6}{\meter/\second}$, $ R = \SI{1.9}{\meter}$ and $ Z = \SI{0.3}{\meter}$.

Boris and BGSDC both converge to the same distribution.
However, BGSDC produces a stable distribution for larger time steps than Boris.
For trapped orbits, the distributions provided by BGSDC with \SI{1}{\nano\second} and \SI{10}{\nano\second} time steps are very similar whereas Boris shows visible differences at a time step of \SI{10}{\nano\second}.
For passing orbits, both methods require slightly smaller time steps to produce stable distributions.
BGSDC has converged for a step size of \SI{5}{\nano\second} whereas Boris requires a smaller step size of \SI{2}{\nano\second}.
\subsection{Results for the JET tokamak: non-collisional case}
For JET, full orbit simulations were carried out in a 2D equilibrium representing typical JET conditions.
We use an artificial source of particles uniformly distributed in pitch angle and inside the plasma volume at a fixed injection energy. 
All runs last $\SI{1}{\second}$ and use 65536 unique particles.
As in the DIII-D runs, we do not include PFCs. 

\paragraph{Numerical drift}
\begin{figure}[t]
	\begin{subfigure}{0.49\textwidth}
		\includegraphics[width=\linewidth]{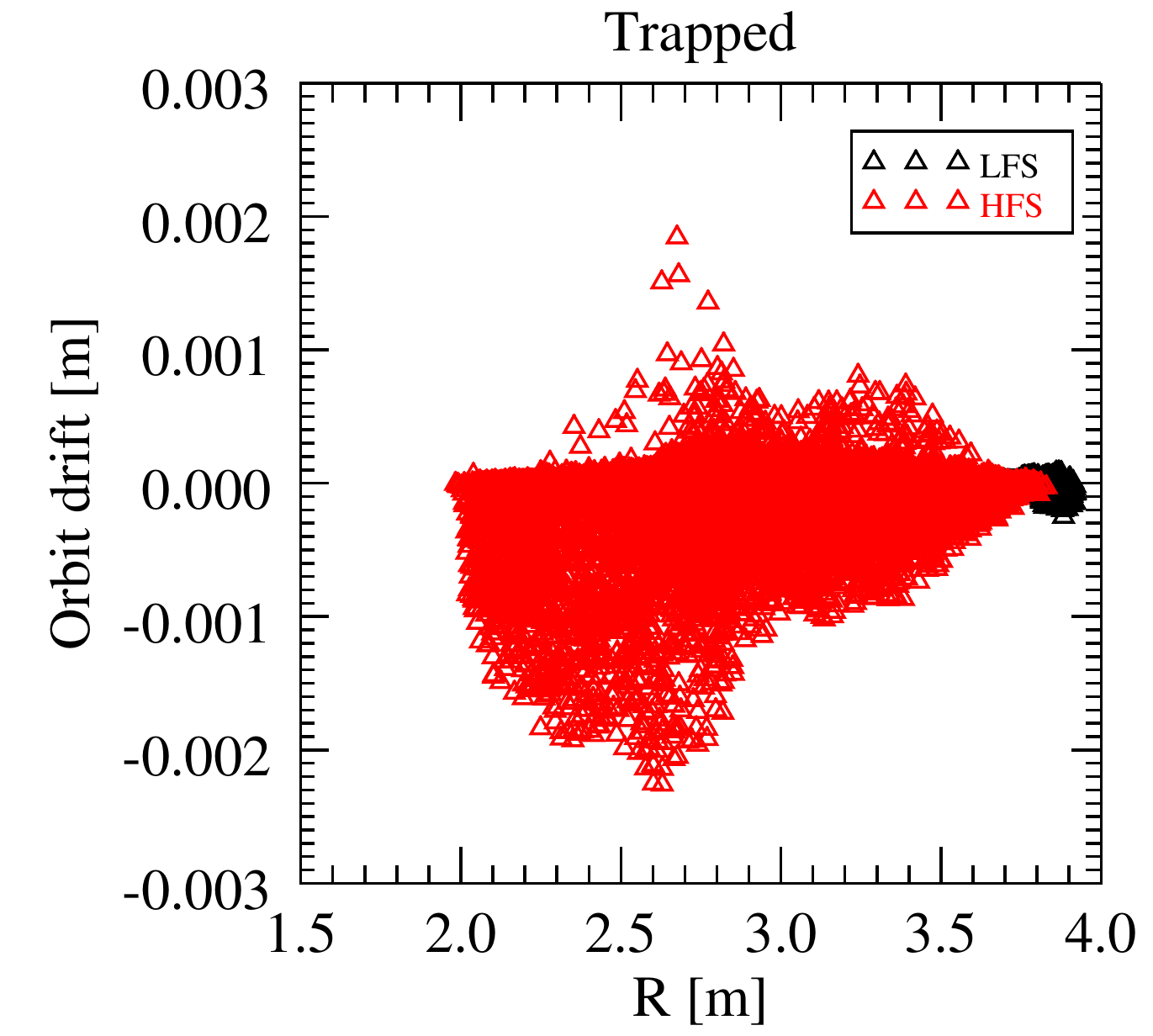}
	\end{subfigure}
	\begin{subfigure}{0.49\textwidth}
		\includegraphics[width=\linewidth]{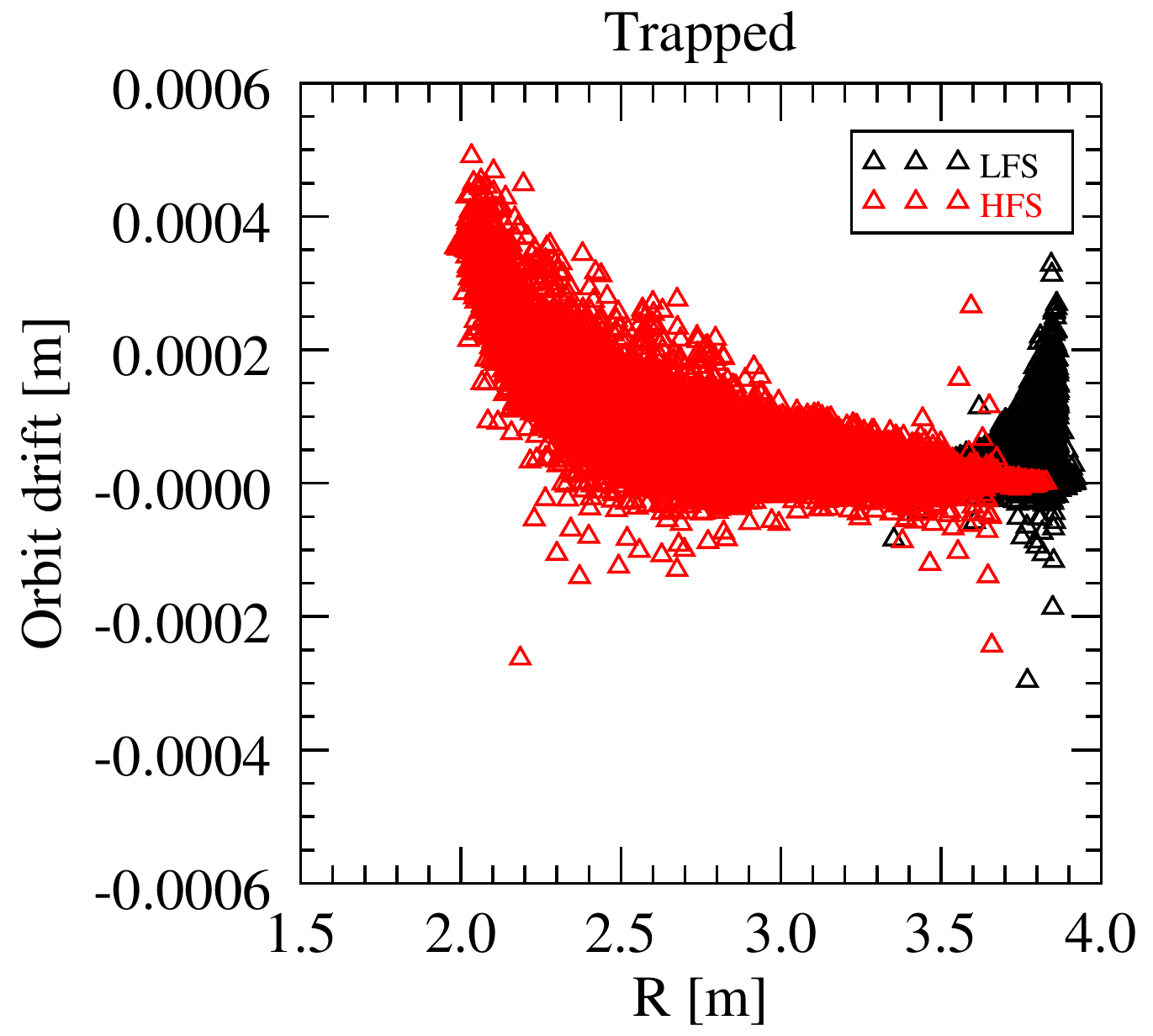}
	\end{subfigure}
	\begin{subfigure}{0.49\textwidth}
		\includegraphics[width=\linewidth]{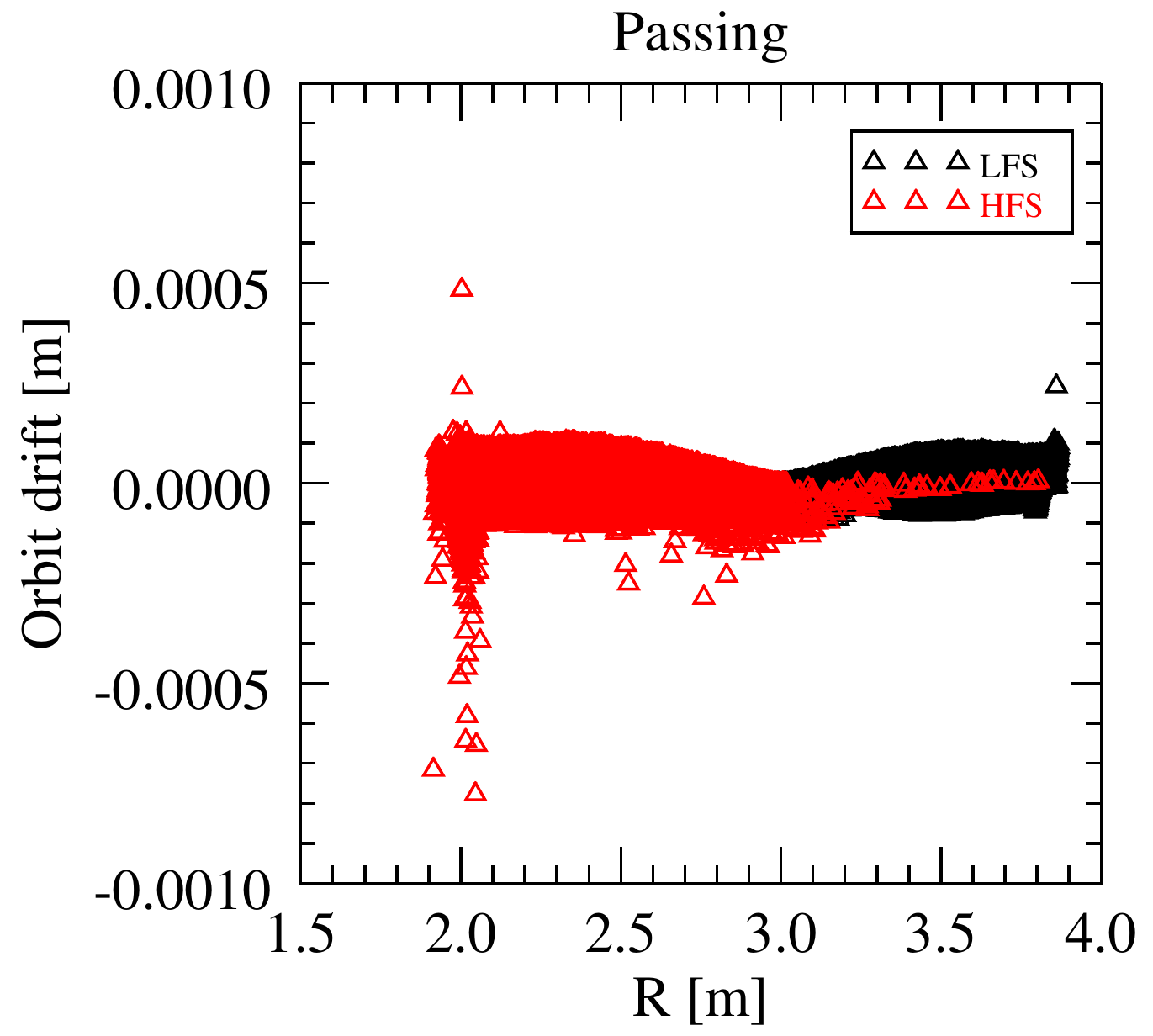}
	\end{subfigure}
	\begin{subfigure}{0.49\textwidth}
		\includegraphics[width=\linewidth]{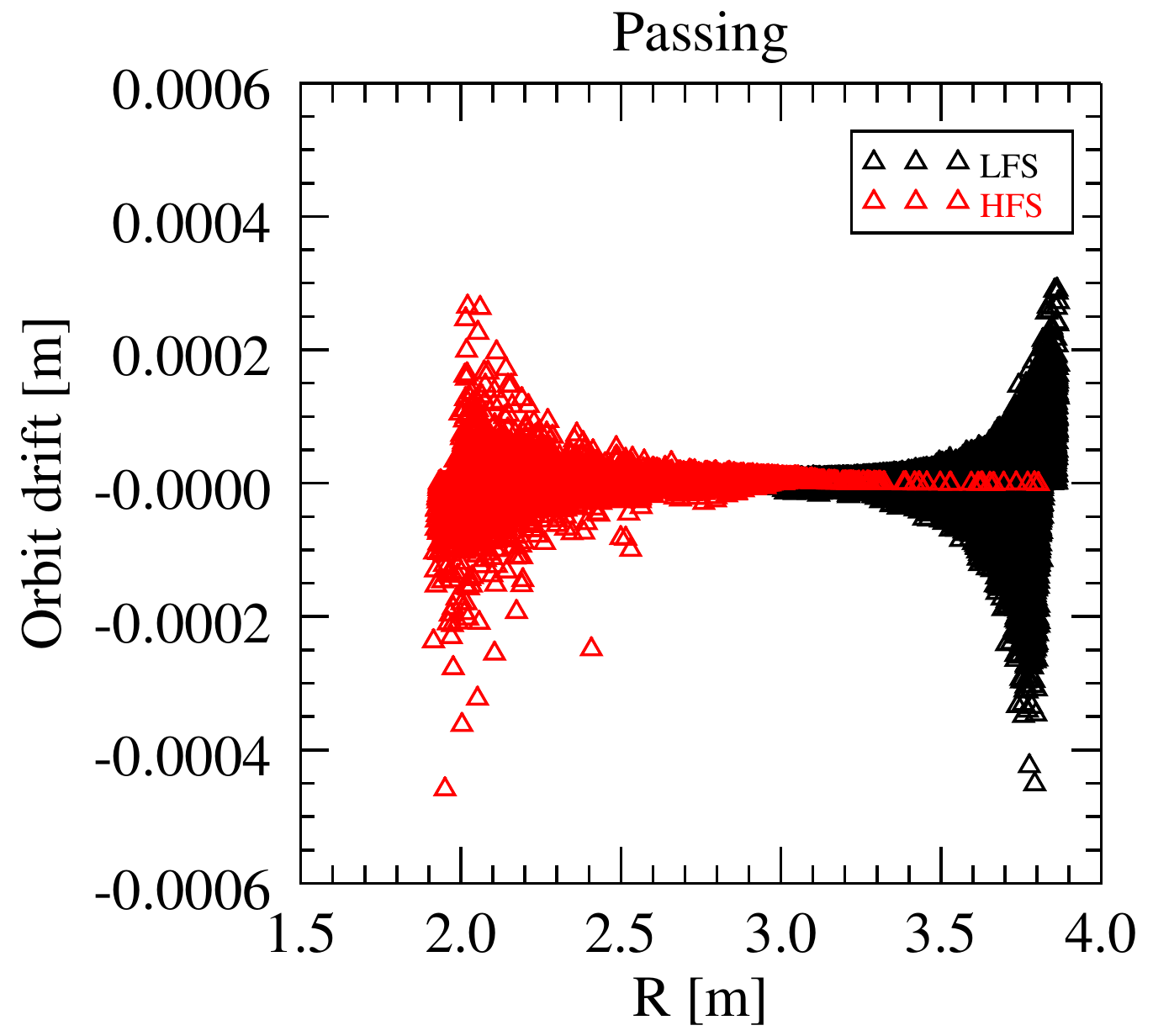}
	\end{subfigure}
	\caption{Particle drift for the JET tokamak after $\SI{1}{\second}$ simulation time. Boris method $\Delta t = \SI{0.5}{\nano\second}$ (left) and BGSDC(1,4) $\Delta t = \SI{2}{\nano\second}$ (right)}
	\label{fig:jet_ptcldrift}
\end{figure}
Fig.~\ref{fig:jet_ptcldrift} shows the numerical orbital drift for all particles at the end of simulation at $t_{end} = \SI{1}{\second}$. 
Results from the Boris method with time step $\SI{0.5}{\nano\second}$ are shown on the left and from BGSDC(1,4) with $ \SI{2}{\nano\second} $ on the right. 
Again, the upper graphs show drift for trapped particles while the lower graphs show drift for passing particles and values for LFS are marked in black while values for HFS are marked red.
For trapped particles BGSDC(1,4) shows a maximum deviation of $ \SI{5e-4}{\meter}$ whereas Boris with a 4 times smaller step shows a $ \SI{2e-3}{\meter}$ deviation. 
Please note that the $y$-axes scale changes.
For passing particles both methods deliver comparable accuracy with BGSDC showing somewhat lower peak values.

\begin{figure}[t]
	\begin{subfigure}{0.49\textwidth}
		\includegraphics[width=\linewidth]{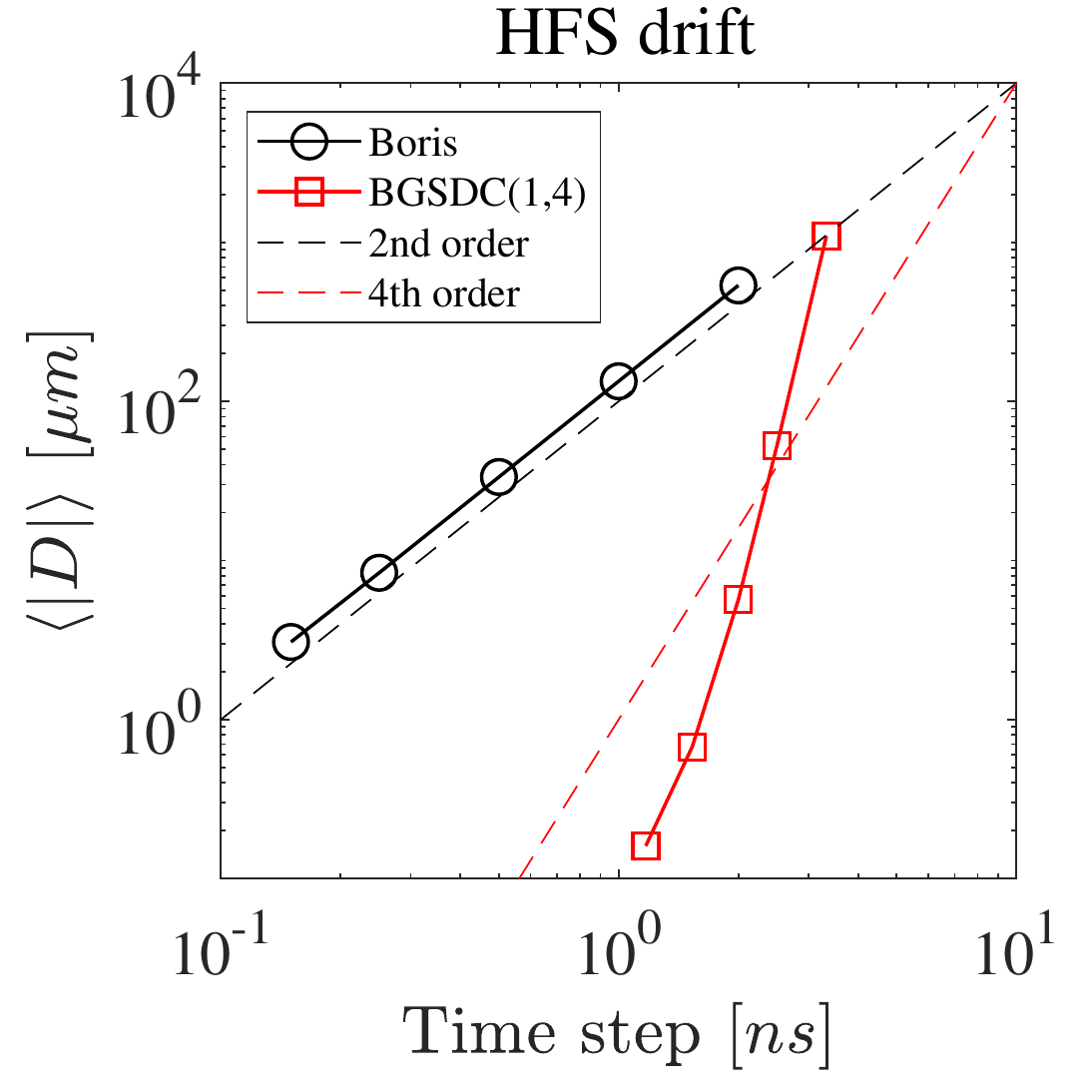}
	\end{subfigure}
	\begin{subfigure}{0.49\textwidth}
		\includegraphics[width=\linewidth]{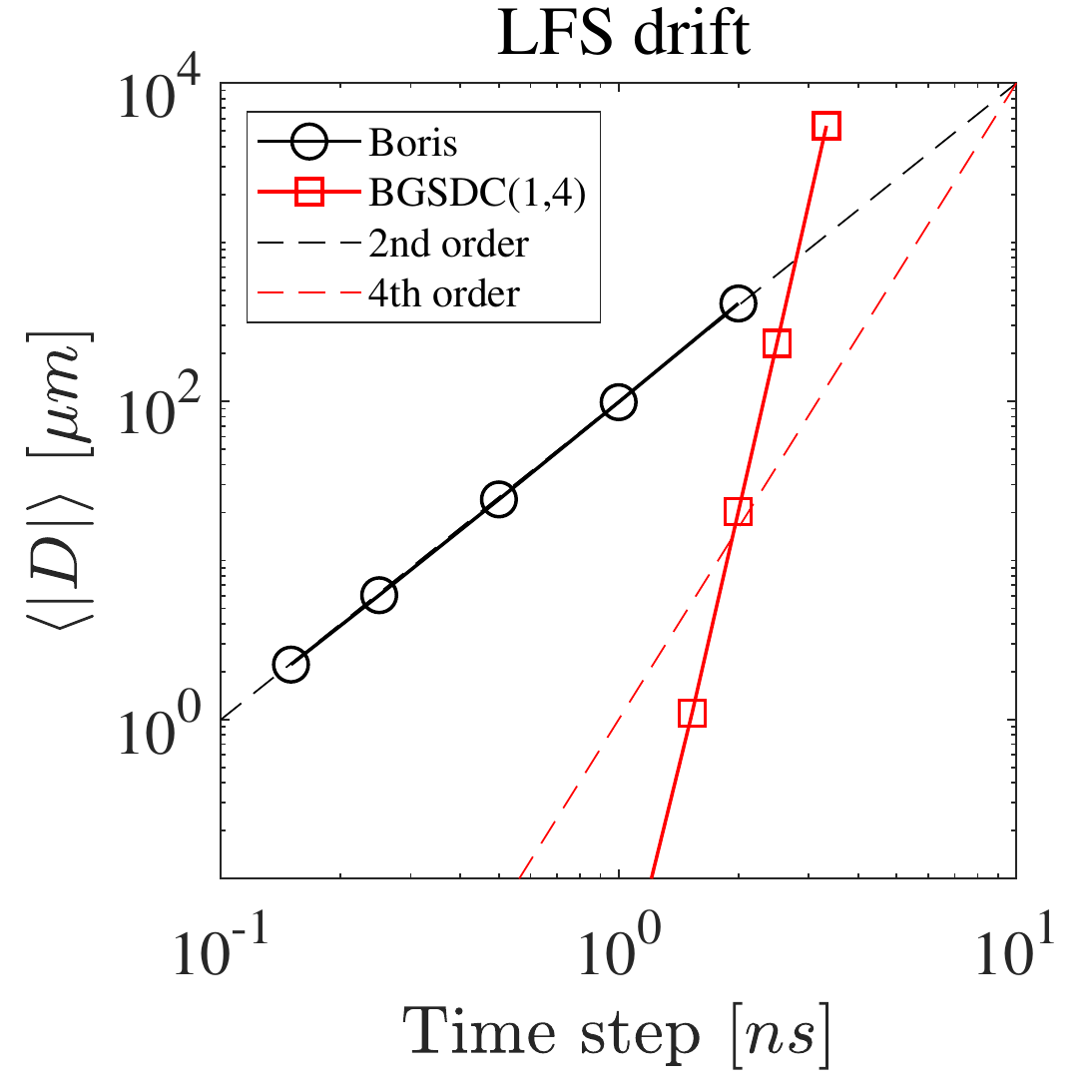}
	\end{subfigure}
	\begin{subfigure}{0.49\textwidth}
		\includegraphics[width=\linewidth]{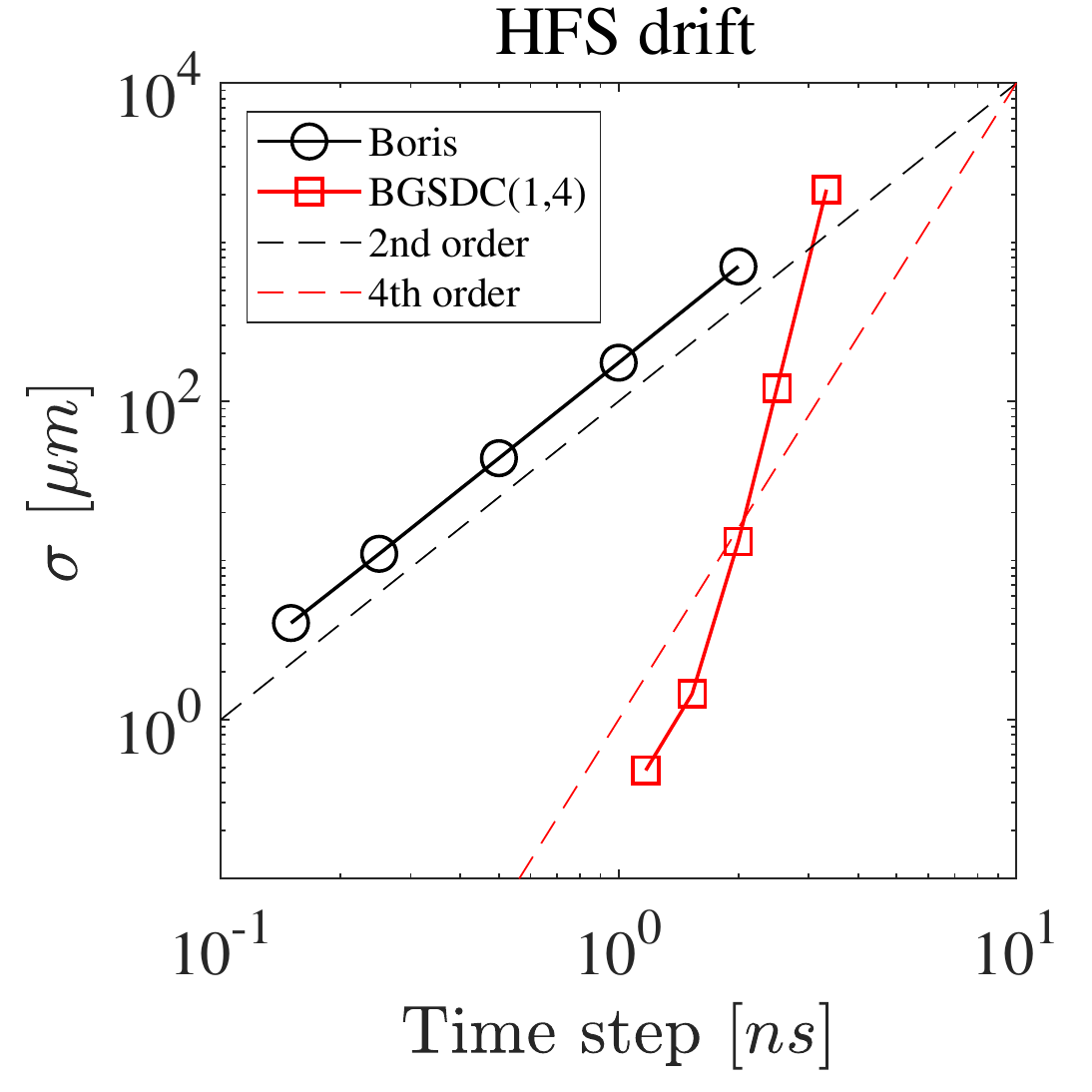}
	\end{subfigure}
	\begin{subfigure}{0.49\textwidth}
		\includegraphics[width=\linewidth]{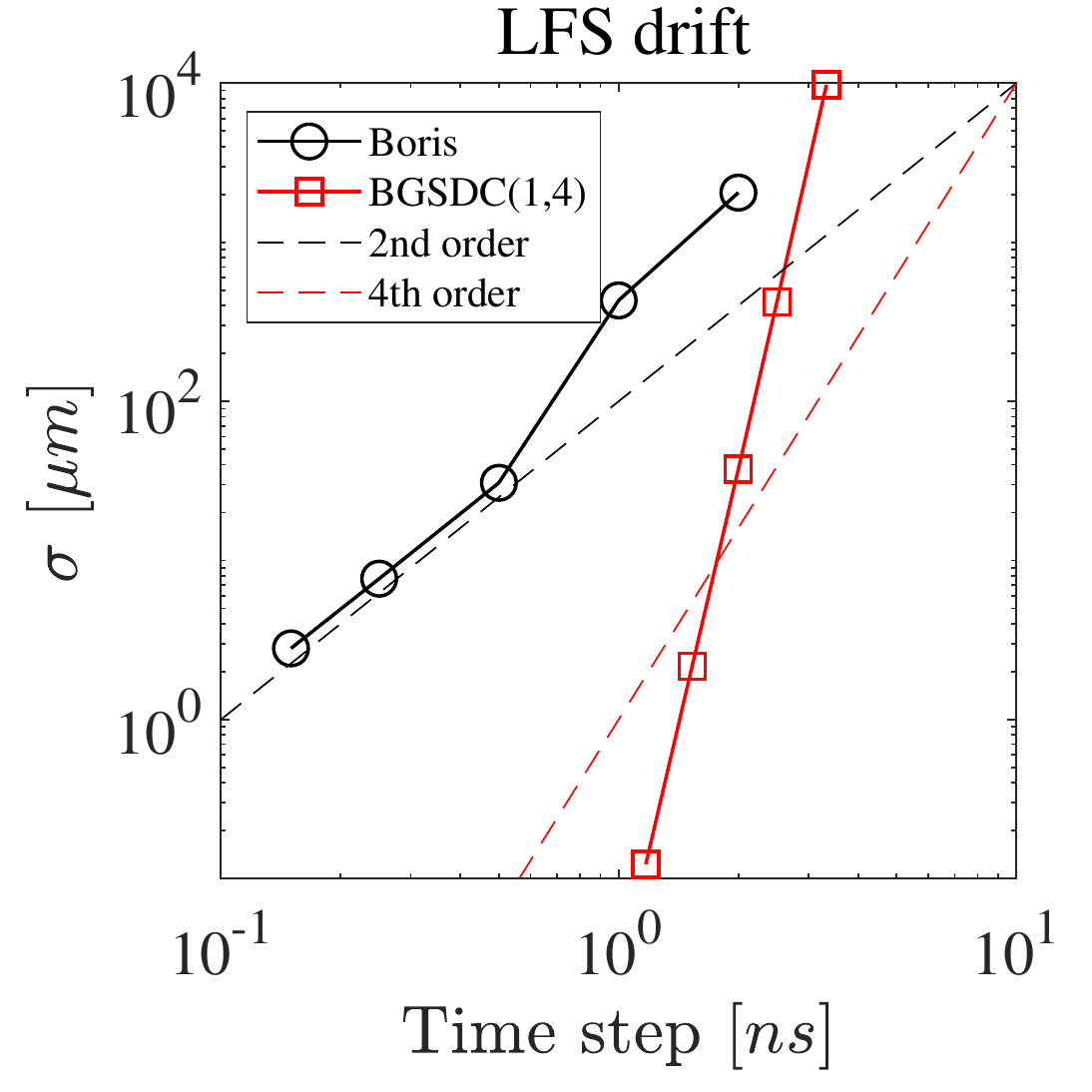}
	\end{subfigure}
	\caption{\revb{Mean (upper)} and standard deviation $\sigma$ \revb{(lower)} of the drift distribution for classical Boris and BGSDC methods for different time steps for the non-collisional regime in JET.}
	\label{fig:jet_sigma}
\end{figure}
Fig. \ref{fig:jet_sigma} shows the \reva{mean (upper) and} standard deviation $\sigma$ (lower) of the drift distribution  for both integrators depending on the time step. 
As for DIII-D, the higher order of accuracy of BGSDC leads to a steeper decrease in $\sigma$ with time step size.
For a fixed $\Delta t$, BGSDC produces much narrower distributions than Boris for both HFS and LFS.

\begin{figure}[ht]
	\centering
	\begin{subfigure}{0.49\linewidth}
	\includegraphics[width=0.99\linewidth]{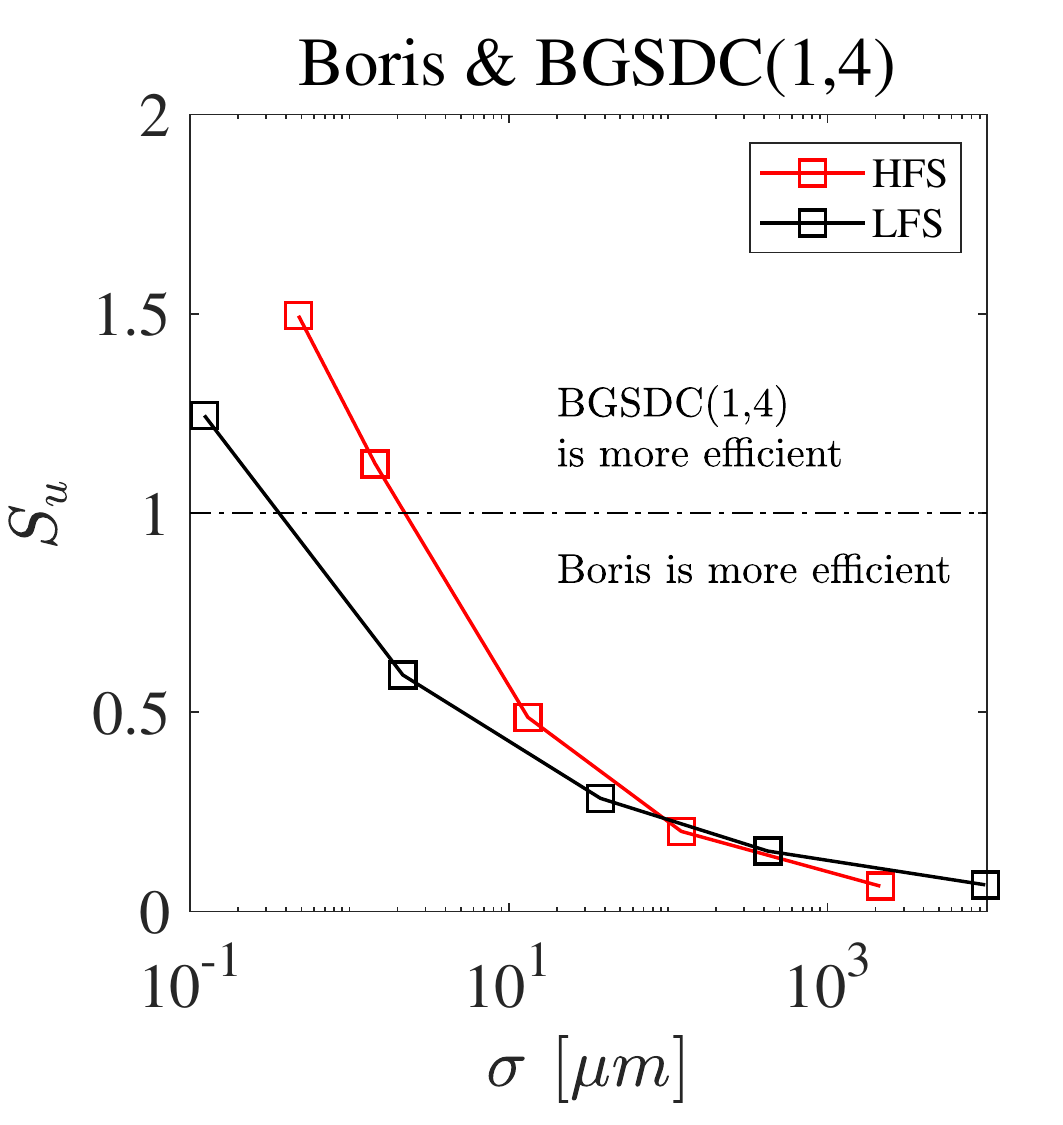}
	\end{subfigure}
	\begin{subfigure}{0.49\linewidth}
	\includegraphics[width=0.99\linewidth]{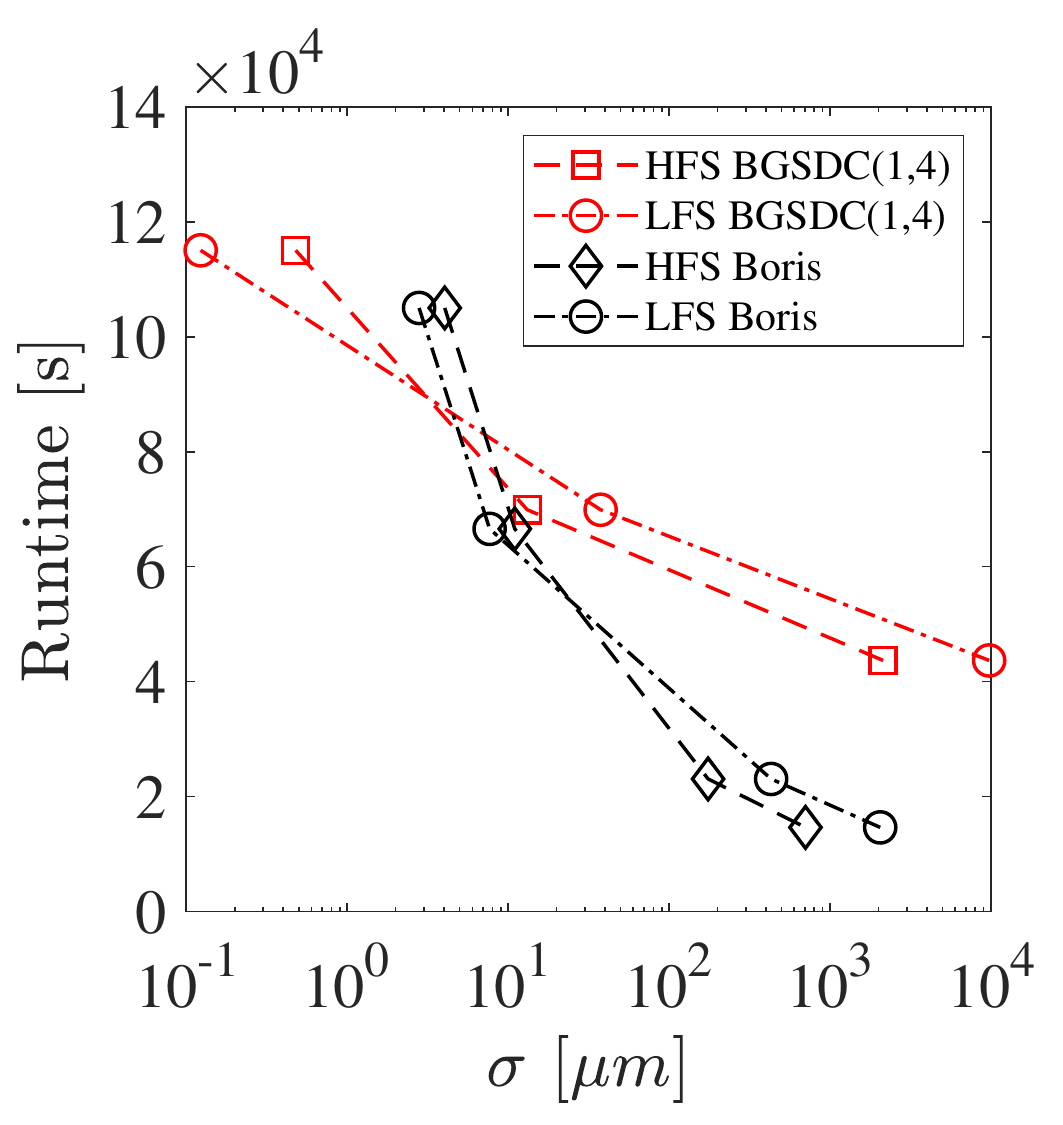}
	\end{subfigure}	
	\caption{Left: Ratio $S_u$ of field evaluations required by Boris to reach some standard deviation $\sigma$ divided by the number of evaluations required by BGSDC(1,4) for the JET reactor. \reva{Right: Runtime required to reach some standard deviation $\sigma$ for Boris (black) and BGSDC(1,4) (red).}}
	\label{fig:jet_speedup}
\end{figure}

%
%
\paragraph{Work-precision}
Fig.~\ref{fig:jet_speedup} (left) shows again the ratio of RHS evaluations for Boris compared to BGSDC required to produce a distribution with a given $\sigma$.
Results are very similar to those for DIII-D.
BGSDC becomes competitive in the range between $\sigma = \SI{0.1}{\micro\meter}$ and $\sigma = \SI{1}{\micro\meter}$ for the HFS and LFS with better gains for HFS.
\reva{Fig.~\ref{fig:jet_speedup} (right) shows runtimes needed to reach a given $\sigma$.
For JET, the break-even point comes earlier than for DIII-D.
BGSDC is faster than Boris for distributions with $\sigma$ around \SI{3}{\micro\meter} or less for both HFS and LFS.}

\subsection{Results for the JET tokamak: collisional case}
\begin{figure}[ht!]
	\begin{subfigure}{0.245\textwidth}	\includegraphics[width=\linewidth]{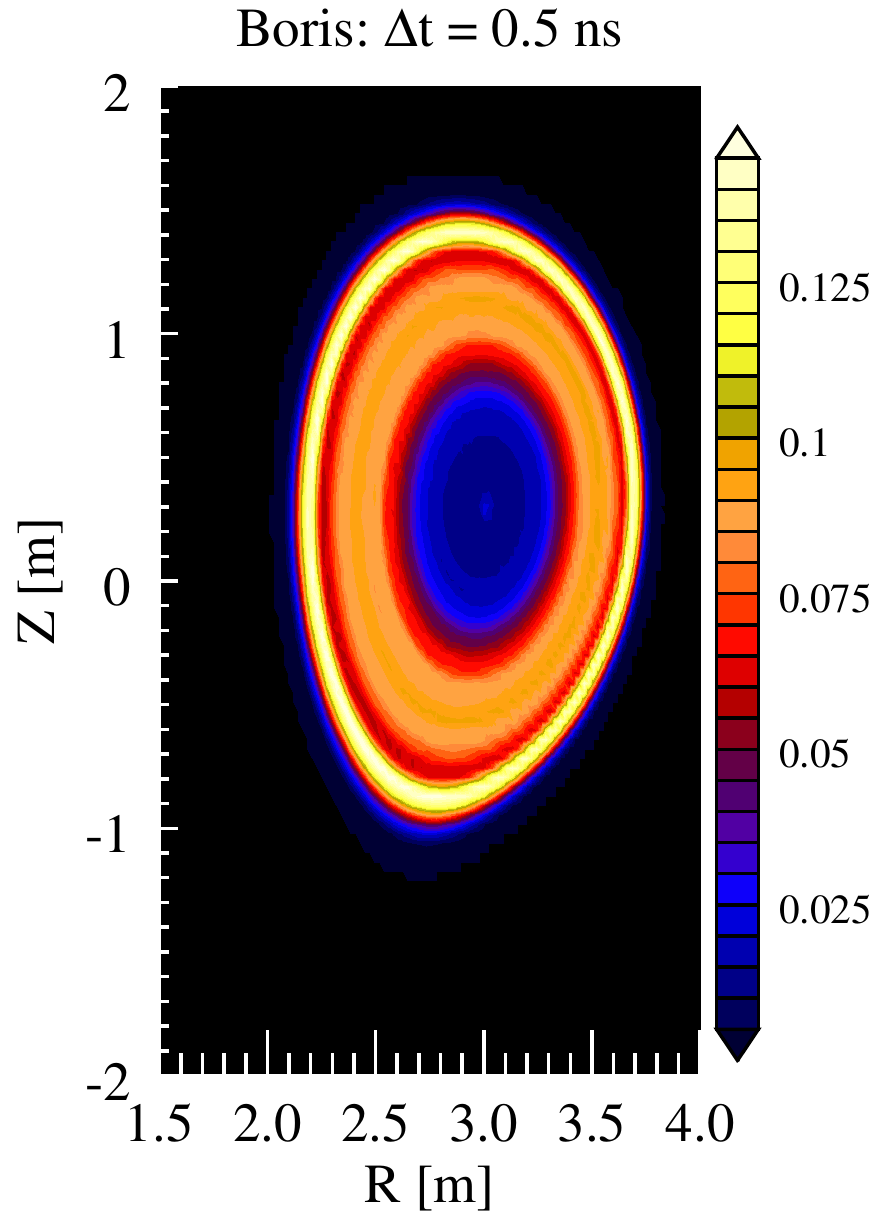}
	\end{subfigure}
	\begin{subfigure}{0.245\textwidth}
	\includegraphics[width=\linewidth]{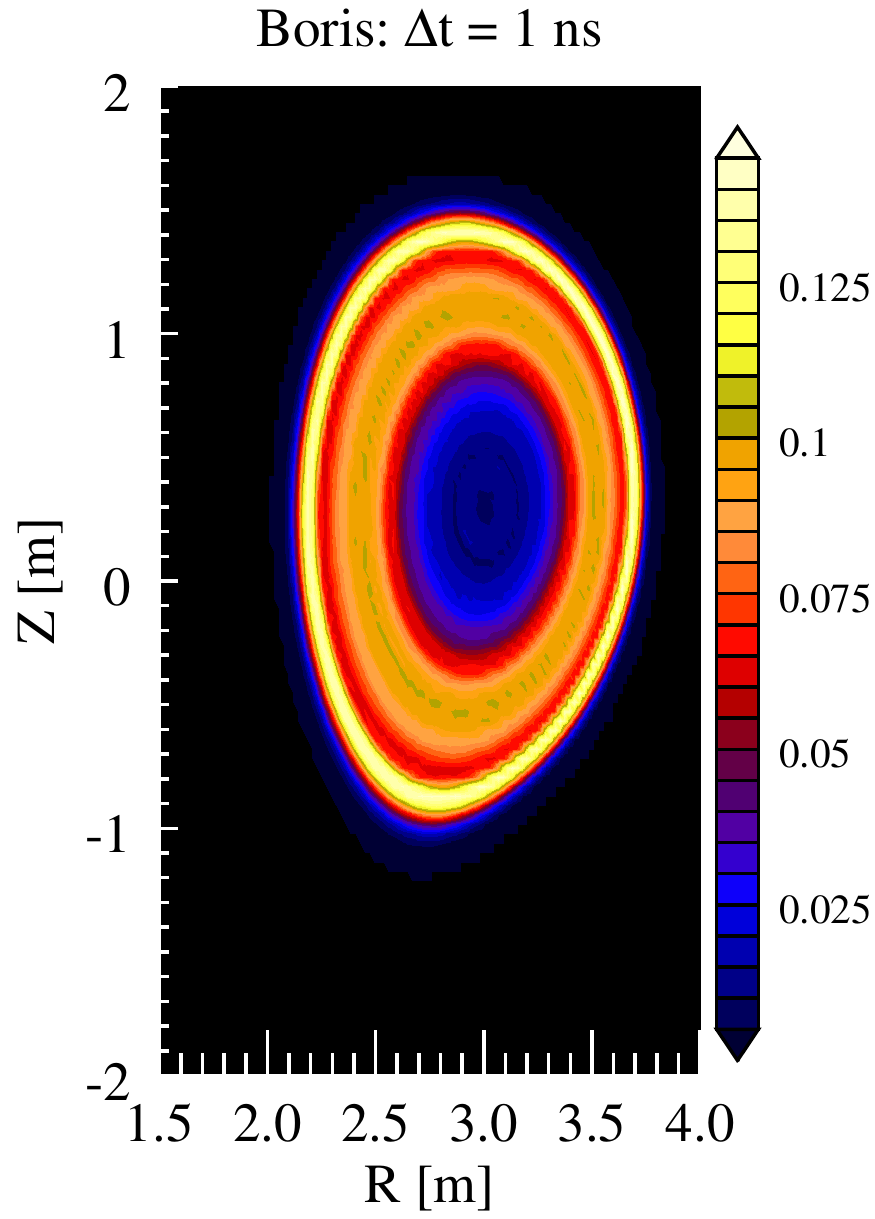}
	\end{subfigure}
	\begin{subfigure}{0.245\textwidth}
	\includegraphics[width=\linewidth]{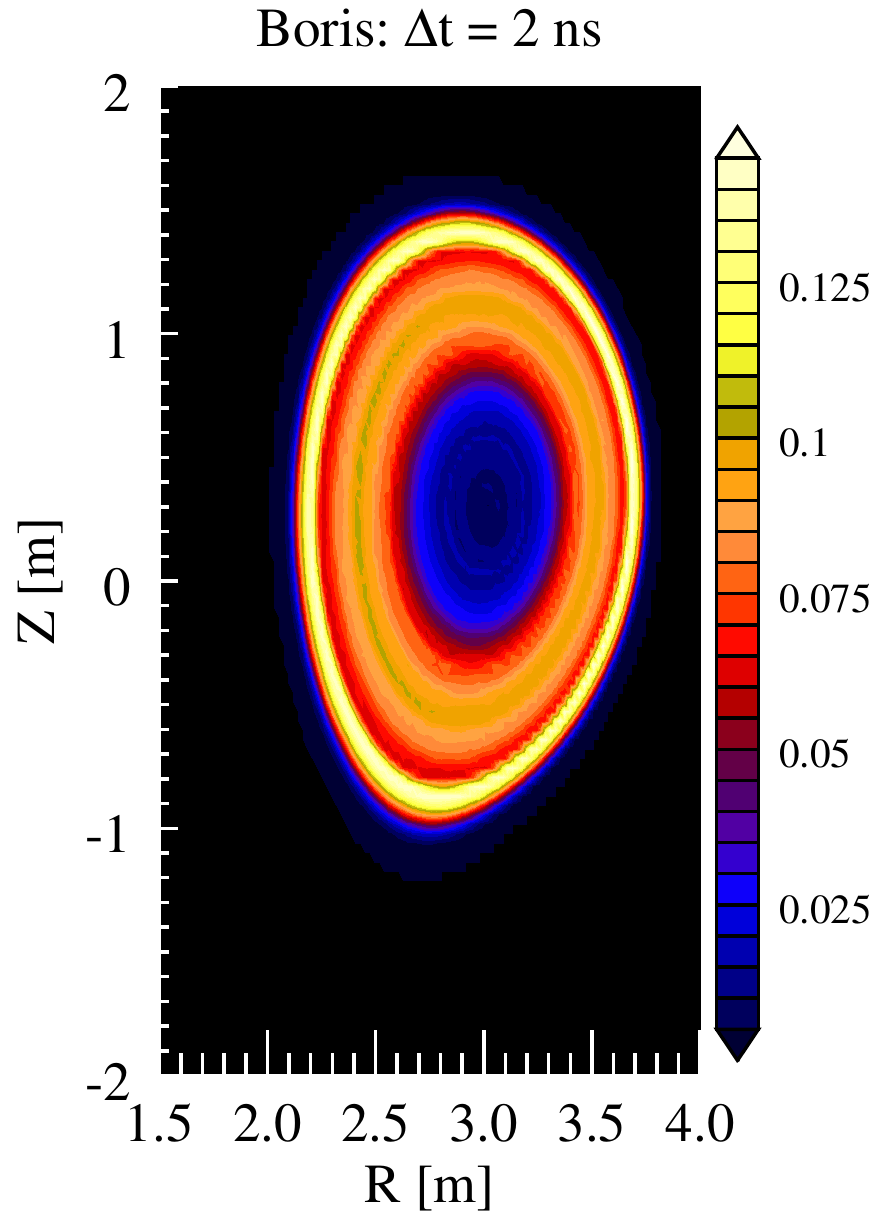}
	\end{subfigure}
	\begin{subfigure}{0.245\textwidth}
	\includegraphics[width=\linewidth]{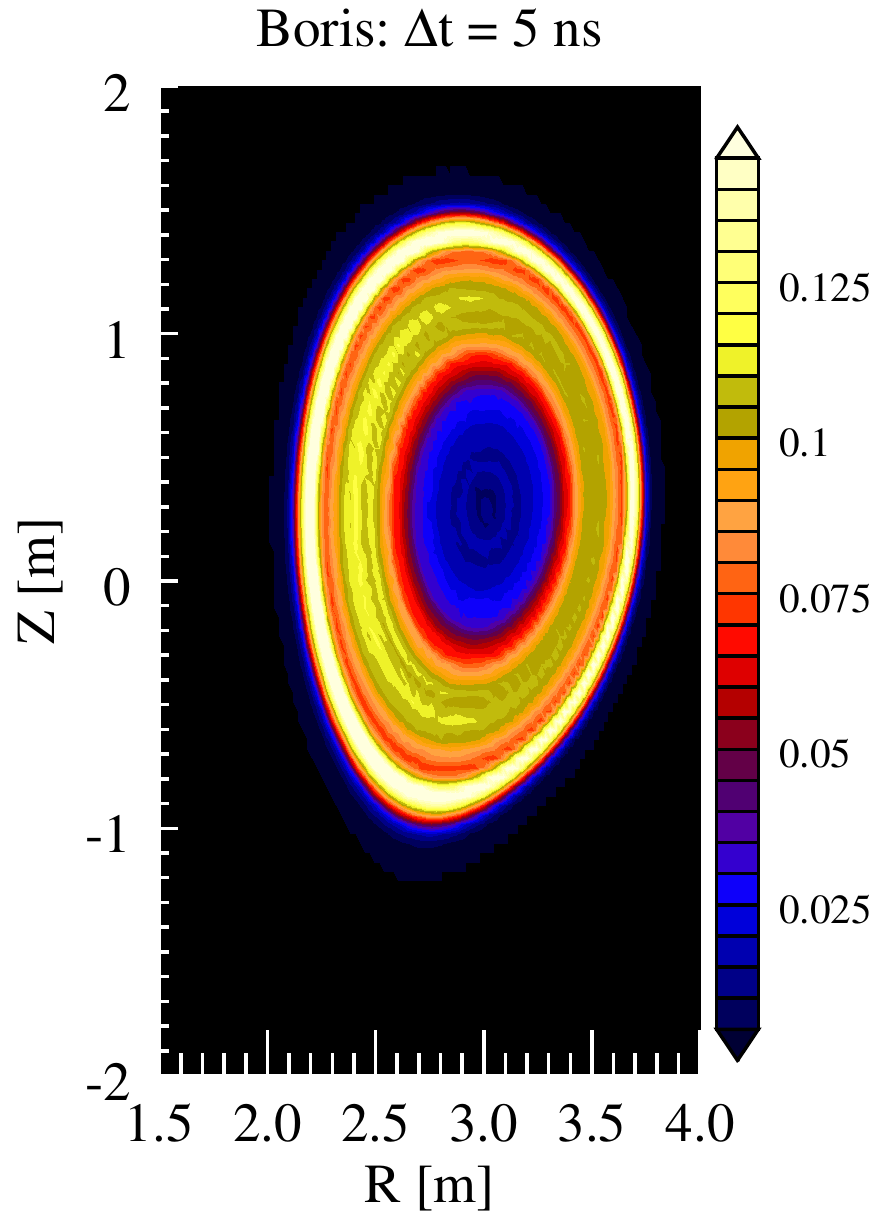}
	\end{subfigure}
	\par\bigskip 
	\begin{subfigure}{0.245\textwidth}	\includegraphics[width=\linewidth]{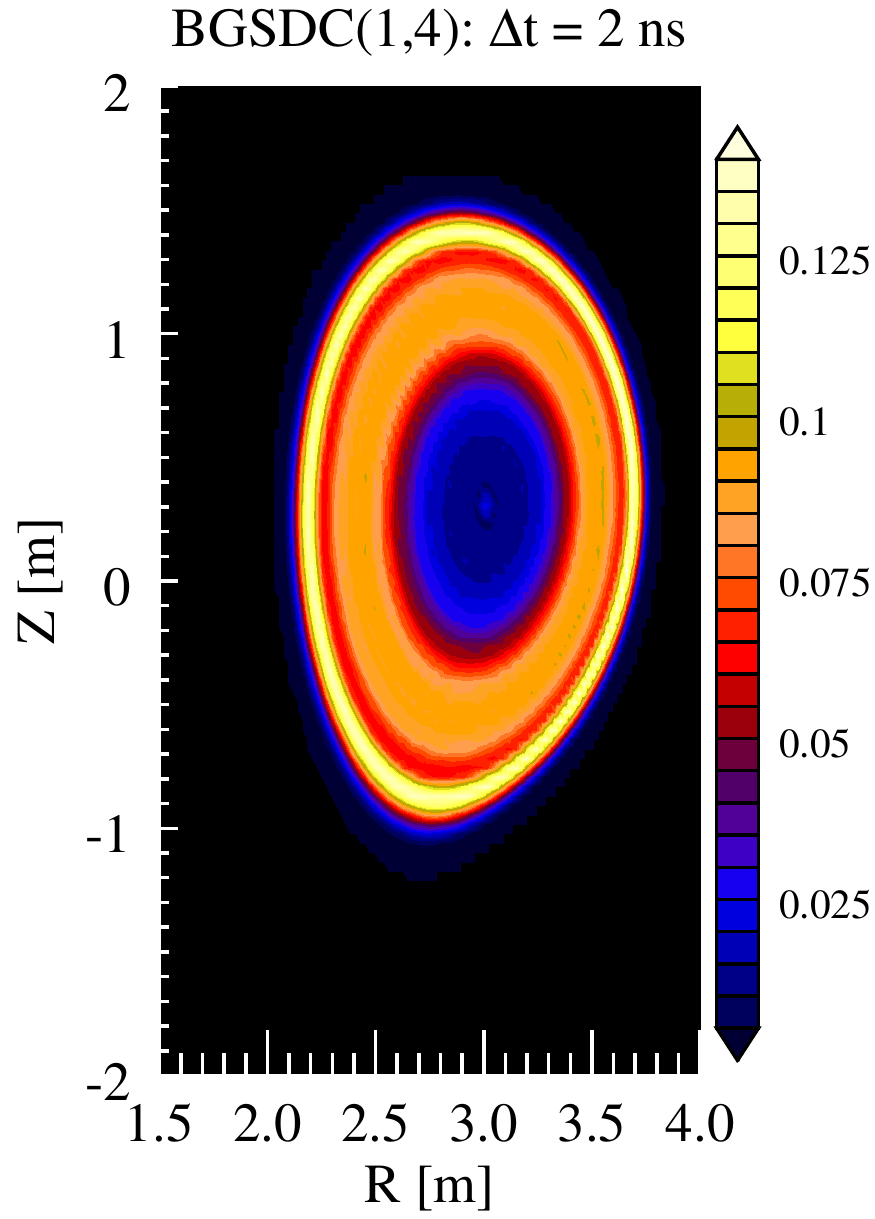}
	\end{subfigure}
	\begin{subfigure}{0.245\textwidth}
	\includegraphics[width=\linewidth]{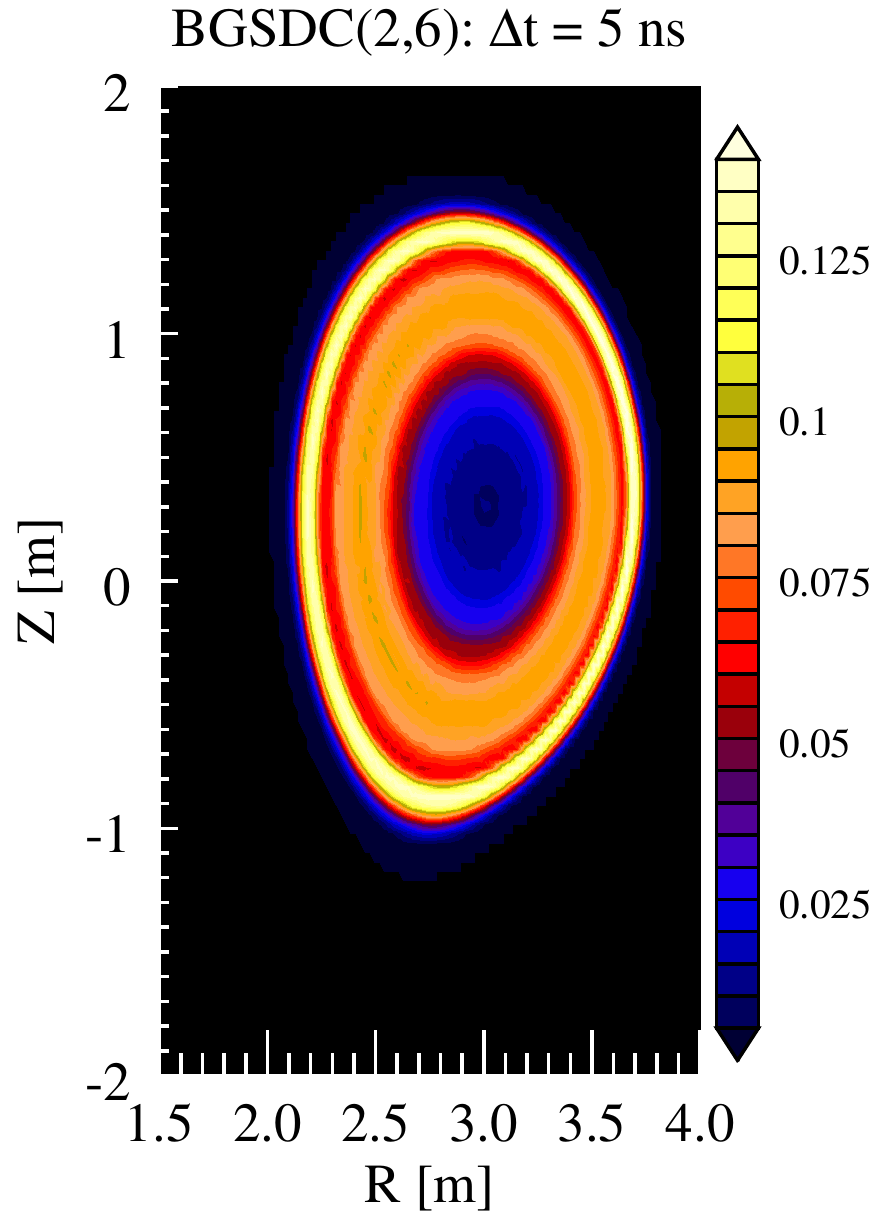}
	\end{subfigure}
	\begin{subfigure}{0.245\textwidth}
	\includegraphics[width=\linewidth]{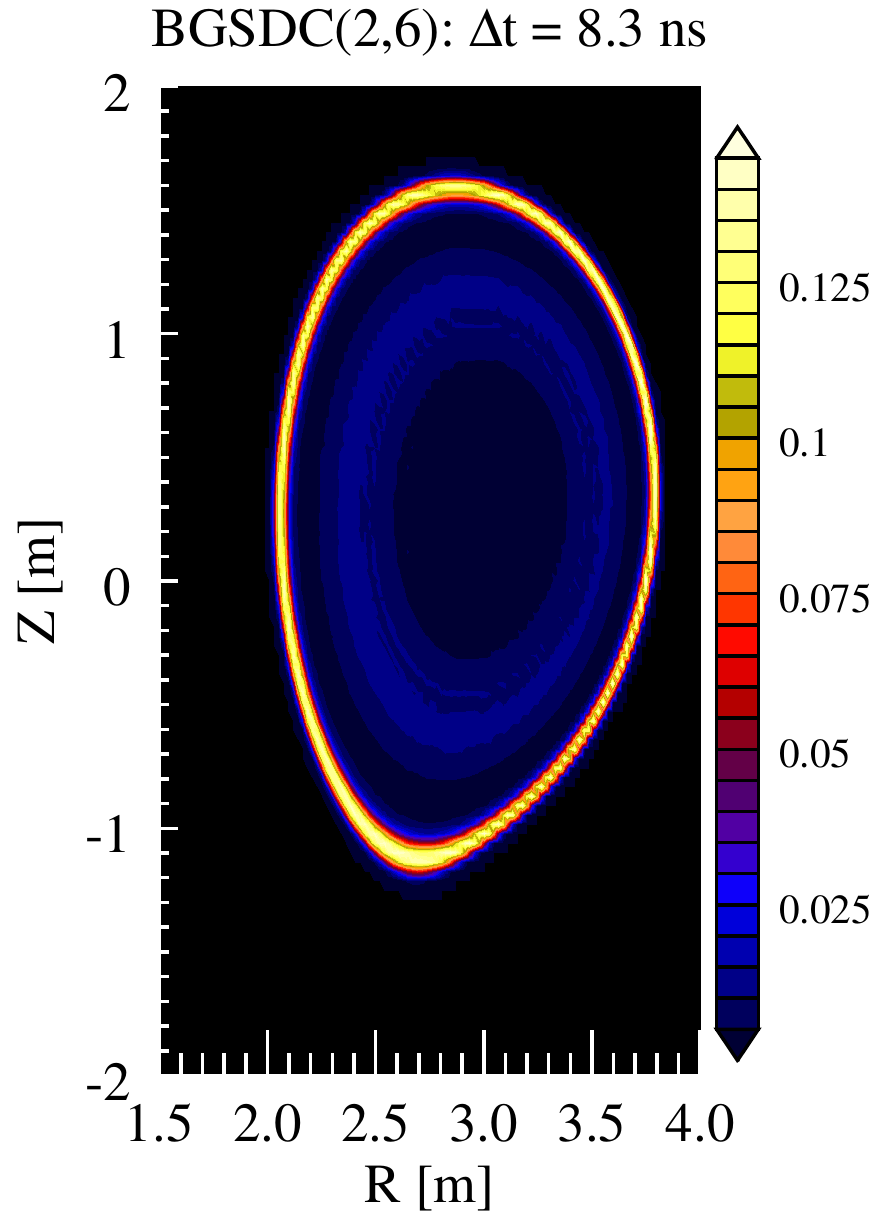}
	\end{subfigure}
	\begin{subfigure}{0.245\textwidth}
	\includegraphics[width=\linewidth]{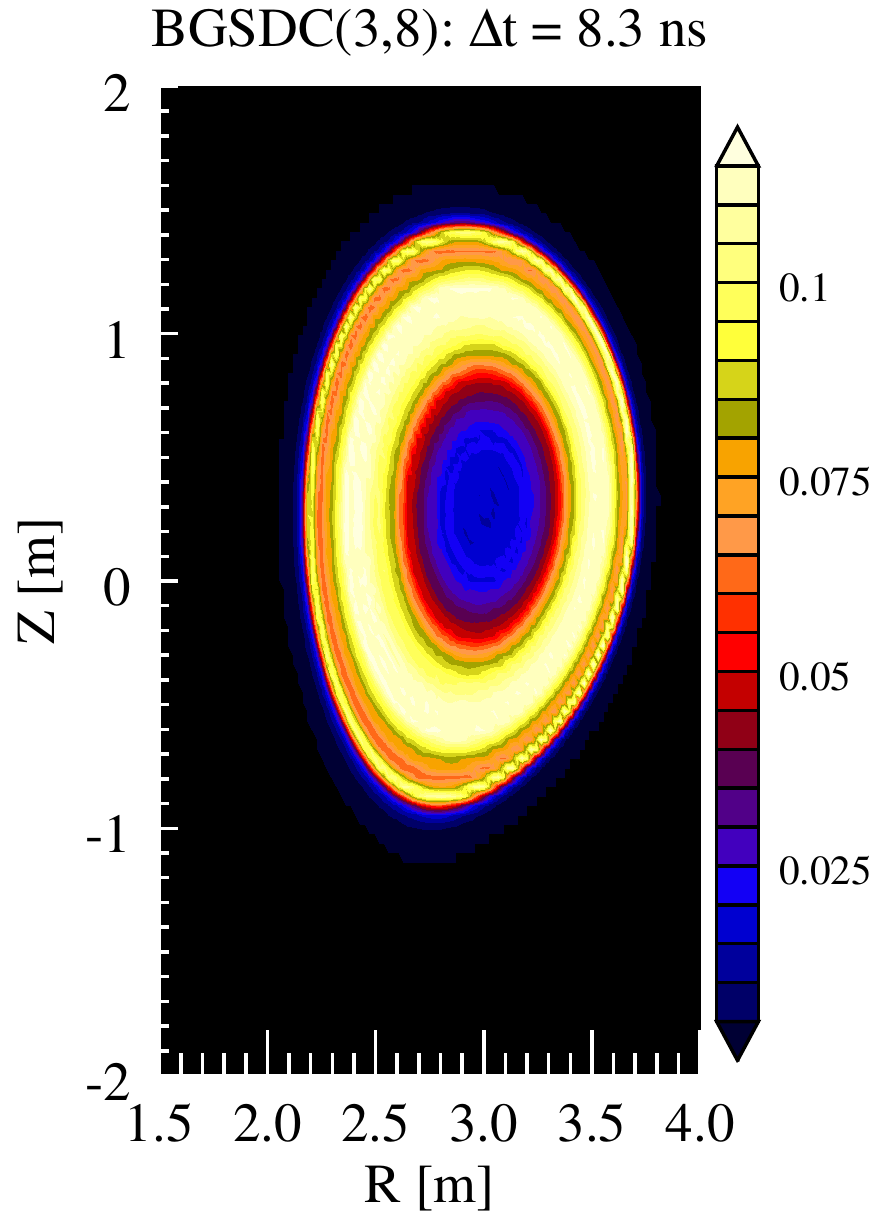}
	\end{subfigure}
	\caption{\reva{2D profiles of the distribution function for fixed velocity and pitch angle.}}
	\label{fig:jet_2dprof}
\end{figure}

For the collisional case we launch one particle 131072 times from the same position and track it until simulated time $\SI{1}{\second}$. 
Fig.~\ref{fig:jet_2dprof} shows 2D profiles of the distribution function in a minor cross-section of the JET tokamak.
BGSDC with $\Delta t = 2$ and $\Delta t =  \SI{5}{\nano\second}$ as well as Boris with $\Delta t = \SI{2}{\nano\second}$ deliver comparable profiles. 
Profiles computed with Boris with time steps $\SI{5}{\nano\second}$ and larger show noticeable differences.

\begin{figure}[ht]
	\begin{subfigure}{0.495\textwidth}	\includegraphics[width=\linewidth]{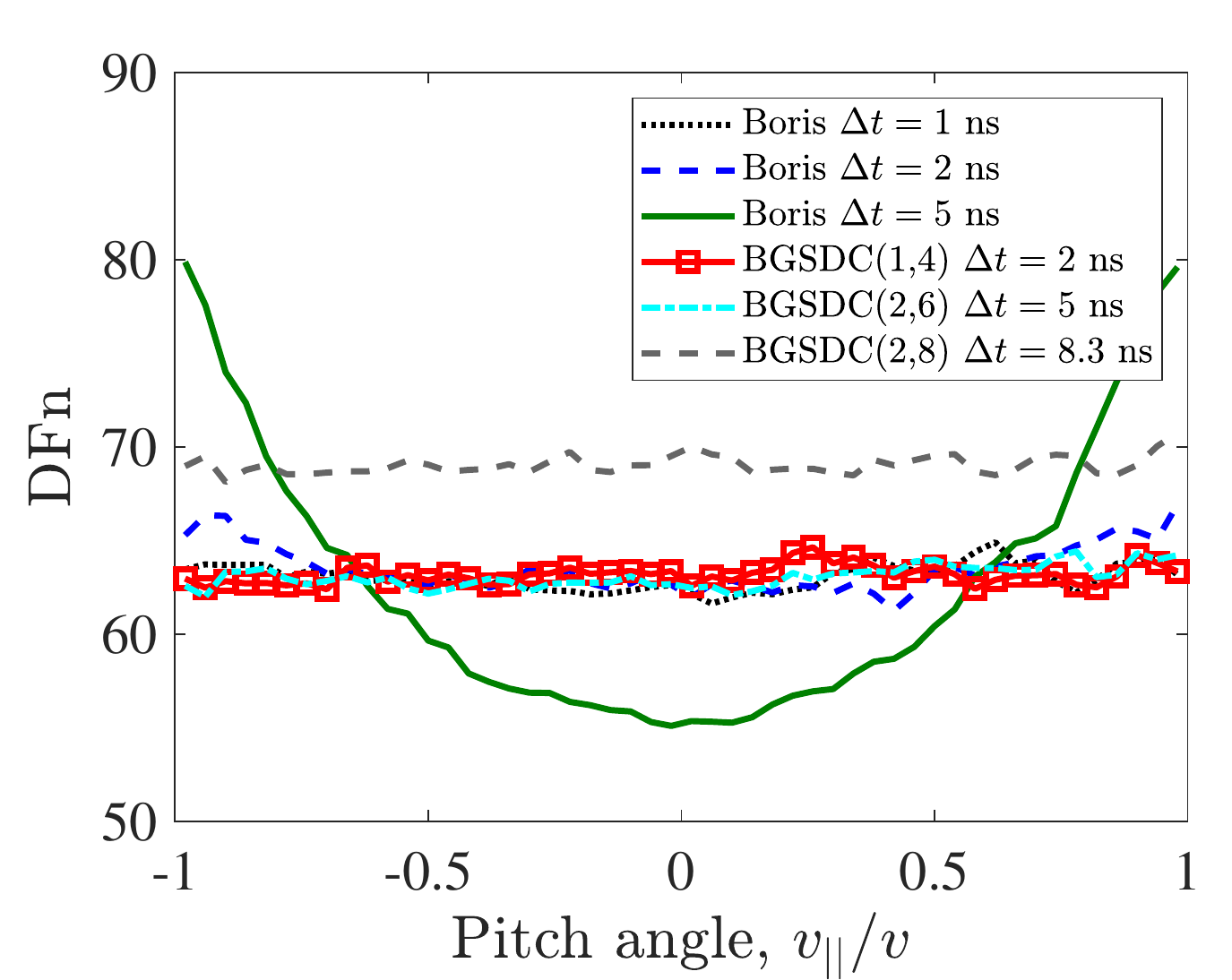}
	\end{subfigure}
	\begin{subfigure}{0.495\textwidth}
		\includegraphics[width=\linewidth]{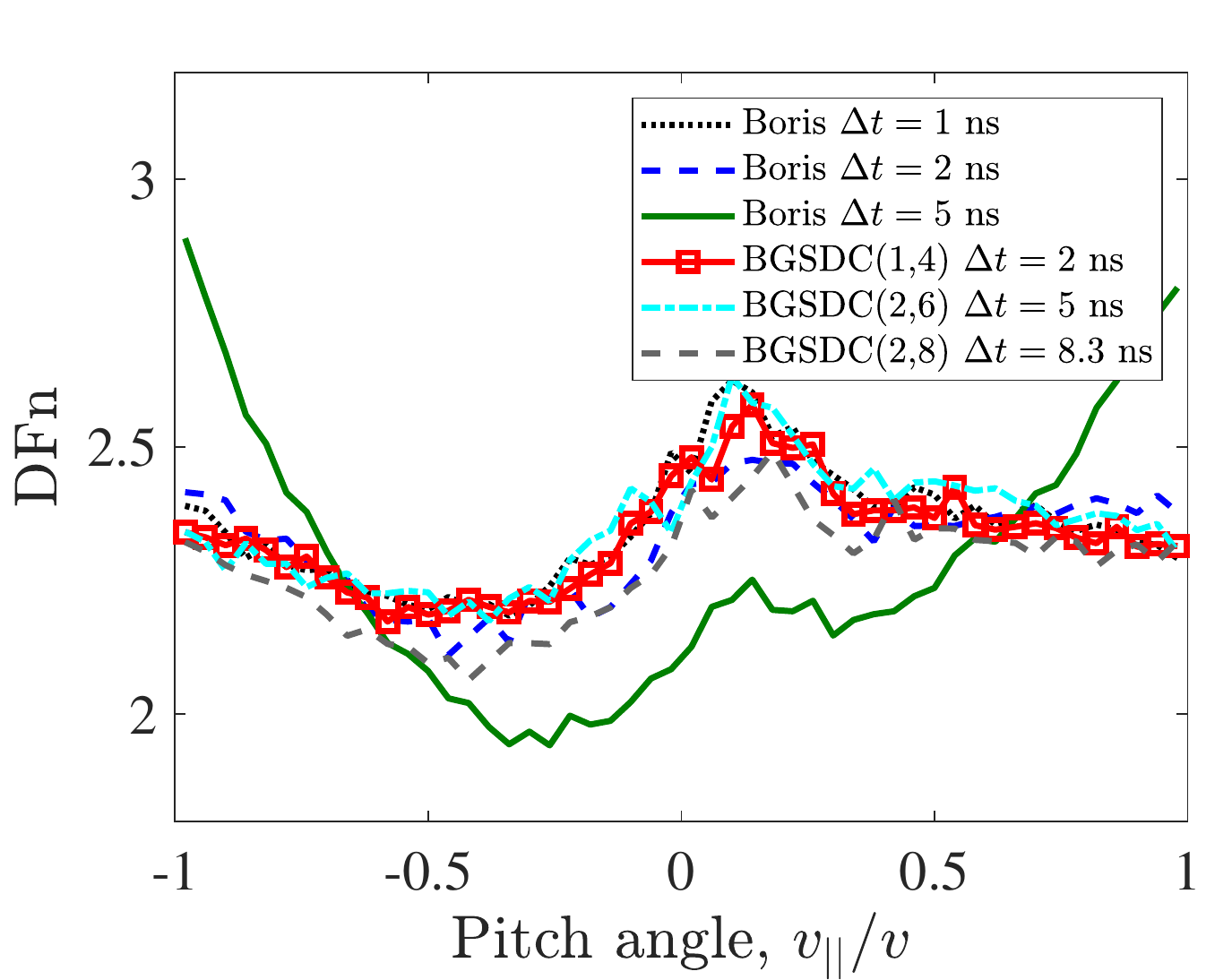}
	\end{subfigure}
	\caption{1D profiles of the distribution function against pitch angle at fixed $v=0.1\cdot10^6, R=2.2, Z=-0.1$ on the left and $v=0.5\cdot10^6, R=2.2, Z=-0.5$ on the right.}
	\label{fig:jet_1dprofL}
\end{figure}

\begin{figure}[ht]
	\begin{subfigure}{0.495\textwidth}	\includegraphics[width=\linewidth]{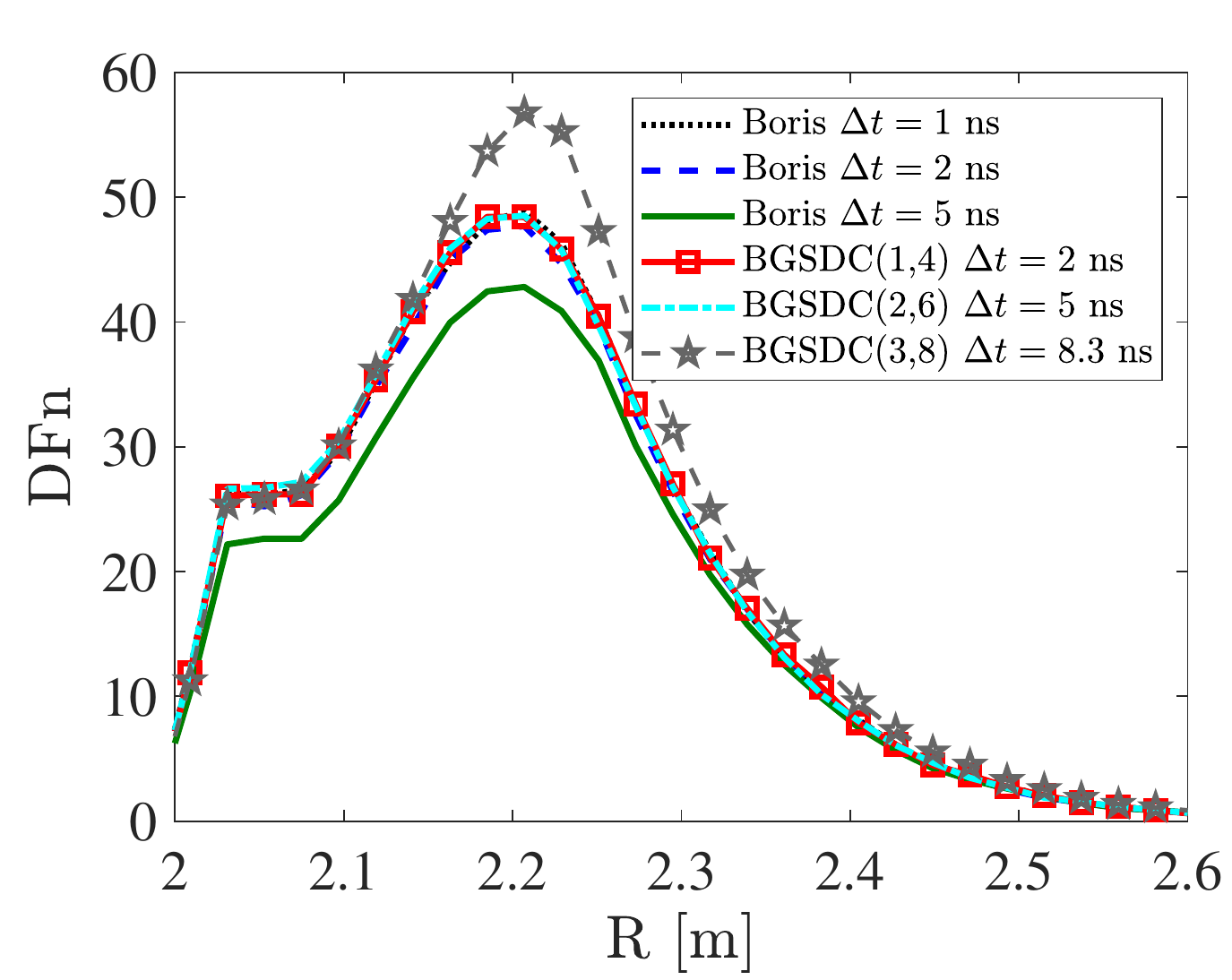}
	\end{subfigure}
	\begin{subfigure}{0.495\textwidth}
		\includegraphics[width=\linewidth]{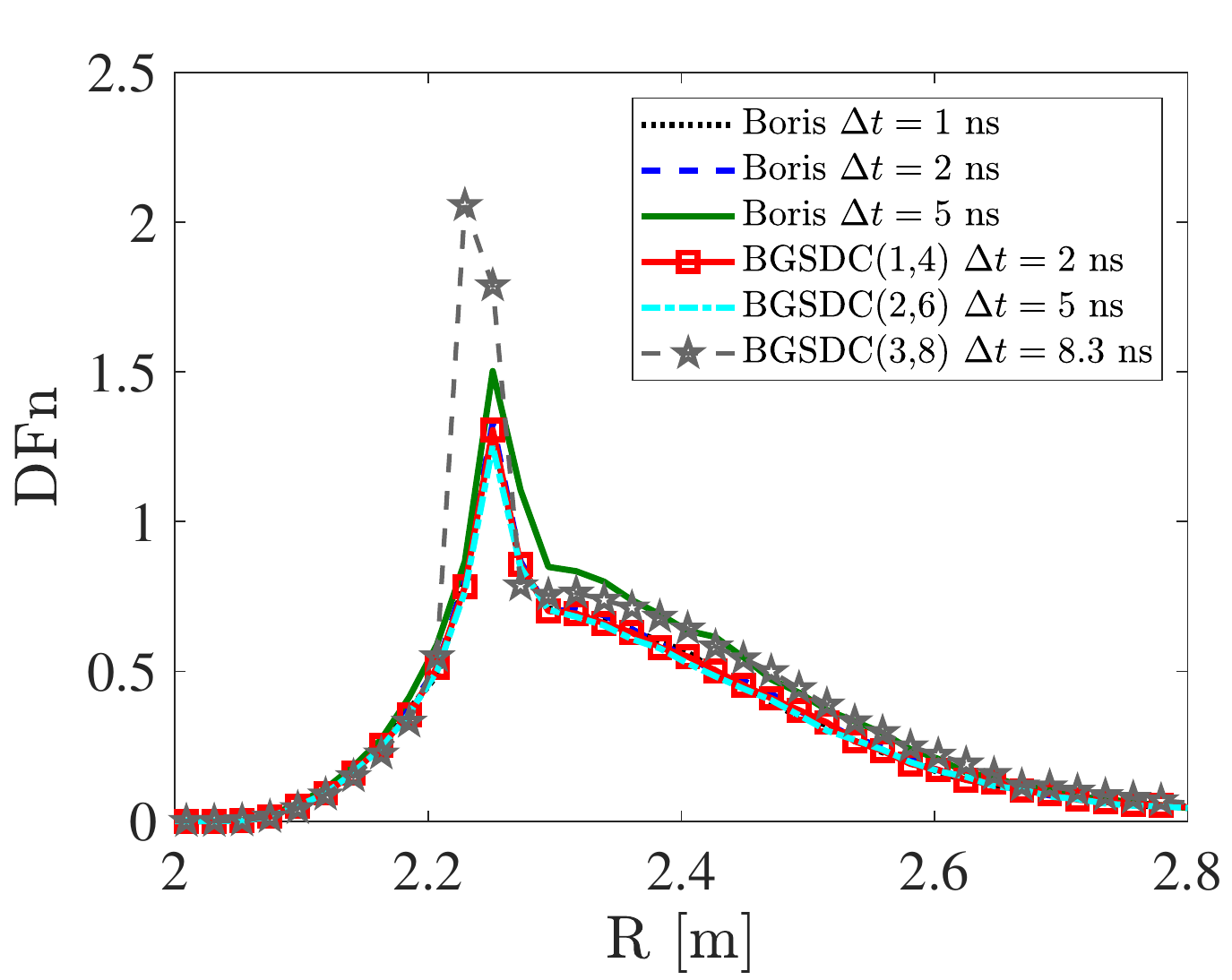}
	\end{subfigure}
	\caption{1D profiles of the distribution function along major radius with $v=0.2\cdot10^6, L=0.2, Z=0.3$ on the left and $v=0.8\cdot10^6, L=-0.9, Z=0.1$ on the right.}
	\label{fig:jet_1dprofR}
\end{figure}

Fig.~\ref{fig:jet_1dprofL} and~\ref{fig:jet_1dprofR} show 1D profiles of the distribution function in a phase box with fixed velocity and position.
In the slice shown in Fig.~\ref{fig:jet_1dprofL}, BGSDC with $\Delta t = 2$ and $\Delta t = \SI{5}{\nano\second}$ and Boris method with $\Delta t = \SI{1}{\nano\second}$ produce similar profiles of $F(L)$. 
Small deviations can be seen on both ends for the blue line (Boris with $\Delta t = \SI{2}{\nano\second}$) in the left picture and in the centre of the right picture. 
Boris with time step $\geq \SI{5}{\nano\second}$ shows significant differences in the $F(L)$ profiles and the radial profiles $F(R)$ shown in Fig.~\ref{fig:jet_1dprofR}. 
Similar results have been observed at different values of $v, R, Z$ and $L$ but are not documented here.

\reva{The shown results suggest that, for the collisional case, BGSDC(2,6) can use a time step about 5 to 10 times larger than the Boris integrator and produce results that look very similar.
This is not enough to offset the fact that it needs 19 instead of one right hand side evaluations.
If the distributions are to be reproduced with better accuracy (probably around 1\% or better), we would expect BGSDC to become competitive, as was observed in the collisionless case.
However, as we do not currently have a useful quantitative measure for accuracy for the collisional case, this analysis is left for future work.}

\section{Conclusions}
We compare the performance of the BGSDC time stepping algorithm against the standard Boris integrator when computing trajectories of fast ions generated by neutral beam injection into a fusion reactor. 
Numerical examples are shown for the DIII-D and JET reactors and include non-collisional and collisional models with plasma.
For the non-collisional case, the model is deterministic and we can use the standard deviation of the numerical drift distribution across all particles as a metric for solution quality.
The results in this paper show that for both DIII-D and JET, BGSDC produces a substantially narrower and thus better distribution \reva{with smaller mean and standard deviation} than Boris at the same time step. 
BGSDC can provide computational gains compared to classical Boris when trajectory simulations need to be of high precision with \reva{means and} standard deviations of the order of \SI{1}{\micro\meter} or lower.
\reva{A recent study of toroidal Alfv\'{e}n eigenmodes (TAE) with LOCUST required time steps of less than \SI{0.5}{\nano\second} to deliver converged results~\cite{FitzgeraldEtAl2020}.
Although no drift distribution was computed in these simulations, in our setup this would correspond to a standard deviation of around $\sigma \approx \SI{10}{\micro\meter}$.
For studies involving higher frequencies and sharper spatial features, even smaller time steps will be required.
Therefore, while most simulations do not yet require such highly accurate distribution, this will likely change in the near future.}

Detailed quantitative assessment of particle trajectories in the presence of collisions is left for future work, as there are no mathematical tools established in the fusion community to define a precise notion of accuracy in these cases.
However, we show that BGSDC and Boris converge to similar distributions and that BGSDC provides stable distributions at larger time steps than Boris.

\section*{References}

\bibliography{sdc,refs}

\end{document}